\documentclass[a4paper,fleqn,usenatbib]{mnras}

\usepackage{newtxtext,newtxmath}

\usepackage[T1]{fontenc}
\usepackage{ae,aecompl}

\usepackage{graphicx}	
\usepackage{amsmath}	
\usepackage{amssymb}	
\usepackage{multirow}
\usepackage{color}
\usepackage{subfig}

\title[Deep learning for galaxy surface brightness profile fitting]{Deep learning for galaxy surface brightness profile fitting}

\author[D. Tuccillo et al.]{D. Tuccillo,$^{1,2}$\thanks{E-mail:diego.tuccillo@obspm.fr} 
M. Huertas-Company,$^{3,1,4}$
E. Decenci\`{e}re,$^{2}$
S. Velasco-Forero,$^{2}$
\newauthor
H. Dom\'{i}nguez S\'{a}nchez,$^{3,1}$
and P. Dimauro$^{1}$
\\
$^{1}$LERMA, Observatoire de Paris, PSL Research University, CNRS, Sorbonne Universit\'{e}s, UPMC Univ. Paris 06, F-75014 Paris, France\\
$^{2}$MINES Paristech, PSL Research University, Centre for Mathematical Morphology, Fontainebleau, France\\
$^{3}$ Department of Physics and Astronomy, University of Pennsylvania, Philadelphia, PA 19104, USA\\
$^{4}$ University of Paris Denis Diderot, University of Paris Sorbonne Cit\'{e} (PSC), 75205 Paris Cedex 13, France
}

\date{Published on MNRAS: 11 December 2017}

\pubyear{2017}

\begin{document}
\label{firstpage}
\pagerange{\pageref{firstpage}--\pageref{lastpage}}
\maketitle

\begin{abstract}
Numerous ongoing and future large area surveys (e.g. DES, EUCLID, LSST, WFIRST), will increase by several orders of magnitude the volume of data that can be exploited for galaxy morphology studies. The full potential of these surveys can only be unlocked with the development of automated, fast and reliable analysis methods. In this paper, we present \textit{DeepLeGATo}, a new method for two-dimensional photometric galaxy profile modeling, based on convolutional neural networks. Our code is trained and validated on analytic profiles (HST/CANDELS F160W filter) and it is able to retrieve the full set of parameters of one-component S\'ersic models: total magnitude, effective radius, S\'ersic index, axis ratio. We show detailed comparisons between our code and GALFIT. On simulated data, our method is more accurate than GALFIT and $\sim 3000$ time faster on GPU ($\sim 50$ times when running on the same CPU). On real data, \textit{DeepLeGATo} trained on simulations behaves similarly to GALFIT on isolated galaxies. With a fast \textit{domain adaptation} step made with the $0.1-0.8$ per cent the size of the training set, our code is easily capable to reproduce the results obtained with GALFIT even on crowded regions. \textit{DeepLeGATo} does not require any human intervention beyond the training step, rendering it much automated than traditional \textit{profiling} methods. The development of this method for more complex models (two-component galaxies, variable PSF, dense sky regions) could constitute a fundamental tool in the era of big data in astronomy. 
\end{abstract}

\begin{keywords}
methods: data analysis; catalogues; galaxies: high-redshift; galaxies: structure
\end{keywords}

\section{Introduction}

The characterization of the structure of galaxies inferred from their surface brightness distribution is a powerful tool in astronomy. The earliest studies on galaxy structural characterization lead to the discovery of the de Vaucouleurs profile (\citealt{Vaucouleurs_1958}) for simple one-dimensional intensity profile fitting. Subsequently, the fitting law was generalized by \cite{Sersic_1968} and increasingly complicated one-dimensional component fitting came in work by \cite{Kormendy_1977} and \cite{Kent_1985}, where galaxies were decomposed into distinct components rather than into a single light profile.

Nowadays the description of the galaxy structure is often obtained with software of {\it{profile fitting}} (or {\it{profiling}}), that fit the surface light distribution of the galaxy with analytic functions (either parametric or non-parametric) in order to obtain a set of simple parameters that would ideally allow the reconstruction of the 2D photometric shape of the galaxy. Computing structural parameters for large samples of galaxies allows to derive more robust scaling relations at low and high redshift (\citealt{Bernardi_2013}, \citealt{vanDerWel_2014}) as well as to test theoretical models. Studies of the scaling relations of different galaxy components also rely on robust and reproducible methods to measure and describe galaxy structure. The images provided by large area surveys like the Sloan Digital Sky Survey (SDSS; \citealt{York_2000}), the Galaxy And Mass Assembly (GAMA; \citealt{Driver_2009}) or at high redshift by CANDELS (\citealt{Grogin_2011}, \citealt{Koekemoer_2011}),  have been effectively used to study the size distribution of galaxies and its dependence on their luminosity  (\citealt{Shen_2003}, \citealt{Lange_2016}, \citealt{Bernardi_2013}). The large databases provide accurate statistics when investigating the distribution of mass and luminosity-surface brightness relation for different classes of galaxies (\citealt{Driver_2007}; \citealt{Kelvin_2014};  \citealt{Kennedy_2016}); \citealt{vanDerWel_2014}.
 
The era of big data in astronomy is marked by the numerous current and future large area surveys like EUCLID, the Large Synoptic Survey Telescope (LSST, \citealt{Abell_2009}, the Wide Field Infrared Survey Telescope (WFIRST), Kilo-Degree Survey (KiDS, \citealt{Jong_2013}) and Dark Energy Survey (DES). These surveys will increase by several orders of magnitude, in a few years, the volume of data that can be exploited for galaxy morphology studies, offering a unique opportunity to constrain models and infer properties of galaxies. In fact, the sheer number of galaxies available with morphological information and photometric or spectroscopic redshifts will allow precise studies of the rarest populations of Active Galactic Nuclei (AGNs) and galaxies, like massive early-type galaxies at high redshift. Improving studies on the co-evolution of their multi-variate distribution functions (luminosity, mass function, stellar mass, etc.). The large volume probed by the new surveys will also make it possible to map the small- and large-scale galactic environment at all redshifts, and to perform, at early cosmological epochs, a statistically significant analysis of the environment effect on the galaxy and AGN properties.
The full potential of these surveys can only be unlocked with the development of  automated, fast and reliable methods to describe galaxy structure. The most popular galaxy fitting codes currently used in literature, i.e. GALFIT (\citealt{Peng_2002}) and Gim2d (\citealt{Simard_2002}) have not been conceived to deal with large amounts of data and several efforts have been made to automatize their use for catalogue compilation in large survey applications. GALAPAGOS, programmed by \cite{Barden_2012}, combines SEXTRACTOR (\citealt{Bertin_1996}) for source detection and extraction, and then makes use of GALFIT for modelling S\'ersic profiles. GALAPAGOS has been proved to be robust in terms of parameter recoverability, however the results of the quality of the fitting depend heavily on the choice of the input parameters. With a similar concept, PyMorph (\citealt{Vikram_2010}) glues together GALFIT and SEXTRACTOR in a single pipeline written in Python. The urgency created by the new generation of surveys has lead to recent efforts to develop new fitting codes like ProFit (\citealt{Robotham_2017}) programmed in C++ and directly conceived to be faster than older profile fitting codes, therefore exploitable for structural analysis of large amounts of data.

Deep learning has revolutionized data analysis in the last few years (\citealt{lecun_2015}; \citealt{Schmidhuber_2015}). In the field of computer vision, convolutional neural networks (CNN) have become the dominant approach for image processing and analysis (\citealt{Krizhevsky_2012}). One of the main benefits of CNNs is that they learn representations automatically from raw inputs, recovering higher level features from lower-level ones, e.g., in images, the hierarchy of objects, parts, motifs, and local combinations of edges. In other words, whereas classical pattern recognition techniques need manual feature engineering to generate the outputs, deep learning automatically builds relevant descriptors from the pixels of the training set, not making any prior assumption on specific features of physical models of the specific problem. Another advantage is that deep learning, using distributed representations, combats the exponential challenges of the curse of dimensionality, making it extremely well suited to big data problems.

In astronomy, several groups have recently explored the application of deep learning methods. Most notably for morphological classification of galaxies, \cite{Dieleman_2015} developed a convolutional neural network (CNN) that reached accuracy  $>99$ per cent in the classification of SDSS galaxies previously classified in the context of the Galaxy Zoo project (\citealt{Lintott_2008}). A similar level of accuracy was obtained by \cite{Huertas-Company_2015} who, using a convolutional neural network, retrieved for over 50,000 unclassified CANDELS FIELDS galaxies, the probabilities of having a spheroid or a disk presenting an irregularity, being compact or a point source. CNNs have also been used for star-galaxy classification (\citealt{Kim_2017}) of SDSS data, automated spectral feature extraction (\citealt{Wang_2017}), and unsupervised feature-learning for galaxy SEDs (\citealt{Frontera_2017}).

In this work, we explore for the first time the possibility of applying CNNs for two-dimensional light profile galaxy-fitting. A deep learning approach to this problem may be extremely valuable for applications on large surveys, because it does not require any hand-made tuning previous to the application of the algorithm, thus automating the processes and greatly cutting the analysis times. In particular, we developed a CNN to decompose the galaxy structure of one-component H-band HST/CANDELS galaxies in terms of their total magnitude, radius, S\'ersic index and axis ratio. Although we obtain good and reliable results that may be already used for astronomical applications, we consider this work as a \textit{proof of concept} on the concrete possibility to apply deep learning methods for this class of astronomical analysis. 

The layout of the paper is as follows. In section 2 we present the artificial image simulations  and the real data that we use through this work, in sections 3 and 4 we describe our method and present our algorithm, that we name \textit{DeepLeGATo}.  In section 5 we compare the results obtained from \textit{DeepLeGATo} and GALFIT on a sample of 5000 simulated galaxies. In section 6 we discuss the  \textit{domain adaptation} of our code and we repeat the  comparison between \textit{DeepLeGATo} and GALFIT  on a  sample of 1000 real galaxies from the HST/CANDELS field. Finally, in section 7, we discuss the conclusion of our work and the planned future development of the method here presented.

\section{Data}
In this work  we used two sets of data. A set of  55,000 images of artificially simulated HST/CANDELS -like galaxies (section \ref{simData}), and 5,000 stamps of galaxies from HST/CANDELS (section \ref{realData}). In this section, we describe the simulations and the data, while in the following sections we will discuss how we used these data to train and test our profile fitting code.

\subsection{Real data}
\label{realData}

The Cosmic Assembly Near-IR Deep Extragalactic Legacy Survey (CANDELS) survey is the largest project ever undertaken by the Hubble Space Telescope imaging data products and consist of five multi-wavelength sky regions fields, each with extensive multiwavelength observations. The core of CANDELS data consists of imaging obtained in the near-infrared by the Wide Field Camera 3 (WFC3/IR) camera, along with images in the visible-light obtained with the Advanced Camera for Surveys (ACS) camera. The CANDELS/Deep survey covers $\sim125$ square arcminutes within GOODS-N and GOODS-S, while the remaining consists of the CANDELS/Wide survey, achieving a total of $\sim800$ square arcminutes across GOODS and three additional fields (EGS, COSMOS, and UDS). A full description of the CANDELS observing program is given by \cite{Grogin_2011} and \cite{Koekemoer_2011}.

We randomly selected 5000 HST/CANDELS galaxies, with the only condition of having magnitude, radius, S\'ersic index and axis ratio spanning in the range of the parameters used to build the simulated data (see Table \ref{table1}). 

For all selected galaxies, there are structural parameters available measured with GALFIT/GALAPAGOS from \cite{vanderWel_2012}, that is focused on the WFC3 data only, i.e. on the three filters F105W, F125W, and F160W. In particular, we select all the galaxies of our sample in the F160W filter. 
In this work we also make use of the catalogue from (Dimauro et al. 2017), presenting the bulge-disc decompositions for the surface brightness profiles of 17.600 H-band selected galaxies in the CANDELS fields ($F160W < 23$, $0 < z < 2$) in 4 to 7 filters covering a spectral range of $430-1600nm$.

\begin{figure*}
\centering
\subfloat{
  \includegraphics[width=88mm]{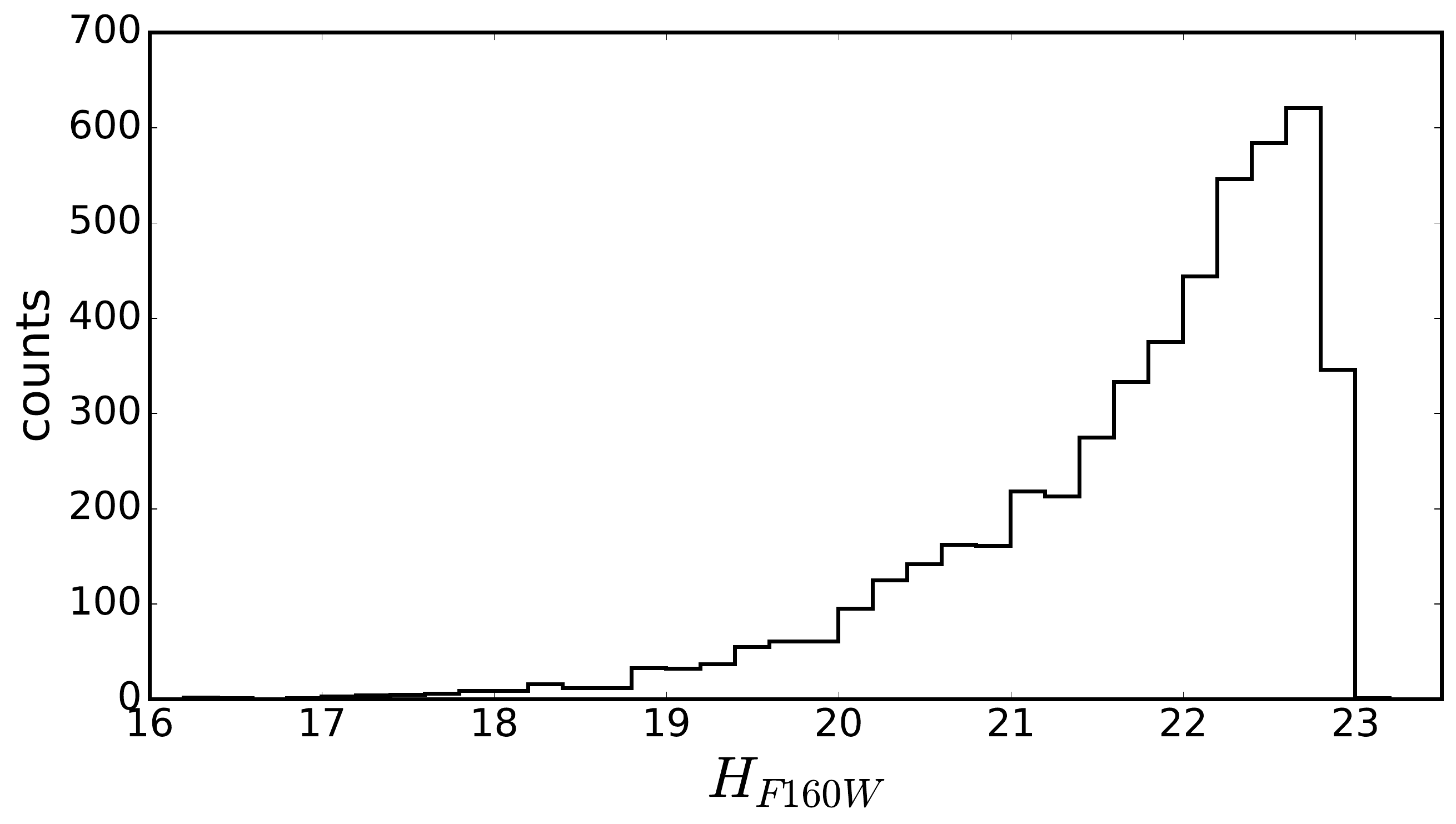}
}
\subfloat{
  \includegraphics[width=88mm]{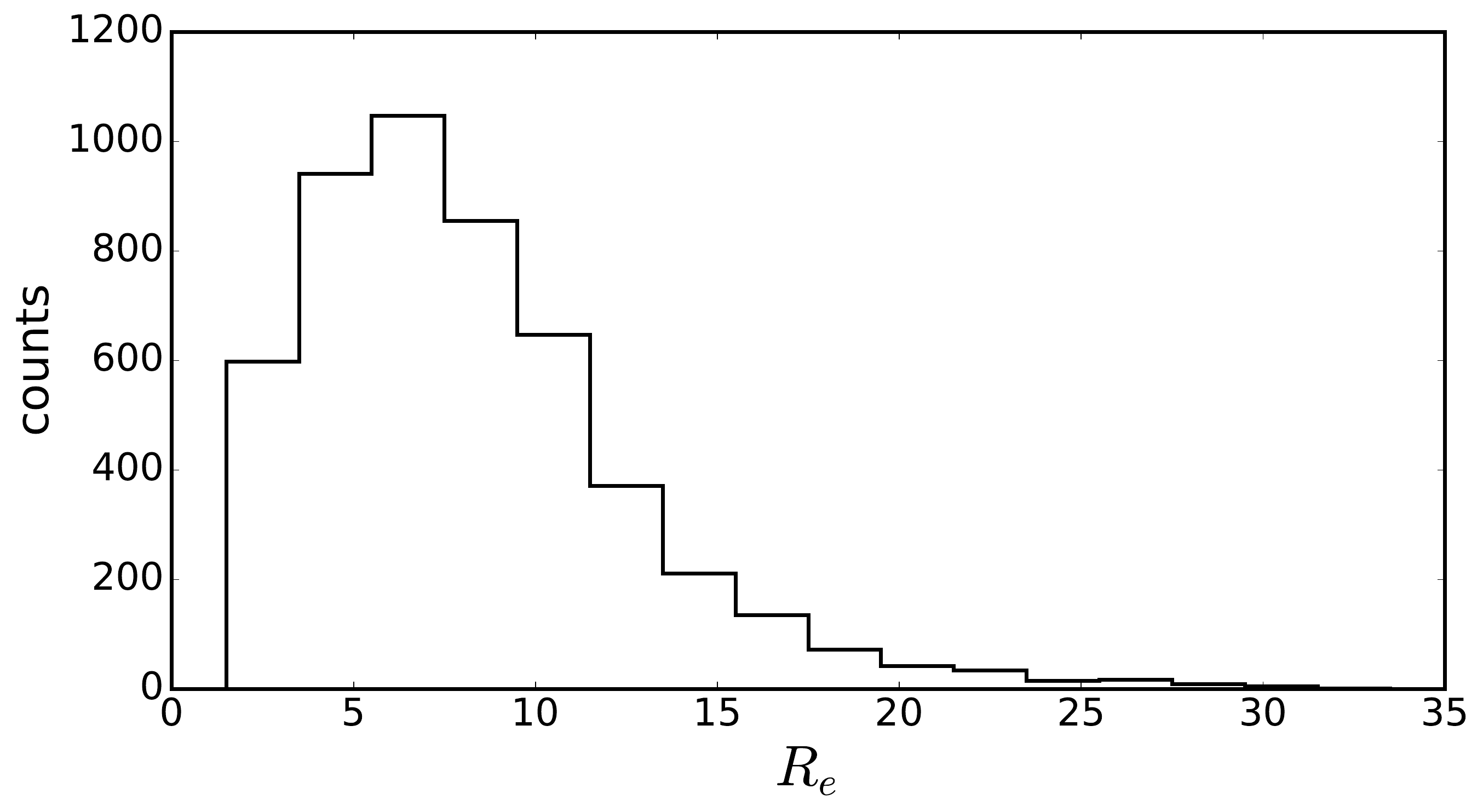}
}
\hspace{0mm}
\subfloat{
\includegraphics[width=88mm]{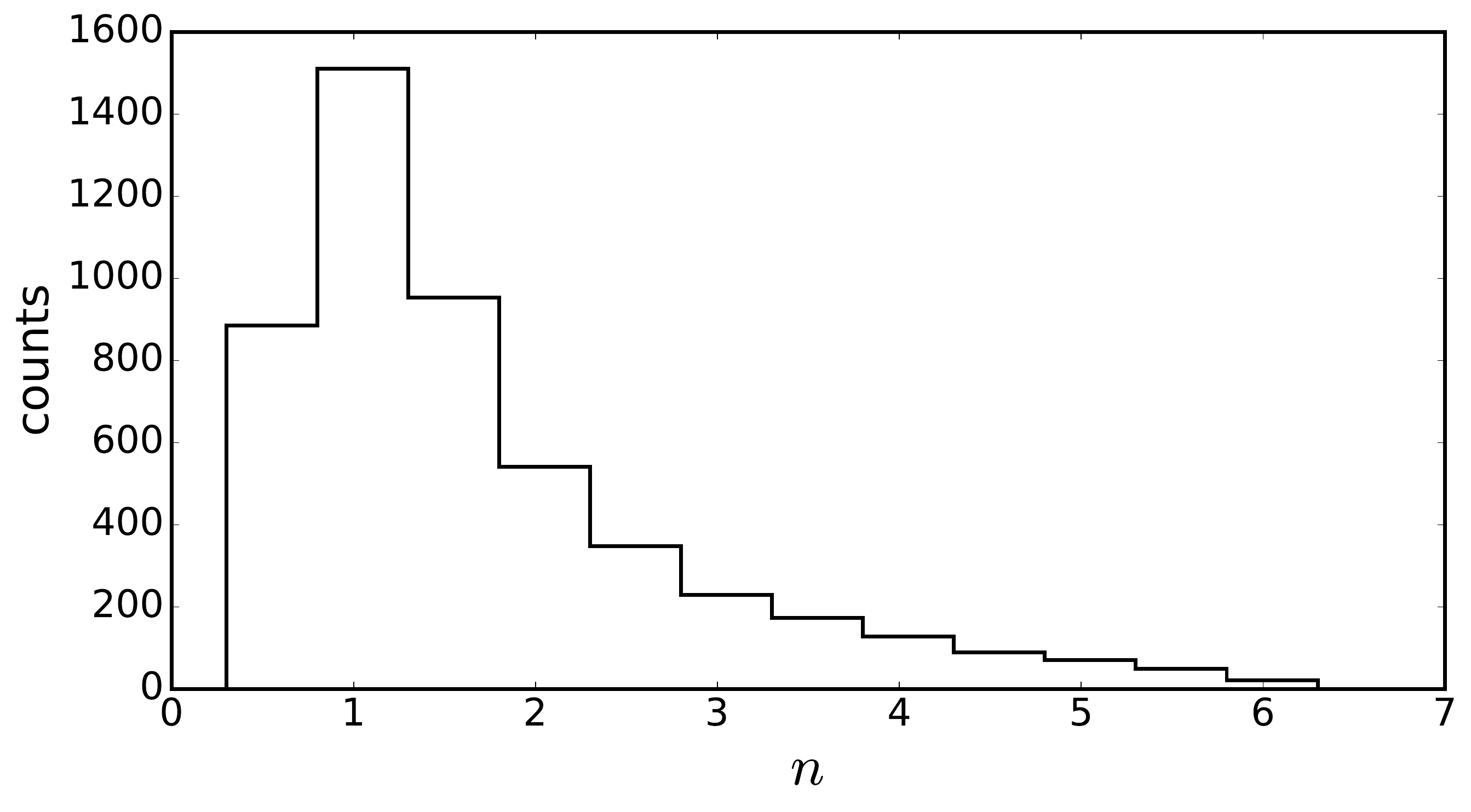}
}
\subfloat{
\includegraphics[width=88mm]{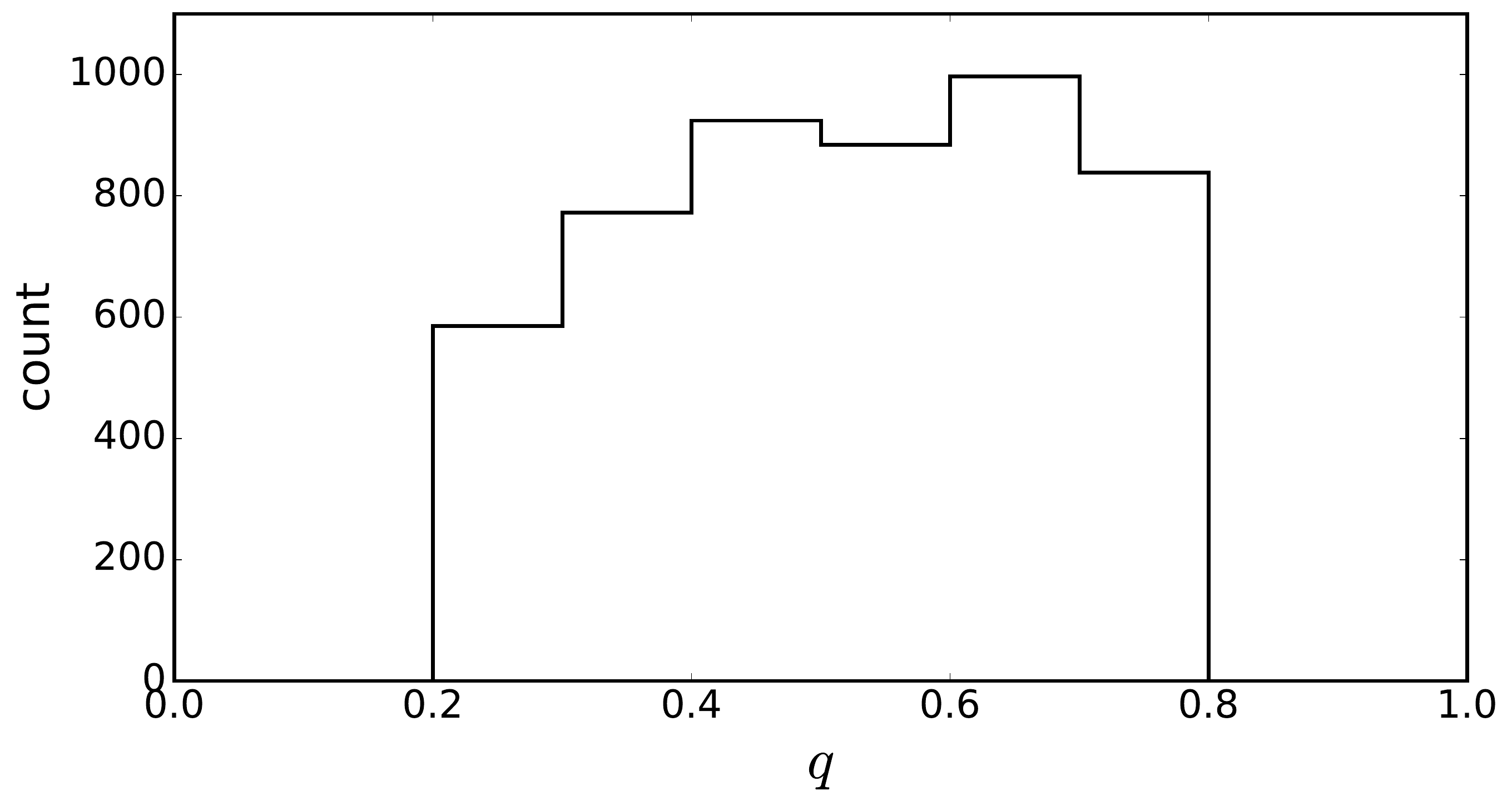}
}
\caption{Distributions in Magnitude ($H_{F160W}$), half light radius in units of pixels ($R_e$), S\'ersic index ($n$) and axis ratio ($q$) of the 5000 HST/CANDELS galaxies used in this work. See section \ref{realData}}
\label{histRealData}
\end{figure*}

In Fig. \ref{histRealData} we show the distributions of the GALFIT- derived parameters of our sample.  As we will discuss in the next sections, we used different subsets of real galaxies in order to test our code and to transfer the learning acquired on simulated data, to the case of real galaxies (process known as \textit{domain adaptation}).
 
\subsection{Simulated data}
\label{simData}

We simulated 55,000 galaxies using  one-component Sersic models. The algorithm that we used for the simulations is the 1.4 version of \textit{GalSim} (\citealt{Rowe_2015}), an open-source software whose bulk of the calculations is carried out in C++. In order to simulate realistic galaxy images, we used real PSF and real noise from the HST/CANDELS F160W filter (H band). Our dataset was obtained uniformly varying magnitude, radius, axis ratio,  S\'ersic index and position angle of the galaxies. All the stamps were generated in order to have a size of $128 \times 128$ pixels and a pixel scale of $0.06^{''}$.  

The total surface brightness of the galaxies is given from the integral over the galaxy area of the flux intensity distribution $I(r)$ measured in units of luminosity per unit area at position (x,y), i.e.  $2\pi \int I(r) r dr$. As conventionally, we express the surface brightness in units of magnitudes per square arcsec, that is related to the physical surface brightness profile through $ mag \propto -2.5 \log_{10} I(r)$. 
The radial surface brightness profile of a galaxy is described by the S\'ersic function given by:
\begin{equation}
I(r) = I(0) \exp[-b_n (r/r_e)^{1/n})]
\end{equation}

where $r_e$ is the effective radius of the galaxy, $b_n$ is a free parameter which ensures the correct integration properties at $r_e$ and $n$ is the S\'ersic index, that describes the brightness concentration curvature of the galaxy. 

As summarized in Table \ref{table1}, we simulated galaxies ranges $16 \le mag\le 23$ and S\'ersic  index $0.3 \le  n \le 6.2$. The radius of the galaxies varies in the interval $ 1.5 \le Radius (pixel) \le 31.6 $ ($ $0\farcs09$ \le Radius (arcsec) \le 1\farcs9$). The axis ratio($q$) of the galaxies ranges within $0.2$ and $0.8$.

\begin{table*}
\centering
\caption{We simulated 55,000 realistic one-component F160W filter (H band) HST/CANDELS galaxies, having structural parameter within these ranges of values.}
\label{table1}
{
\begin{tabular}{c c c c c}
\hline
 & Magnitude & Radius & S\'ersic index & axis ratio \\
 & (AB) & (Pixels) &  &  \\ 
 \hline
Range  & $16-23$ & $1.5-31.6$ & $0.3-6.2$ & $0.2-0.8$ \\
\hline
\end{tabular}}
\end{table*}

As we will see in the next sections, we used 50,000 simulated galaxies to train and validate our code, and a separated set of other 5000 simulated galaxies to compare the performance of our method with GALFIT.

\section{Introduction to Convolutional Neural Networks}

Deep artificial neural networks are a particular subclass of artificial neural networks that have allowed a dramatic progress in image and natural language processing in the last years. In this section, we give a very brief introduction to these methods, and we refer to  \cite{Schmidhuber_2015} for a thorough review of the field.

A standard Artificial Neural Network (ANN) consists of many simple, connected processors called neurons, each of them producing a sequence of real-valued activations. Neurons are organized in layers, with the neurons of the input layer directly activated from the features of the input data, and the neurons of the hidden layers activated through weighted connections from previously activated neurons. Analogous to biological systems, where learning involves adjustments in the synaptic connections that exist between neurons, in artificial neural networks the learning process consists in finding the weights that infer the desired output from the features of the input data. Therefore the tensorial output $\mathbf{h}^l$ of a given layer $l$ of the network is given by:
 
 \begin{equation}
 \mathbf{h}^l = f( \mathbf{W}^l \cdot  \mathbf{h}^{l-1} + \mathbf{b}^l)
 \end{equation}
 
where $f$ is a non linear function called \textit{activation function}, $\mathbf{W}^l$ is the weight matrix and $\mathbf{b}^l$ is a biases vector. The biases correspond to the additive constants of the linear combinations that feed each neuron. The network is trained to obtain a specific output for a given input minimizing the \textit{loss function} and thus optimizing the parameters of weights and biases. We refer to \cite{Lecun_1989} for their complete description. Briefly, CNNs use convolution layers in place of general matrix multiplication in at least one of the layers, operation usually denoted as $S(i,j) = (I \ast K)(i,j)$ with $K$ being the {\textit{kernel}} of the convolution function, and the output $S(i,j)$ known as {\textit{feature map}}.  This results in local connections, with each neuron connected to a subsample of the input instead of all the inputs. After the convolution in each neuron follows an \textit{activation function} operation like in ordinary ANNs. Typical components of a CNN include one or more {\textit{max-pooling}} layers, that are subsampling layers where  the  feature  map is down-sampled. The max-pooling operation is usually obtained  by applying a max filter to (usually) non-overlapping subregions of the initial representation, that reduces the output dimensionality while keeping the most salient information. Other typical elements are the {\textit{dropout layers}} (\citealt{Nitish_2014})  which are a regularization technique introduced to reduce overfitting. The {\it{fully connected}}  layer is the final layer of a CNN, where each neuron is completely connected to the other neurons. In a CNN, each layer applies different filters, and each one of these filters detects a specific feature of the inputs. For example, when the inputs are images, different filters can learn from the raw pixels the edges of the images, then other filters can detect simple shapes of the images, that are then used to detect higher-lever features. The last fully connected layer may then be a classical ANN that uses the recovered information for classification or regression tasks. The implementation of the convolution greatly improves a machine learning system for processing grid-like topology data, like 2D grids of pixels.

\section{Method description: \textit{DeepLeGATo}}

Our surface brightness fitting code was developed as a convolutional neural network used for regression tasks, that in our problem consists in the prediction of galaxy structural parameters given their 2D fits images as input. We called it \textit{DeepLeGATo}, standing for Deep Learning Galaxy Analysis Tool. We implemented our code using \textit{Keras}  (\citealt{chollet_2015}) on top of \textit{Theano}  (\citealt{Bastien_2012}), two frameworks commonly used to build Deep Learning models. Our architectures were inspired by VGG-net (\citealt{Simonyan_2014}), the main difference lies in the use of a noise layer. We trained and tested many different CNN architectures and we got the best results with two of them, the first of them will be described in subsection \ref{Archi1} and the second in \ref{Archi2}. 

In subsection \ref{trainVal} we describe how we trained and validated our models, in the next section we discuss in detail the performance of the networks on a sample of 5000 galaxies excluded from the training and validation.

\subsection{Architecture 1}
\label{Archi1}

The architecture 1 of our model is schematically illustrated in the upper image of Fig \ref{archiPlot}. First, we apply zero-centered additive Gaussian noise to the input images (128x128 pixels), then they are processed by two 2D convolution layers with a 4x4 filter size and, finally, subsampled by a 2x2 max pooling and a dropout layer. The Gaussian layer, used both in this architecture and in the other one, add robustness to the model by decreasing its sensitivity to small changes of the input data.
Other three units follow with the same configuration of convolutional and max pooling layers, but with a growing dimensionality of the output space (i.e. the number of filters in the output) and a reducing filter size (4x4, 3x3, 2x2) in the convolutions. Only the first of these three units is followed by a dropout layer. In this architecture we have therefore a total of eight convolutional layers, four max pooling layers and two dropout layers. Each one of the convolutional layers is followed by a rectified linear unit (ReLU) step. The output of all these units is then processed through three fully-connected layers with decreasing number of neurons (128, 64, 1).

\subsection{Architecture 2}
\label{Archi2}

\begin{figure*}
\centering
\subfloat[Architecture 1]{
\includegraphics[width=140mm]{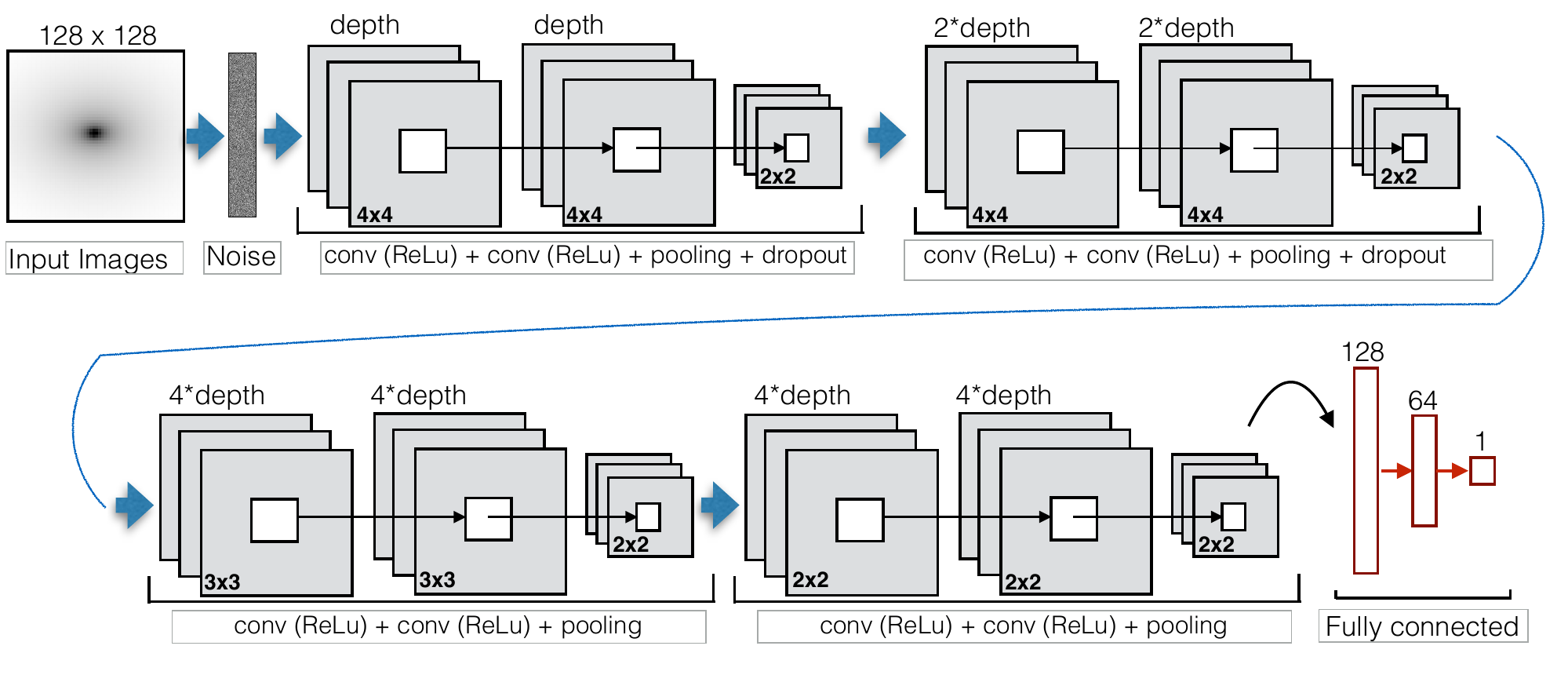}
}
\\
\hspace{0mm}
\subfloat[Architecture 2]{
\includegraphics[width=140mm]{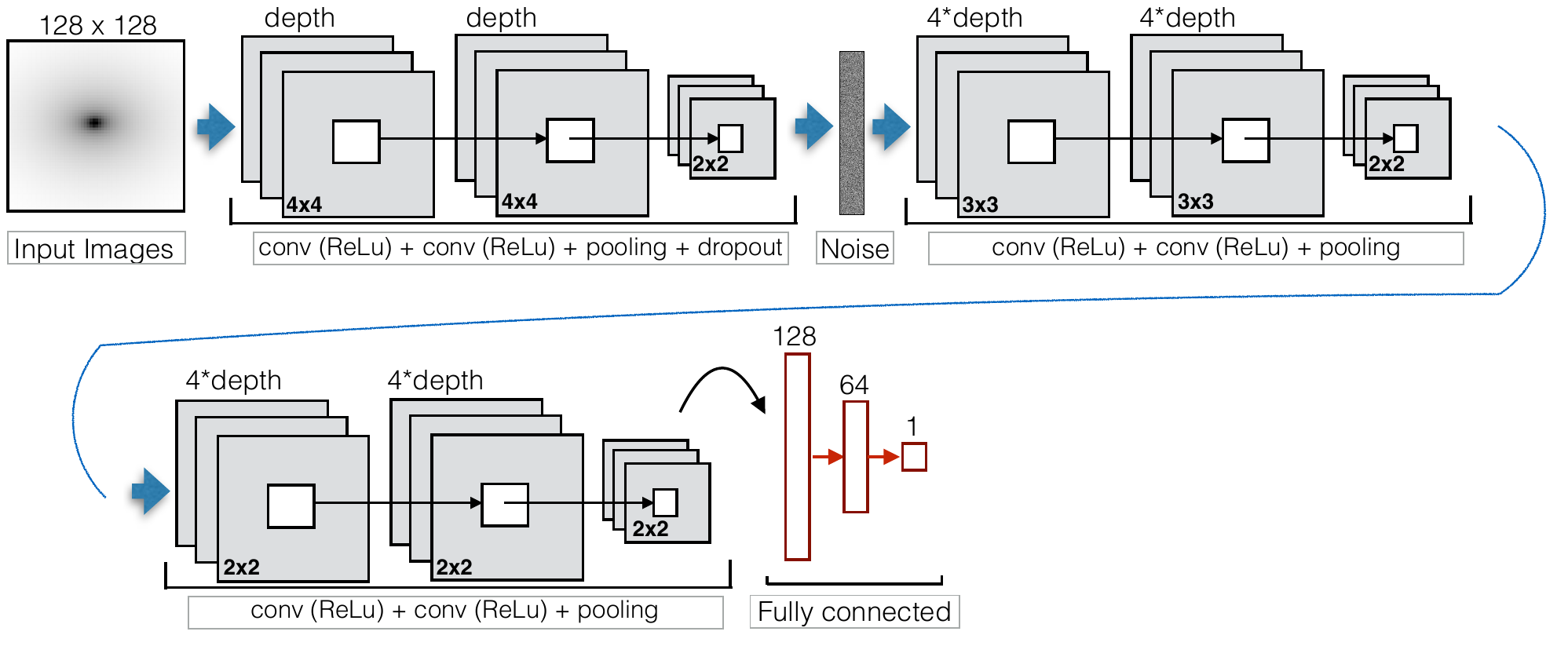}
}
\caption{The upper image (a) shows the scheme of Architecture 1 used by \textit{DeepLegato}. The input image is processed adding Gaussian noise, then the first two blocks are composed of 2 convolution layers with ReLU activation functions, followed by max pooling layers and dropout. The second two units are similar to the previous, but they do not apply dropout. The last block is composed of three connected layers, which outputs the estimation of the evaluated parameter. See section \ref{Archi1} for details. The bottom image (b) shows Architecture 2 used by \textit{DeepLegato}. Here the first block is composed of two convolution layers with ReLU activation function, followed by a max-pooling layer and a dropout. After the addition of a Gaussian noise layer, two units follow with similar architecture as the first one, but they do not apply dropout. The last block is composed of three connected layers, which outputs the estimation of the evaluated parameter. See section \ref{Archi2} for details }
\label{archiPlot}
\end{figure*}

The architecture 2 of our model is illustrated in the bottom image of Fig \ref{archiPlot}. Considering the architecture 1 as the reference, here the input images (128x128 pixels) are first processed by the block composed of two 2D convolution layers with a 4x4 filter size, and the 2x2 max pooling followed by the dropout layer. After this step, the zero-centered additive Gaussian noise is applied. Other two units follow with the same configuration of convolutional and max pooling layers, but with a growing dimensionality of the output space. Only the first of these three units is followed by a dropout layer. In this architecture we have therefore one block less respect to the architecture 1, for a total of six convolutional layers, three max pooling layers and one dropout layer. Also here, each one of the convolutional layers is followed by a rectified linear unit (ReLU) step, and the block of three fully connected layers is identical to architecture 1.

\subsection{Training and validation}
\label{trainVal}

Our CNNs have been trained and tested on the simulated data described in section \ref{simData}. Each parameter is estimated independently, i.e. each CNN is trained and tested using only one parameter at a time as target. The size of the training set was 50k, divided in the proportion of 4/5 for the training and 1/5 for the validation. For the regression problem we used a mean absolute error cost function on a normalized version of variables, i.e., values are in the interval [-1,1]. For weights updates we used an adaptive moment estimation (ADAM, \citealt{Kingma_2014}) 

In order to increase the size of the training set and make the model more robust and invariant to specific transformations, we performed \textit{data augmentation} on the training set.  Different types of \textit{image augmentation} effectively improve the quality of generalization of CNNs (\citealt{Krizhevsky_2012}). In particular, we applied:

\begin{itemize}
\item Random shifts for all parameters, used in order to make the model insensitive to centering. In particular the images were randomly shifted of 0.05 times the total width of the image.  
\item Random horizontal and vertical flips of the images for all parameters.
\item Random zoom in the the images within the range $[0.7,1.3]$ only for the regression of the radius and the S\'ersic index  . 
\end{itemize}

During the training we always initialized the weights of our model with random normal values. We \textit{warm up} the training (\citealt{Huang_2016}) of the CNN for 10 epochs, using no exponential decay rate and a starting learning rate of $0.001$. During the pre-warming and the subsequent proper fit, we use the ADAM optimizer, which improves and stabilizes the learning rate. After the \textit{warm up} phase the networks are trained using an \textit{early stopping method}, for a maximum number of 300 epochs. The early stopping method consists in stopping the training if a monitored quantity does not improve for a fixed number (called \textit{patience}) of training epochs. The quantity that we monitored and minimized was the \textit{mean absolute error} of the regression on the parameter of the validation sample, and the patience was fixed to 20.

\section{Tests on simulated data}
\label{test_simData}

The two CNNs architectures that we have described in the previous section have been chosen from among several different architectures on the basis of their good performance on the validation set. However, in order to finally judge those two architectures, we selected a third test-set of simulated galaxies excluded from the training and validation of the models. 
This way, we evaluate the two models without incurring the risk of meta-training that makes the test set work as a second training set. In subsection \ref{p_simdata} we describe the metric used for this test and the comparison of the two architectures performance. In subsection \ref{galfit_simdata} we compare the performance of best CNNs with the predictions of GALFIT run on the same data test.

\subsection{CNNs performance on simulated data}
\label{p_simdata}

The test set includes 5000 of the simulated galaxies described in section \ref{simData}. As metric to evaluate the regression, we used the \textit{coefficient of determination} $R^2$  between the predicted parameter and the ground truth is given from the galaxy simulations. The $R^2$ is a standardized measure of the degree of the regression accuracy defined as: 

\begin{equation}
R^2 = 1 - \frac{\sum_i^n (y_i - f_i)^2}{ \sum_i^n (y_i - \bar{y})^2}
\end {equation}

where $f_i$ is the predicted value of the true (or input) value of variable $y_i$, and  $\bar{y}$  is the mean of the whole set of $n$ input data. Essentially, $R^2$ measures how much better we can do in predicting $y$ by using our model instead of just using the mean as a predictor. When the predicted values come from a least-squares regression line, the coefficient of determination measures the proportion of total variation in the response variable that is explained by the regression line. In fact, in the second term of the above equation, the numerator is the squared sum of the regression errors and the denominator measures the deviations of the observations from their mean. The objective of ordinary least squared regression is to get a line which minimizes the sum squared error. The default line with minimum sum squared error is an horizontal line through the mean. So in the case a least-squares regression line is used to predict the values of $y$, then $0 \le R^2 \le 1$, with $R^2= 0 $ if the regression line explains no value and  $R^2 = 1$ if it explains the 100\% of the variation in the response variable. 
On the other hand, if we use a different model $f$ to predict the values of $y$, then $R^2$ naturally range between $- \infty \le R^2 \le 1$, with negative values meaning that an horizontal line at the mean $\bar{y}$ actually explains the data better than the model $f$.  The $R^2$ can be used as an alternative to the \textit{mean squared error}, and the two measurements are related by $R^2 = 1- \frac{MSE}{\sigma_y^2}$, where $MSE$ is the mean squared error and $\sigma_y^2$  is the squared standard deviation of the dependent variable. 

In Table \ref{Architectures} we show the comparison of the CNNs accuracies on the basis of the $R^2$ value. It is interesting that a particular network architecture works very well for a particular parameter (e.g., architecture 1 for S\'ersic index and axis ratio; architecture 2 for magnitude and half-light radius), while it is not so efficient for other. We conclude that the use of only one architecture for all the parameters does not allow to obtain the best results since some parameters need more level of abstraction than others.

\begin{table*}
\centering
\caption{The coefficient of determination of the two CNN Architectures for the parameters of the light profile fitting, for 5000 simulated galaxies excluded from the training and validation of the methods.}
\label{Architectures}
{
\begin{tabular}{l c c c}
\hline
\multicolumn{1}{c}{} &
\multicolumn{3}{c}{$R^2$ simulated data } 
\\
\hline
\multicolumn{1}{c}{Parameter} &
\multicolumn{1}{c}{Architecture 1} &
\multicolumn{1}{c}{Architecture 2 } &
\multicolumn{1}{c}{GALFIT} 
\\
\hline
Magnitude  & 0.961     &  \bf{0.997}  & 0.983 \\
Radius   & 0.899    &  \bf{0.972}  & 0.877 \\
S\'ersic index   & \bf{0.968}     &  0.881 &  0.607 \\ 
Axis ratio           &  \bf{0.983}   &   0.959        &  0.903  \\
\hline
\end{tabular}}
\end{table*}

\subsection{Comparison with GALFIT}
 \label{galfit_simdata}

In order to compare the performance of our code with standard algorithms used in the literature for galaxy surface brightness fitting, we used GALFIT on the same set of 5000 simulated galaxies used to test the CNNs architectures. The fitted parameters, listed in Table \ref{table1}, are the magnitude, half-light radius measured along the major axis, S\'ersic index, axis ratio (from which we derive the axis ratio).  As PSF image we used the same used to generate the simulations and, following a procedure similar to the one used in  \cite{vanderWel_2012}. we used SExtractor to input initial guesses for some of the parameters used in the GALFIT configuration file. In particular, SExtractor provides initial values for magnitude, half-light radius, axis ratio. A constrain file is also provided to GALFIT to force it to keep the  S\'ersic  index between 0.2 and 6.3, the effective radius between 0.3 and 130 pixels, the axis ratio between 0.0001 and 1, the magnitude, between $-3$ and $+3$ magnitudes from the input value given by the  SExtractor magnitude.

The accuracy of GALFIT is measured using the same metrics, i.e. the \textit{coefficient of determination} $R^2$, and directly compared with the CNNs performance in Table \ref{Architectures}. Additionally, in Figure \ref{PerformanceSim} we compare the best fit given from the CNN and GALFIT on the simulated data as a function of the apparent H band magnitude. In particular, we compute the mean difference between the recovered and input parameters for different magnitude bins. The error bars represent the standard deviation in each bin and reflect thus the uncertainty in the measurements.

For these tests, we run GALFIT on a MacBook Pro running Sierra with 3.1 GHz i7 processors and 16 GB of RAM, and our CNNs on  Nvida Titan X GPU. GALFIT takes approximately 3.5 hours to fit 5000 galaxies. Our code, once trained, take less than 4 seconds on the GPU and about 200 seconds on the same CPU machine where we tested GALFIT.

From these comparisons on the simulated data between GALFIT and our CNNs, we show that our models are considerably faster and provide one component galaxy structure decomposition equally or more accurate than GALFIT in the range of magnitude, effective-radius and the S\'ersic  index used to generate our simulated data.

\begin{table*}
\begin{tabular}{c c}
Results using \textit{DeepLeGATo} & Results using GALFIT\\
\includegraphics[width=8.5cm]{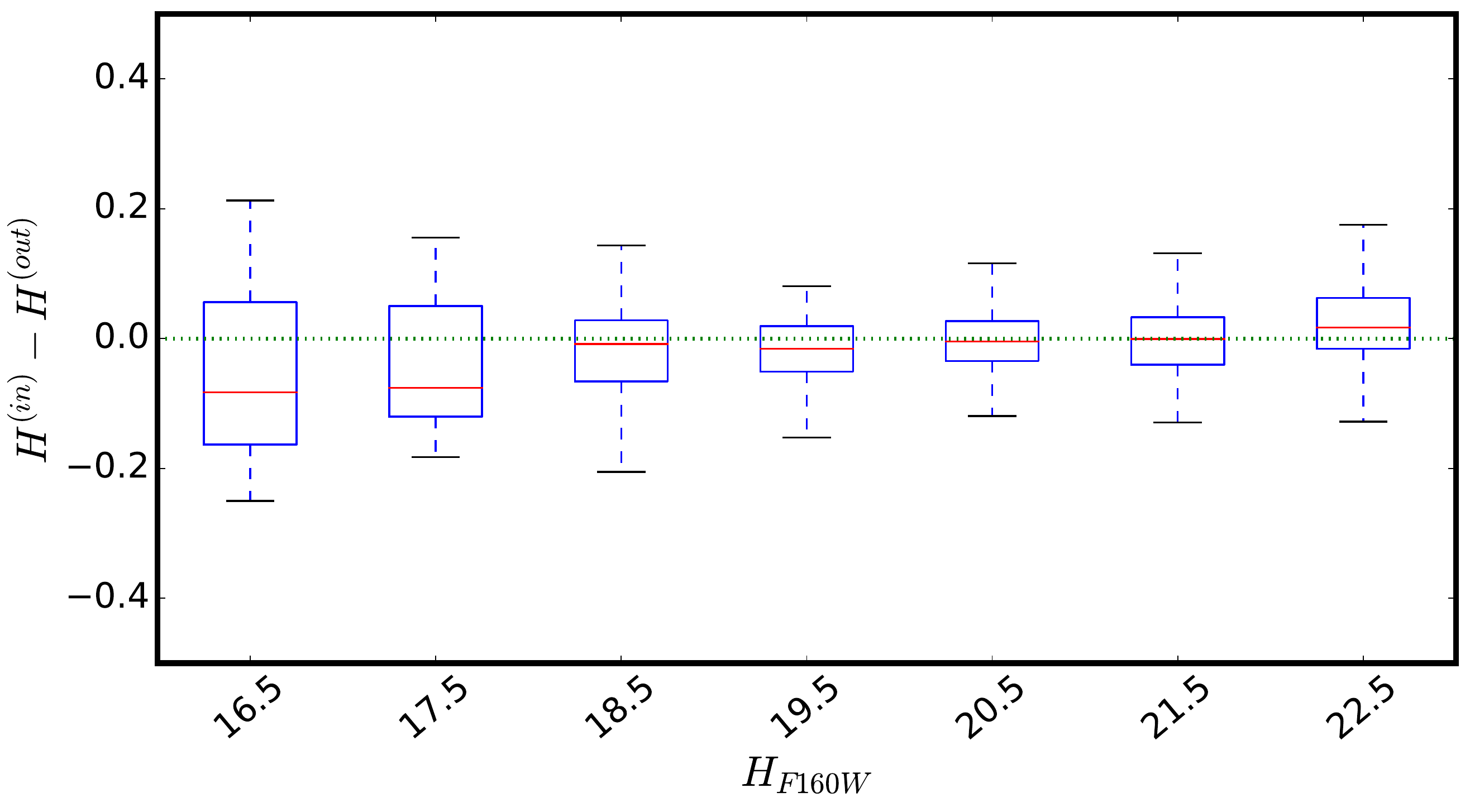} & \includegraphics[width=8.5cm]{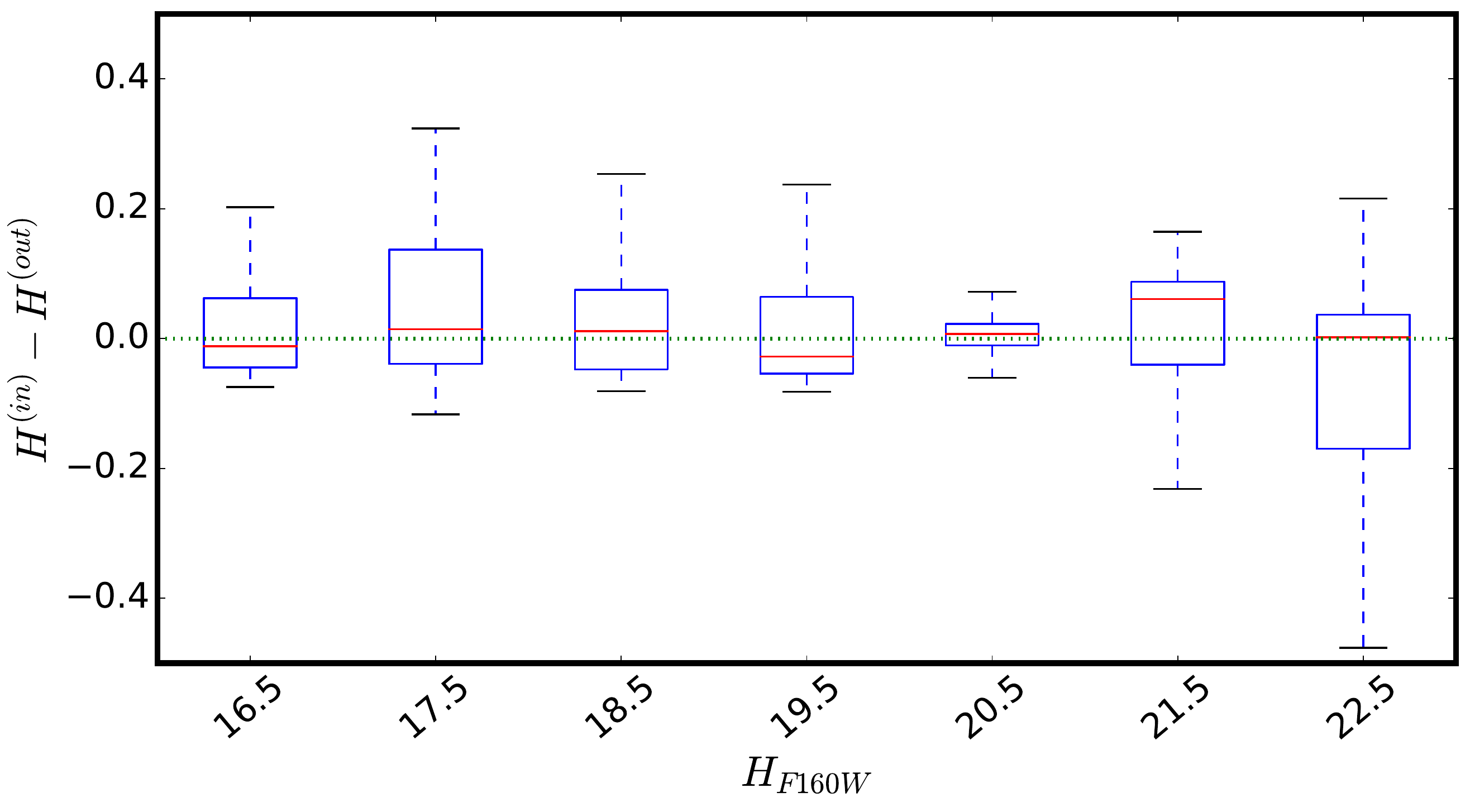} \\

\includegraphics[width=8.5cm]{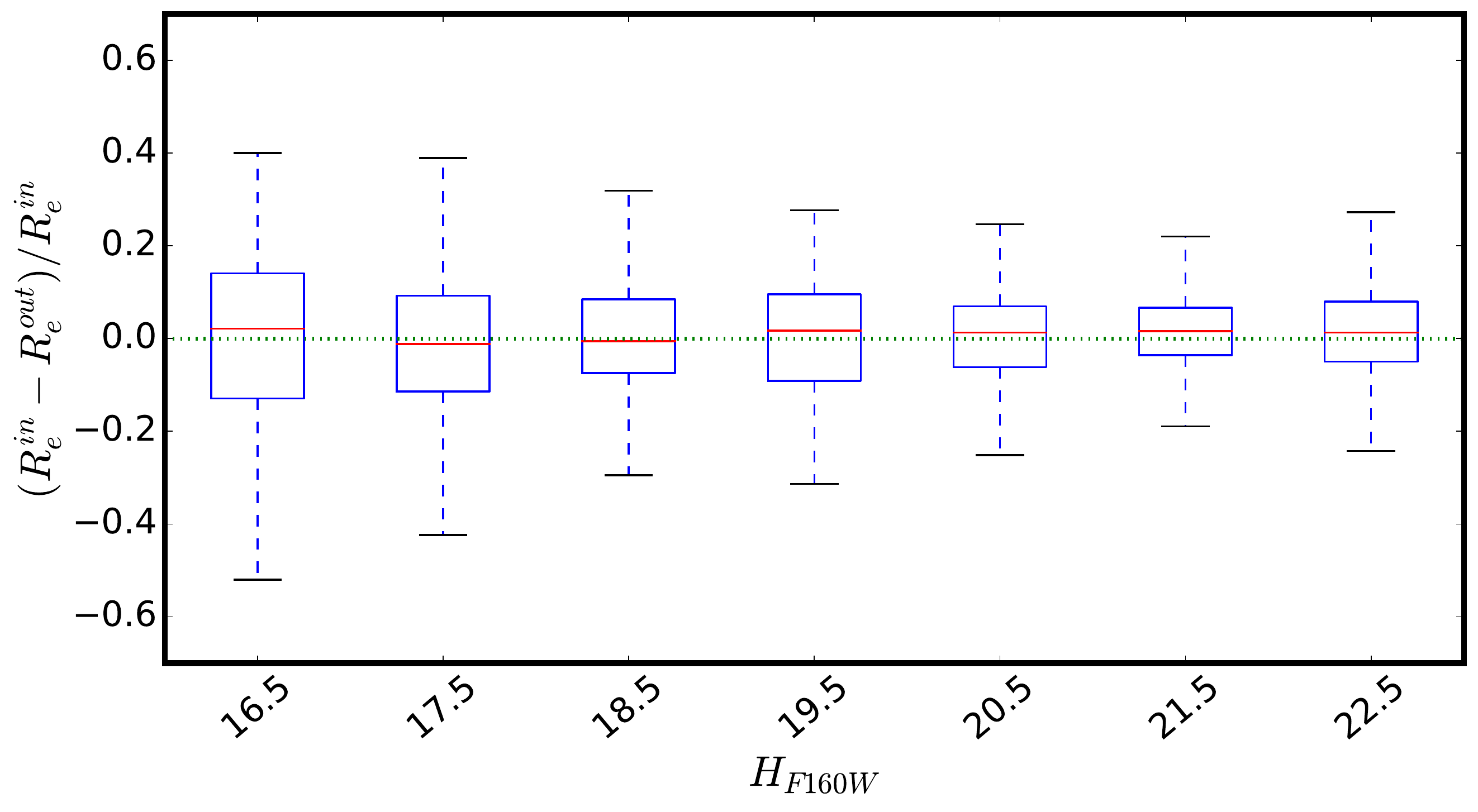} & \includegraphics[width=8.5cm]{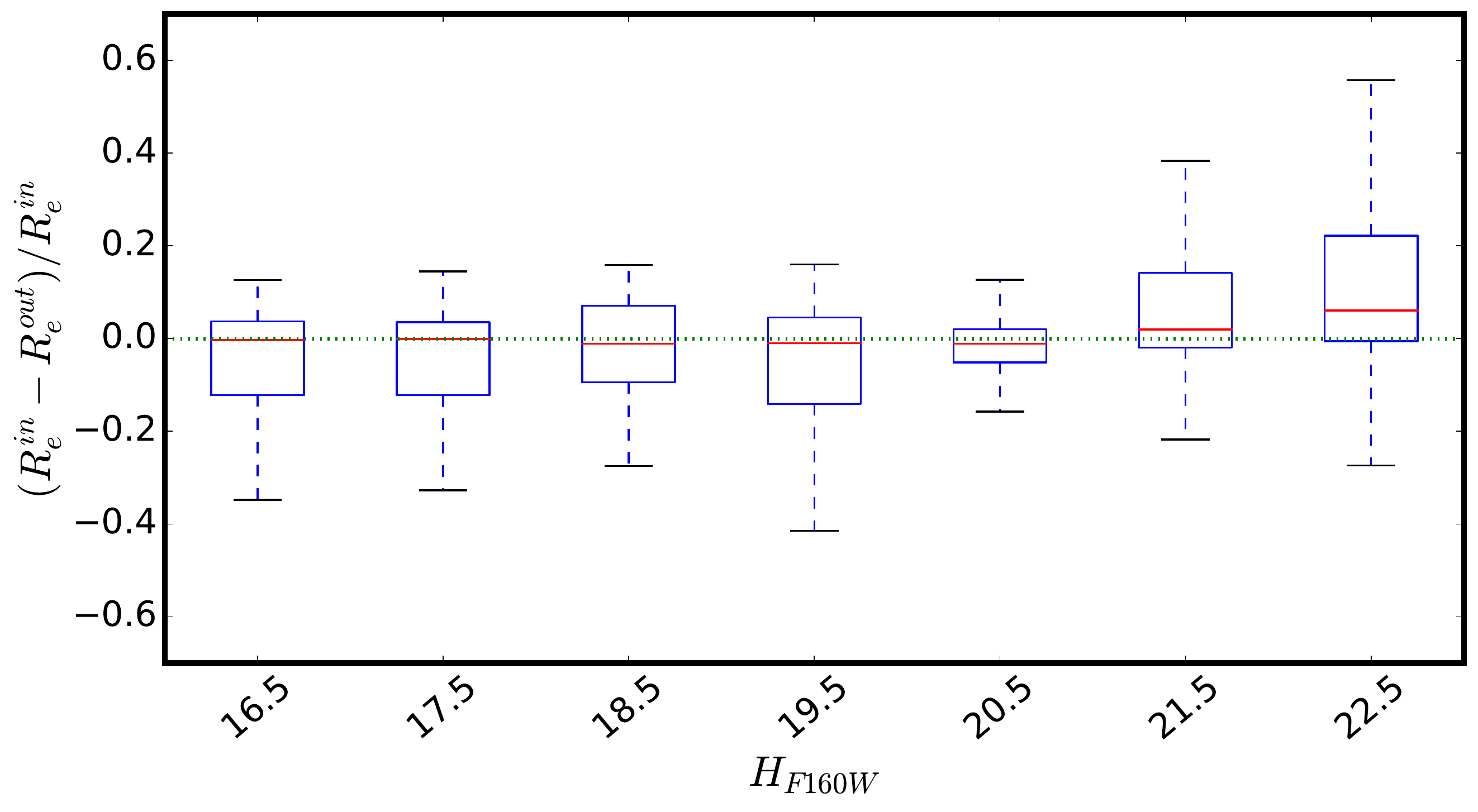} \\

\includegraphics[width=8.5cm]{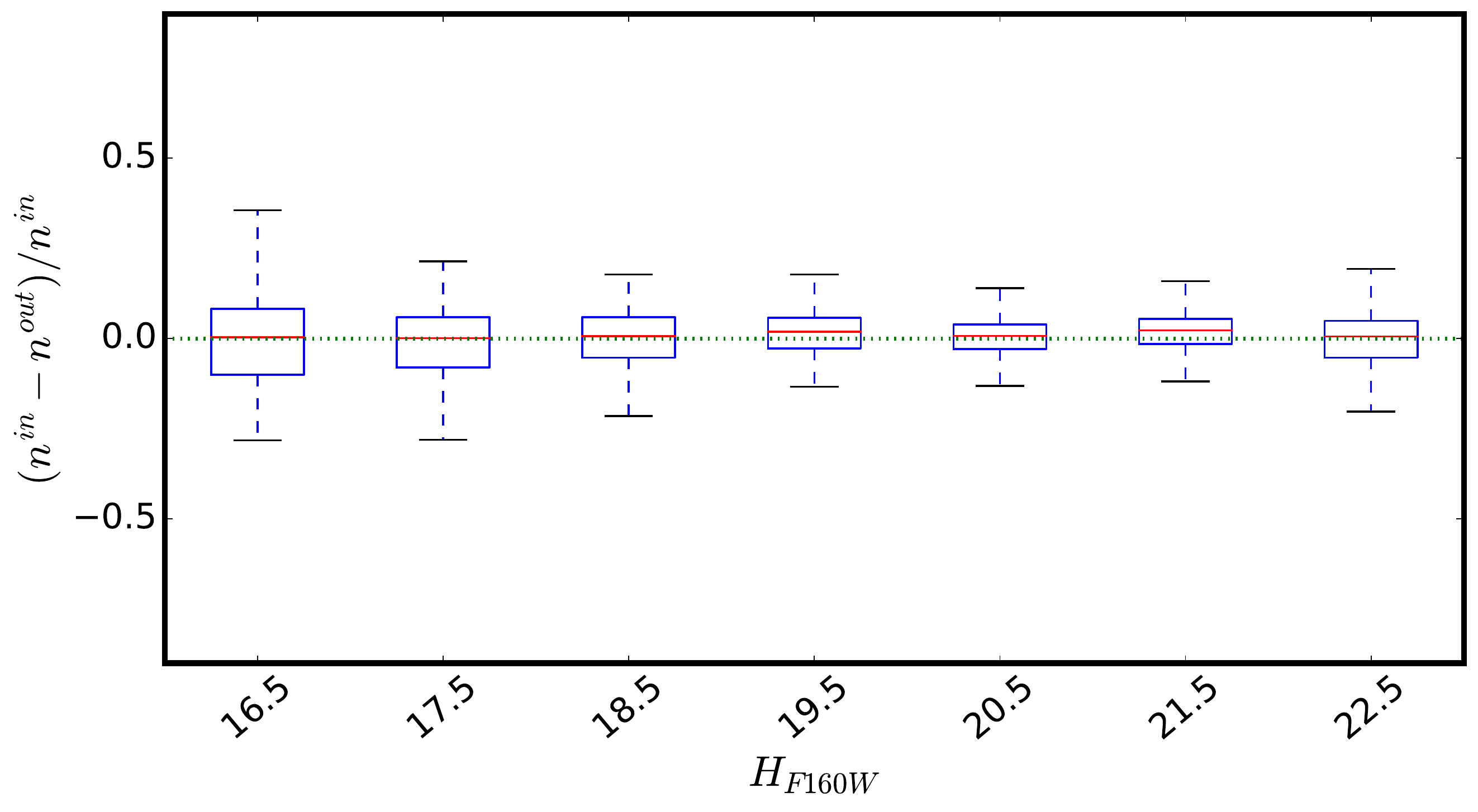} & \includegraphics[width=8.5cm]{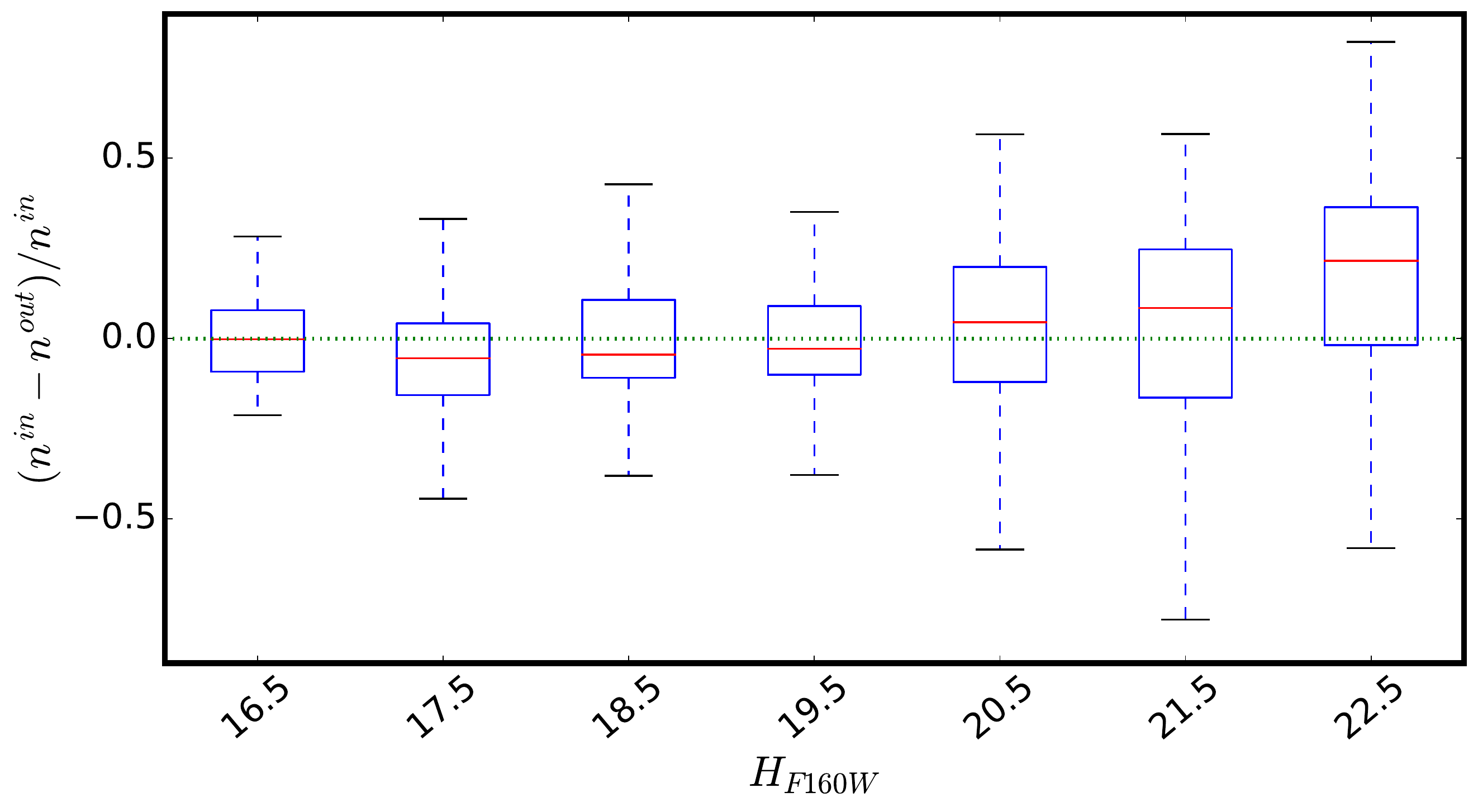} \\

\includegraphics[width=8.5cm]{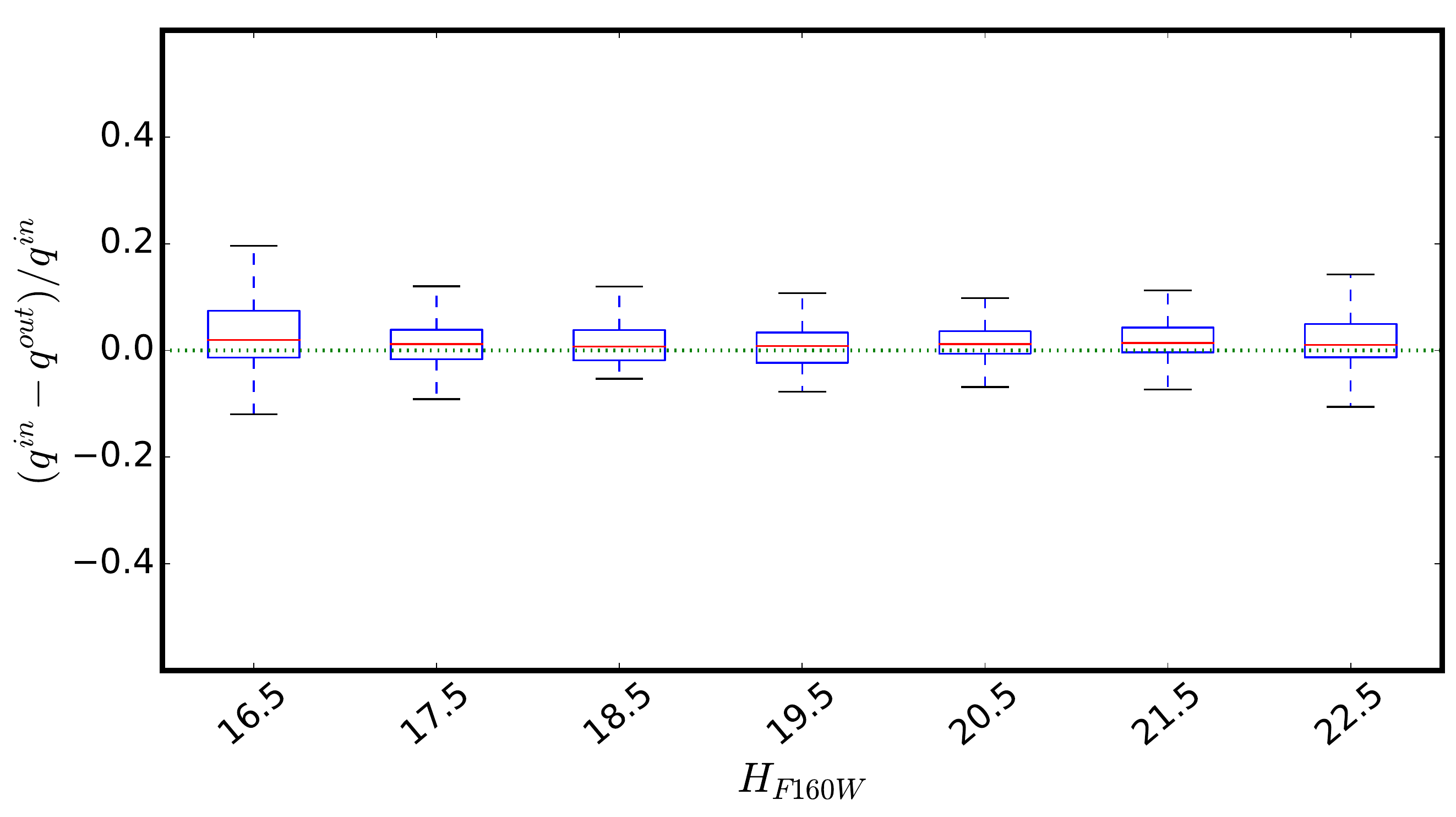} & \includegraphics[width=8.5cm]{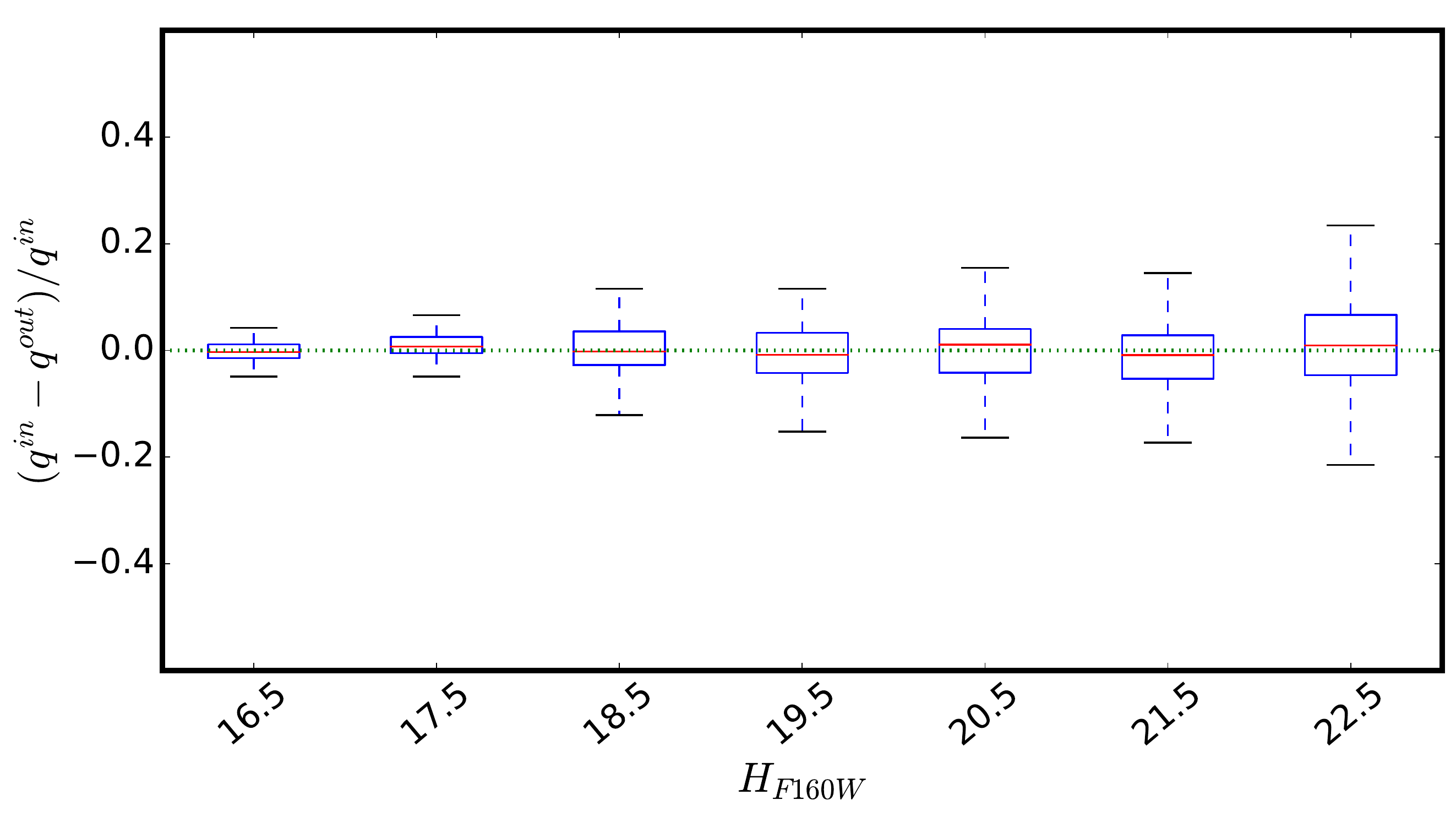} \\

\end{tabular}
\captionof{figure}{Results of profile fitting for the images of 5000 simulated galaxies. The left column shows the results obtained using \textit{DeepLeGATo} (architecture 1 for S\'ersic index and axis ratio; architecture 2 for magnitude and half-light radius), the right column the results obtained using GALFIT. For each galaxy and for each parameter $X$ we calculated the difference between the parameter used to simulate the galaxy ($X^{in}$) and the estimation ($X^{out}$) obtained from profile fitting codes. In the first row we show the results for the magnitude $H_{F160W}$, in the second for the half-light radius $R_e$, in the third for the S\'ersic index $n$ and in the last for the axis ratio $q$. The results are shown in bins of magnitude (bin width = 1 mag) and non parametrically in the form of box plots. As usually, the boxes are delimitated by the first and third interquartiles of the data, while the whisker indicates the range between $\pm1.5$ the interquartile range (IQR). The red line in the box indicates the median of the data.} 
\label{PerformanceSim}
\end{table*}

\section{Tests on real data}

Having discussed in the previous section the performance of our code on simulated data, in this section we show its results on the real data presented in section \ref{realData}.


\subsection{Direct application of the learned system on real data}
\label{directApp}

Testing on real data is a fundamental step for every profiling method. Galaxy simulations are regular by definition and even when analyzing a sample generated with a wide range of structural parameters, we may underestimate the true errors involved in the estimations. Real galaxies may be more asymmetric and difficult to decompose than simulated objects, but above all, the stamps of real galaxies may include the presence of companions that are not included in our simulations. On the other hand, tests on the simulated data presented in the previous section have the great advantage that the \textit{ground truth} is exactly known, i.e. the structural parameters used to generate the data are known by definition. In the case of real data we do not know the authentic \textit{ground truth}. Testing the accuracy of our code on real galaxies, we need to rely on the comparison with estimations done with some other methods. For this reason, we choose galaxies included in the \cite{vanderWel_2012} catalog, and we use their estimations as \textit{ground truth} for our further tests. This will, of course, add additional scatter into the results since the parameter estimations are certainly affected by both random and systematic (often unknown) errors.

\begin{figure*}
\centering
\subfloat[Before domain adaptation (BDA)]{
  \includegraphics[width=85mm]{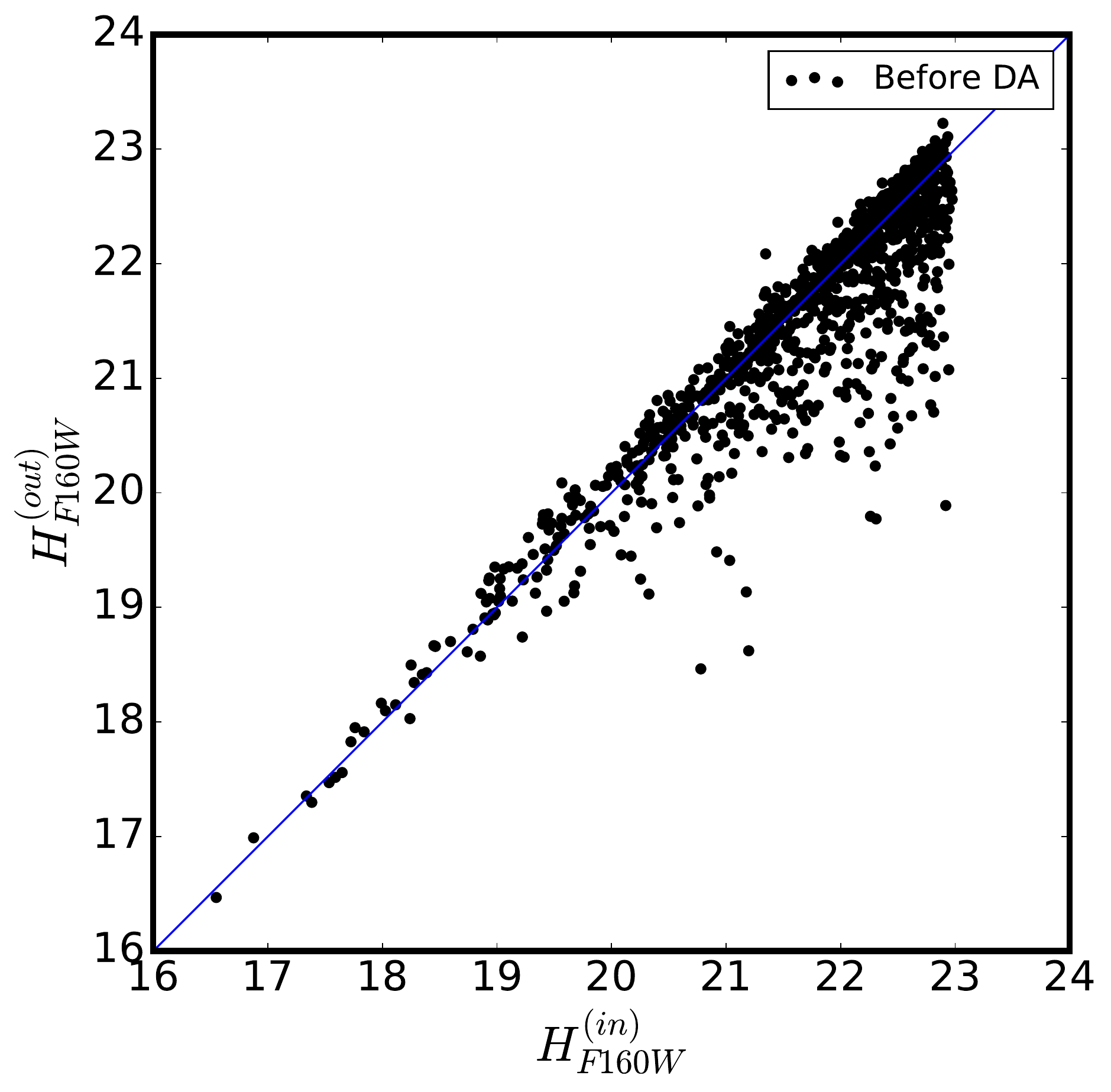}
}
\subfloat[After domain adaptation ]{
  \includegraphics[width=85mm]{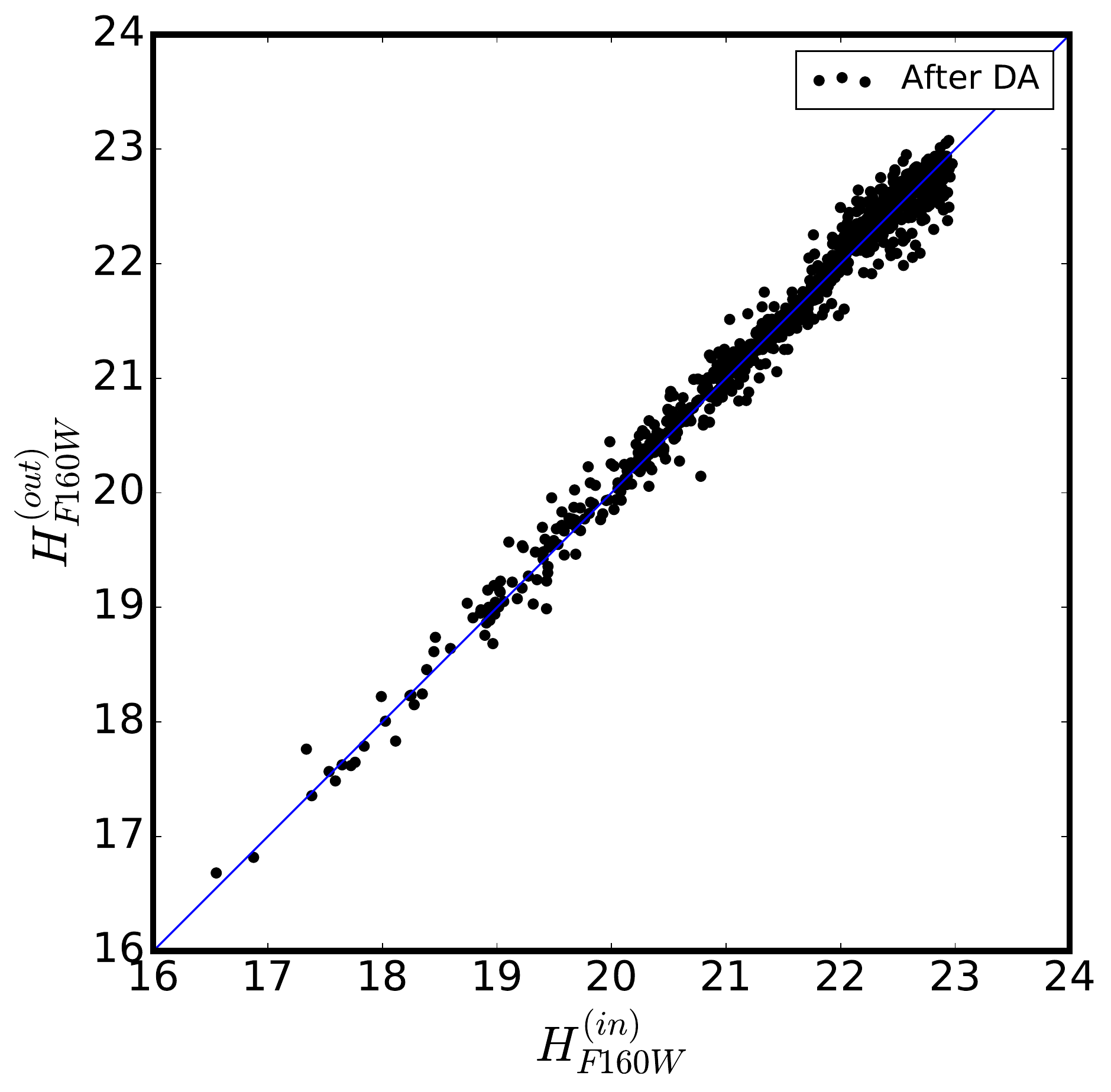}
}
\hspace{0mm}
\subfloat[BDA bright neigbours]{
  \includegraphics[width=60mm]{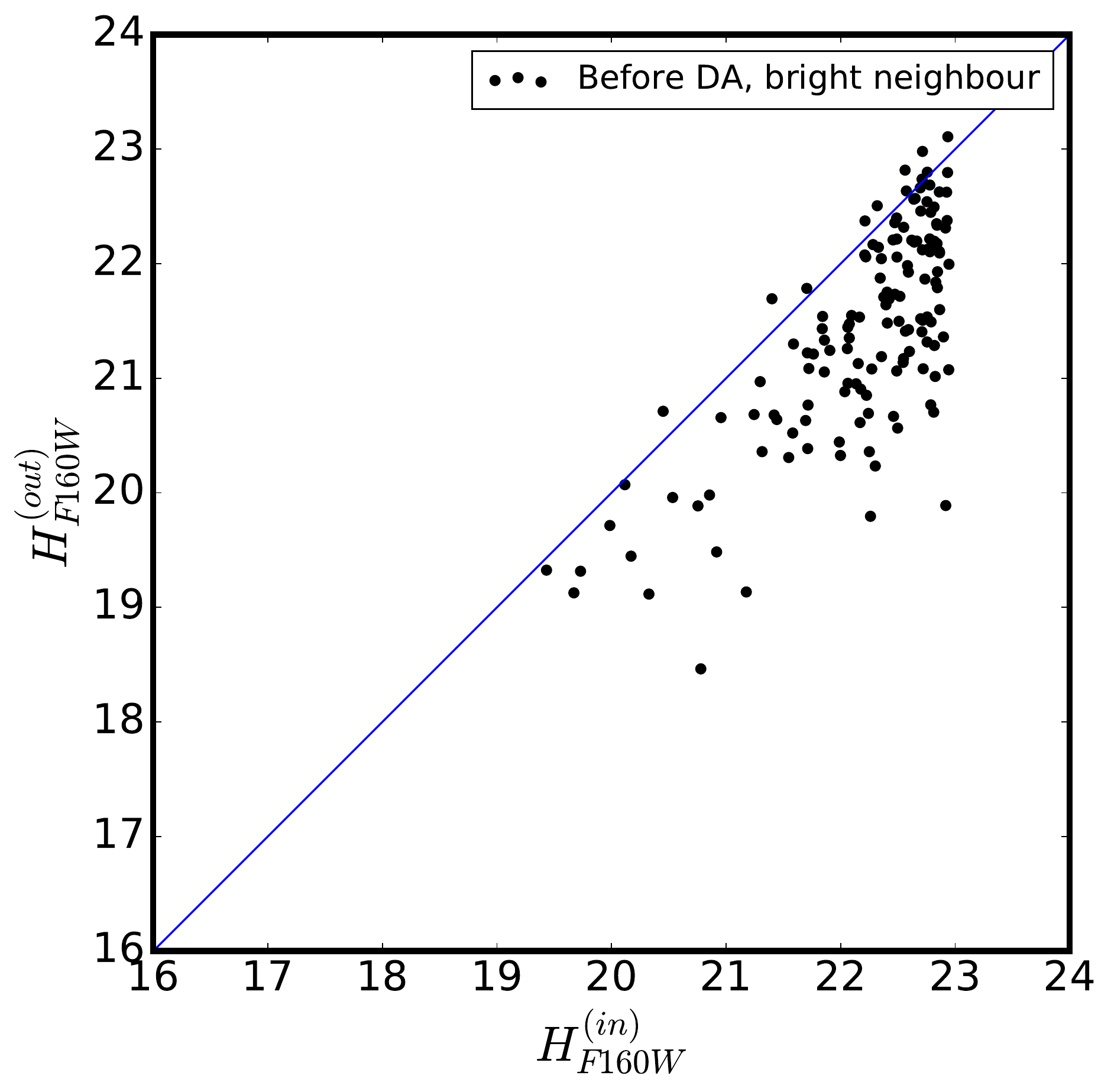}
}
\subfloat[BDA faint neigbours]{
  \includegraphics[width=60mm]{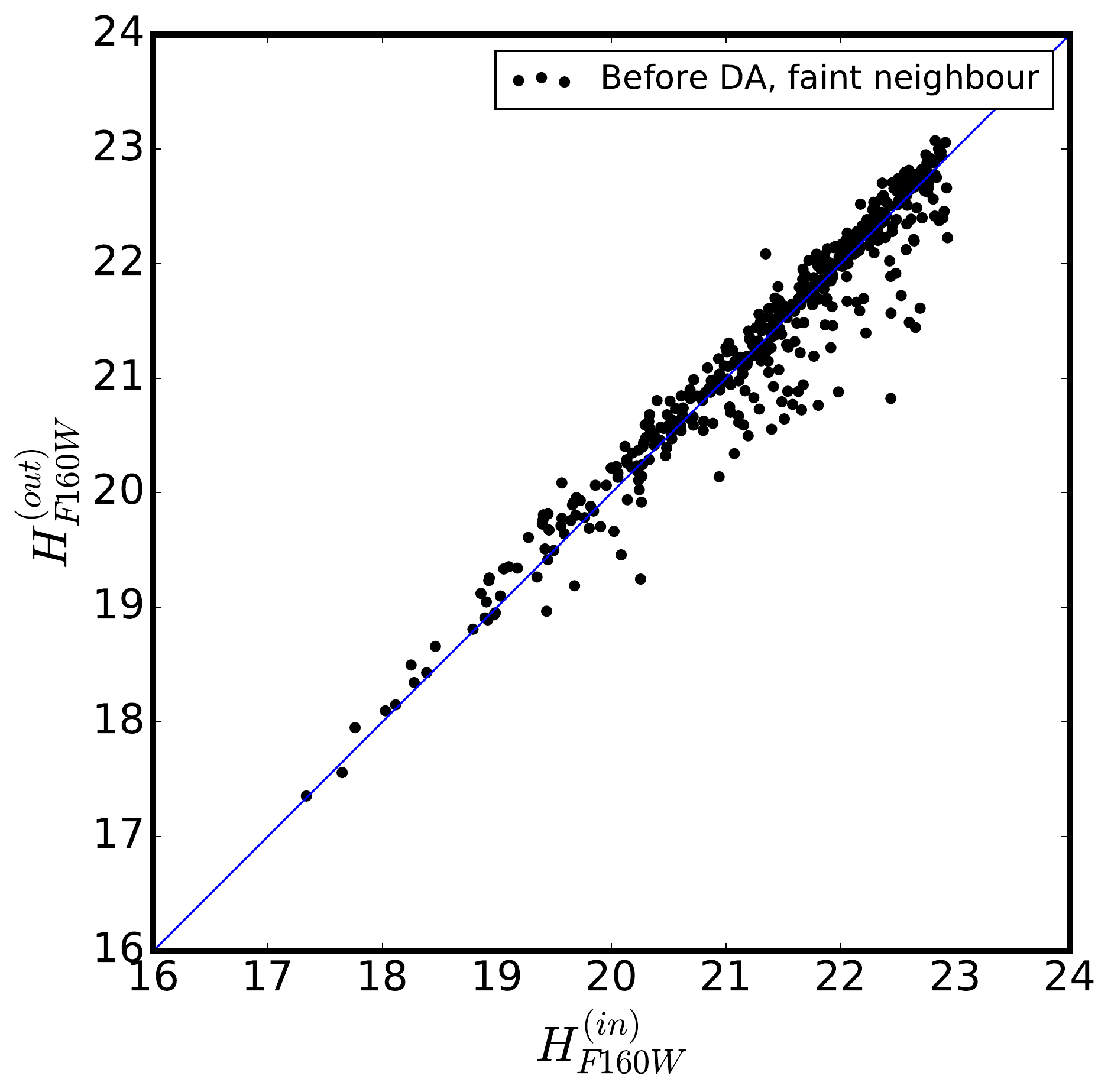}
}  
\subfloat[BDA isolated galaxies]{
  \includegraphics[width=60mm]{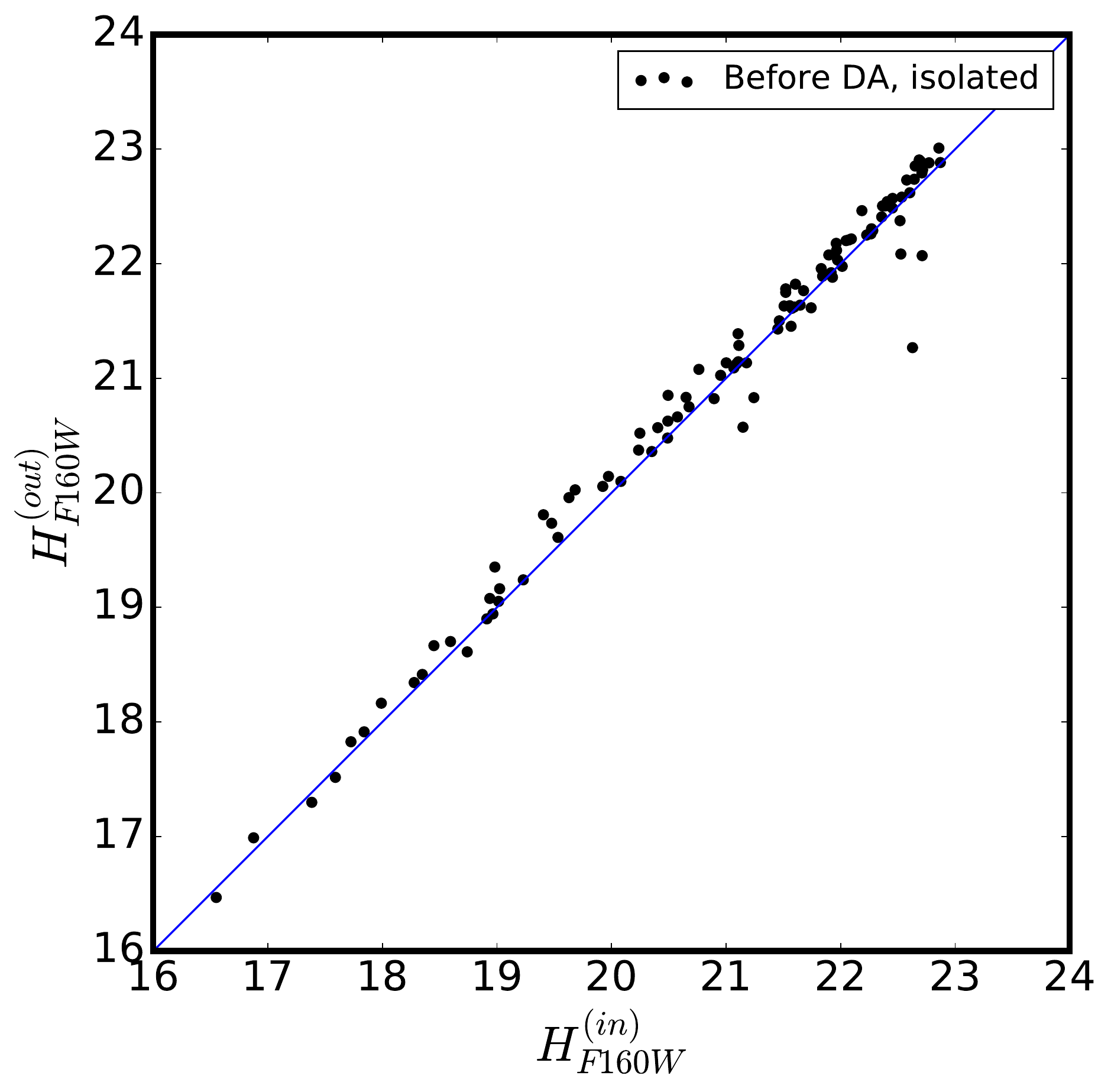}
}  

\caption {We show the results of our CNNs in fitting the magnitude of one thousand real galaxies. On the x-axis we plot the parameter estimation given from the van der Wel et al. (2012) catalogue, while on the y-axis we give the parameter values estimated with DeepLeGATo. The upper panels show the results obtained for the whole sample before the \textit{domain adaptation} step (panel a) and after the \textit{domain adaptation} (panel b). The three bottom panels show: (c)  the results obtained for the 142 galaxies whose brightest companion has at least the 50\% of their flux; (d) the middle panel shows the results for the 450 galaxies whose companion has less than the 10\% of the flux; (e) the third panel shows the results for the 103 isolated galaxies of our test-sample, i.e. without any companion within the stamp}
\label{scatter1} 
\end{figure*}

With this assumption, we run our best CNNs architectures (architecture 1 for S\'ersic index and axis ratio; architecture 2 for magnitude and half-light radius) on 1000 real galaxies introduced in section \ref{realData} and we compare their predictions with those of the \cite{vanderWel_2012} catalog. As in section \ref{test_simData}, we quantify this comparison using the \textit{coefficient of regression} $R^2$. When we fit the images of real galaxies with \textit{DeepLeGATo} trained only on simulations, we do not obtain results as reliable as those obtained on the test-sample of simulated data (see section \ref{galfit_simdata}). We show this result for the magnitude in panel (a) of Figure \ref{scatter1}, and for all the other parameters in the plots (a) of Figures \ref{scatter2}, \ref{scatter3} and \ref{scatter4} of the Appendix A. In those figures, on the x-axis we plot the parameter estimation given from the \cite{vanderWel_2012} catalogue, while on the y-axis we give the parameter values estimated with \textit{DeepLeGATo}. The same result is given in terms of $R^2$ for the first column of Table \ref{table_testRealData}.

\begin{table*}
\centering
\caption{The coefficient of determination of the light profile fitting obtained with different methods for a sample of 1000 HST/CANDELs real galaxies.  In the first column the parameters were obtained with \textit{DeepLeGATo} before of the domain adaptation step (see section \ref{directApp}). In column two we repeat the same test restricting the analysis to 103 \textit{isolated} galaxies, i.e. without neighbour galaxies or stars within the stamp. In column three we apply again \textit{DeepLeGATo} to the whole sample of 1000 real galaxies, but after of the domain adaptation step (see section \ref{DomainAdap}). Finally, in column four we compare the estimations of van der Wel et al. (2012) with those of Dimauro et al. (2017), i.e. comparing two estimations of two different set-up of GALFIT (see section \ref{GalfitCompReal})}
\label{table_testRealData}
{
\begin{tabular}{l c  c c c}
\hline
\multicolumn{1}{c}{} &
\multicolumn{4}{c}{$R^2$ Real data } 
\\
\hline
\multicolumn{1}{c}{Parameter} &
\multicolumn{1}{c}{Before TL} &
\multicolumn{1}{c}{BTL isolated} &
\multicolumn{1}{c}{After TL } &
\multicolumn{1}{c}{2 GALFIT } 
\\
\hline
Magnitude  & 0.795  &  0.979 & 0.980 &  0.984\\
Radius   & -0.431     & 0.630 &  0.813   & 0.860 \\
S\'ersic index   & -0.331 & 0.516         &  0.813  &  0.819 \\
Axis ratio   &  0.773  & 0.915 & 0.934   & 0.914   \\
\hline\end{tabular}
}
\end{table*}

We qualitatively analyzed the reasons of this discrepancy by directly looking at the stamps of the galaxies, both for the simulated data and for the real ones. At first glance, the simulated and real data seem to be quite similar, but the differences are evident when we order the data depending of the difference between the parameters predicted by the CNNs and the ground truth. The first two lines of Figure \ref{stamps} show twelve randomly selected stamps of simulated galaxies. The third and the fourth line show twelve stamps of real galaxies selected between the ones best predicted by our CNNs (with respect to the estimations given in the \cite{vanderWel_2012} catalog). Finally, the last two lines show twelve stamps of real galaxies selected between the ones worse predicted by our CNNs. Notice that for both best and worse predictions, we chose three galaxies for each of the four parameters. From these images we can see that the CNNs purely trained on simulated data are able to recover the parameters of real galaxies that are as regular and isolated as the simulated. The main issue our method encounters when fitting real data appears to be the presence of bright companions in the stamp, since our simulation includes only one object per stamp.The CNNs can still give accurate predictions when the real galaxies show smooth asymmetries and in the stamps there are other fainter and smaller galaxy companions. They give unreliable results when the stamps include several brighter companions.

\begin{table*}
\begin{tabular}{c c c c c c}
\hline
\multicolumn{6}{c}{Stamps (128x128 pixels) of simulated galaxies}\\
\hline
\includegraphics[width=2.67cm, height=2.5cm]{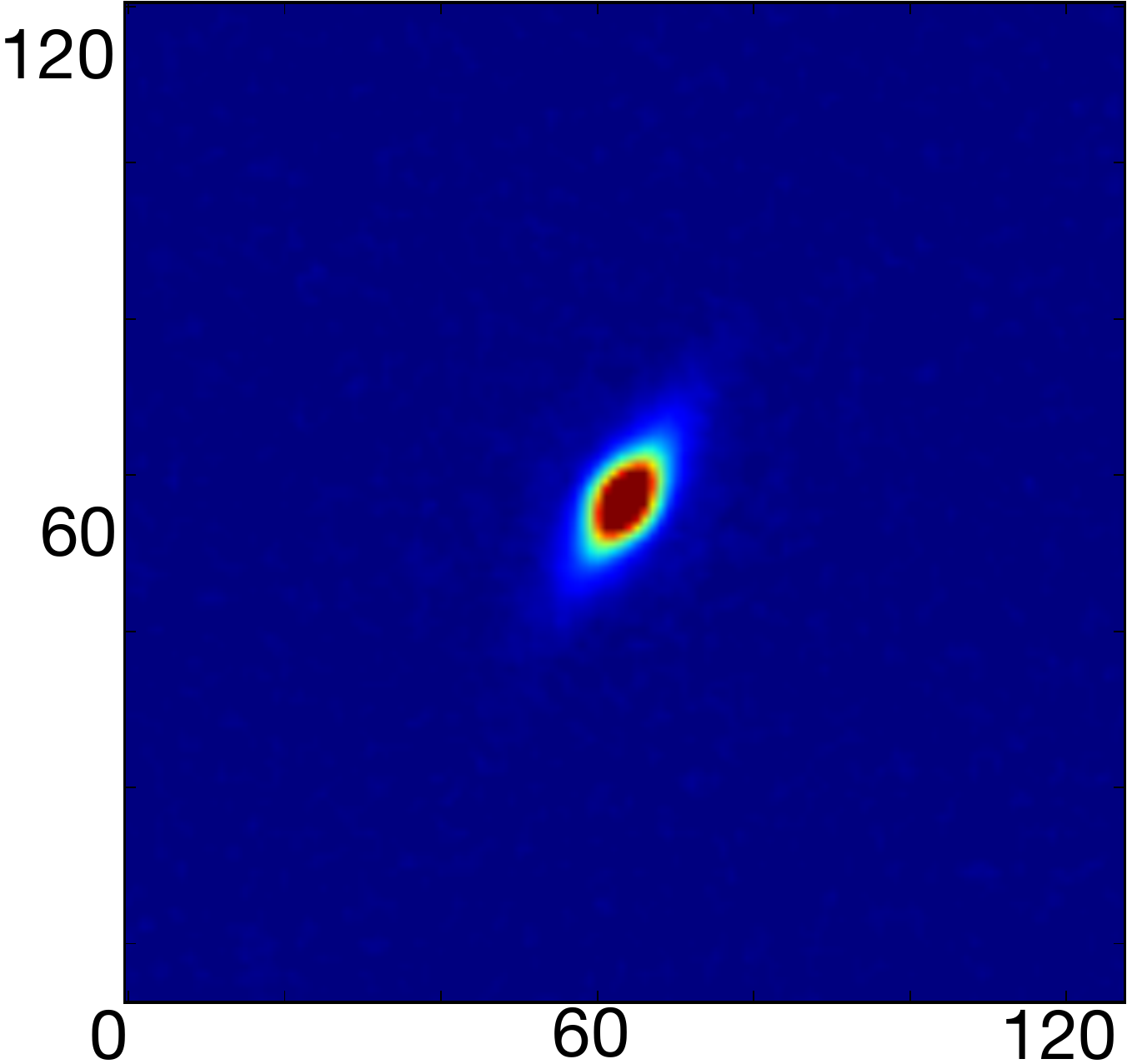} & \includegraphics[width=2.67cm, height=2.5cm]{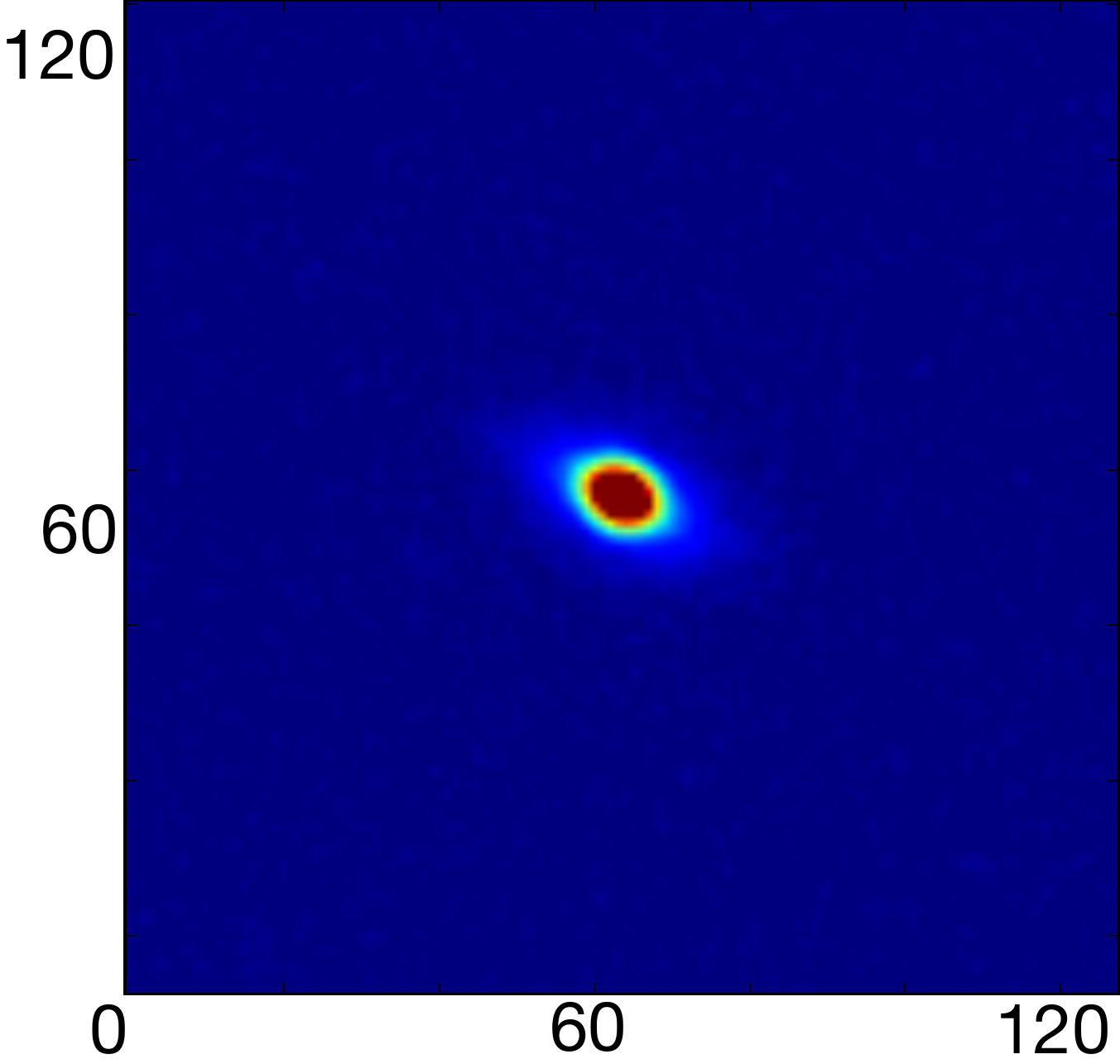} & \includegraphics[width=2.67cm, height=2.5cm]{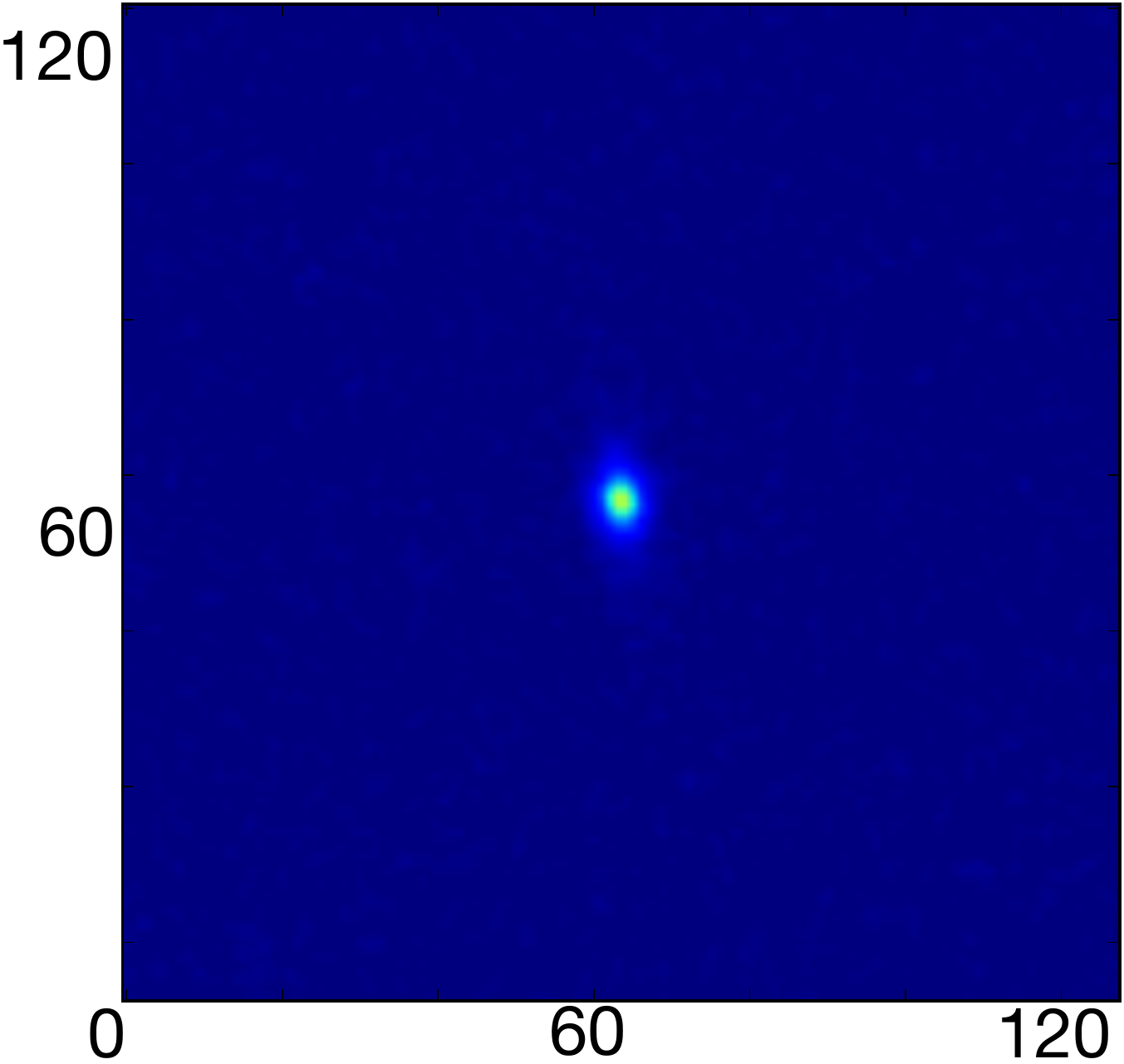} & \includegraphics[width=2.67cm, height=2.5cm]{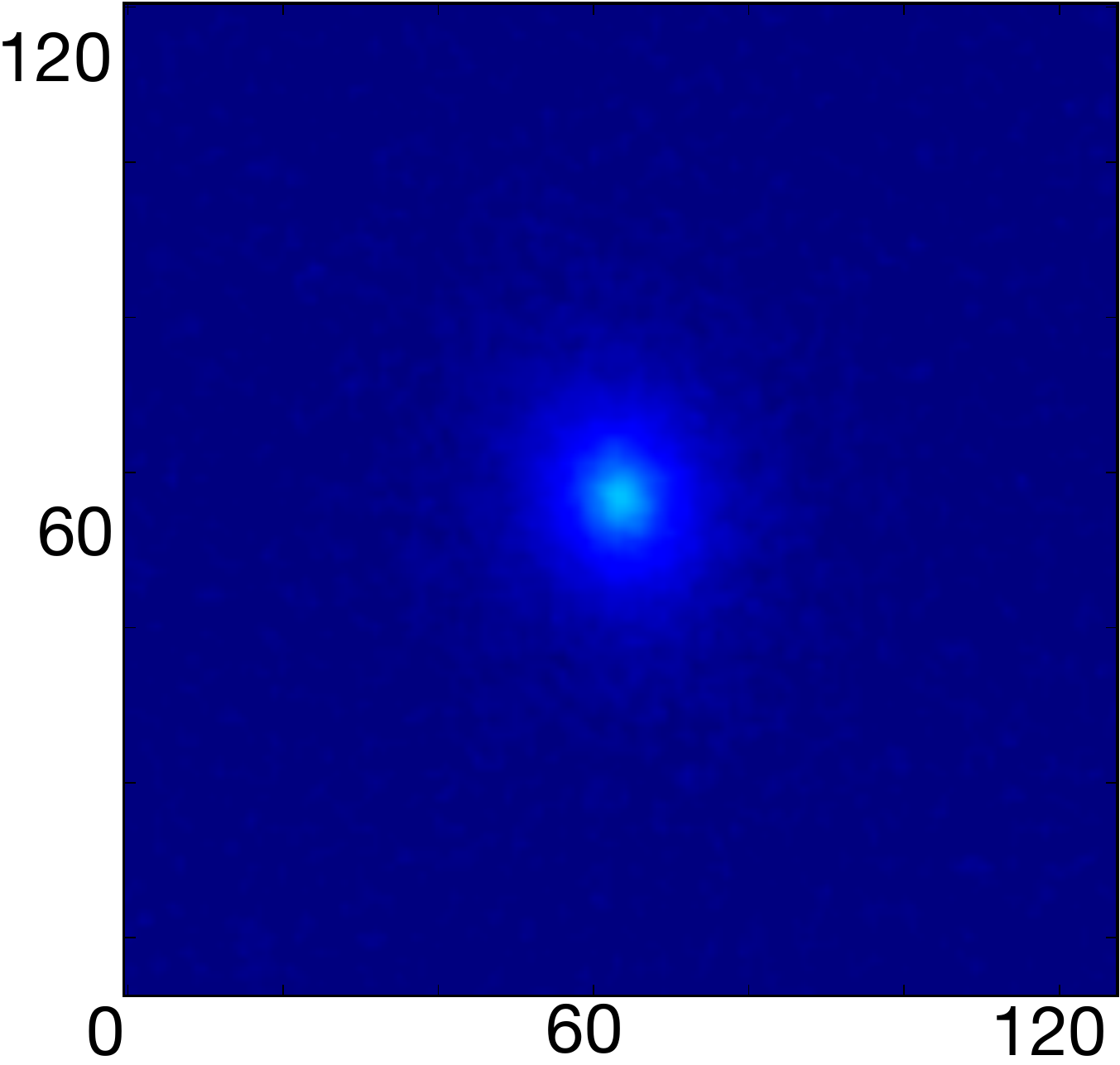} &
\includegraphics[width=2.67cm, height=2.5cm]{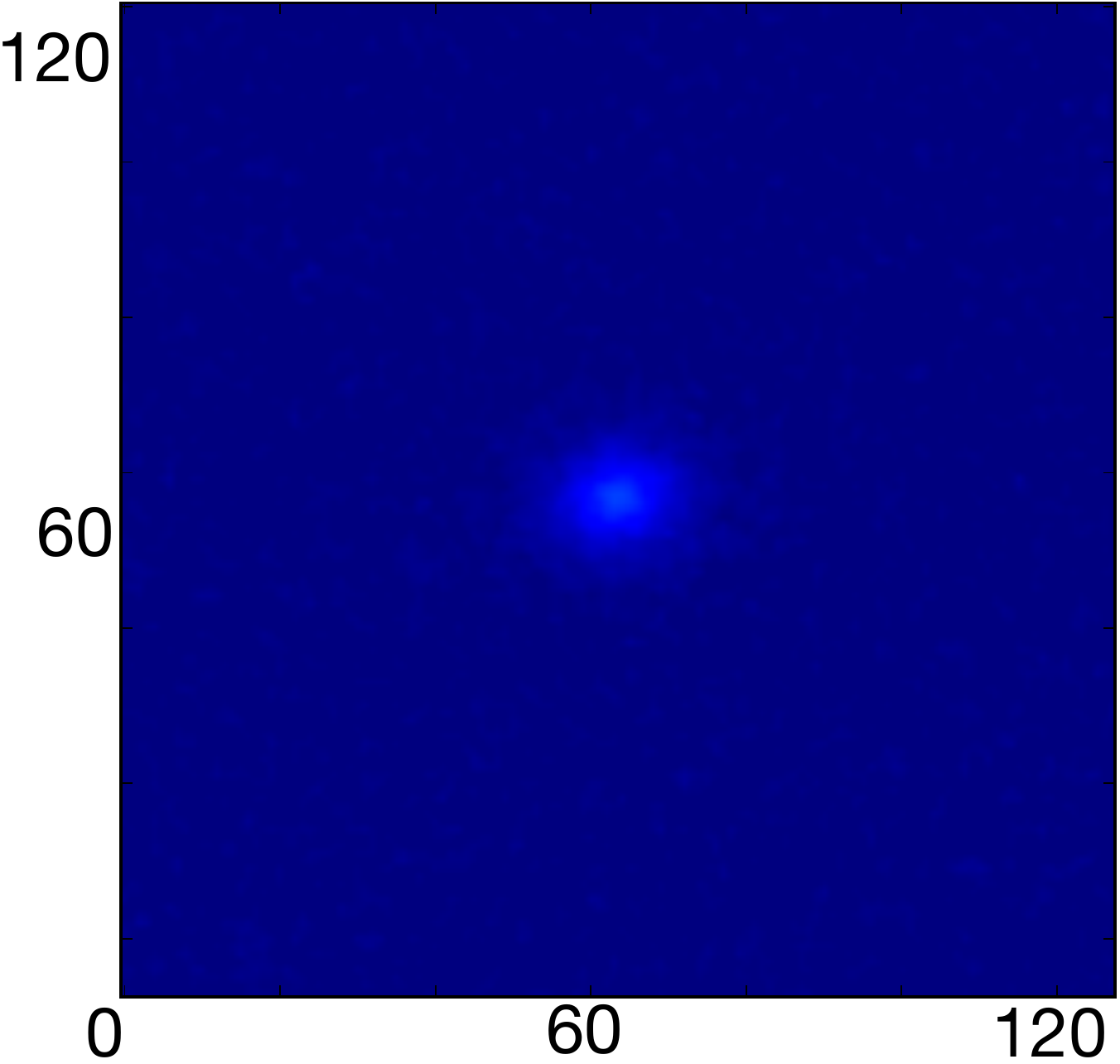} & \includegraphics[width=2.67cm, height=2.5cm]{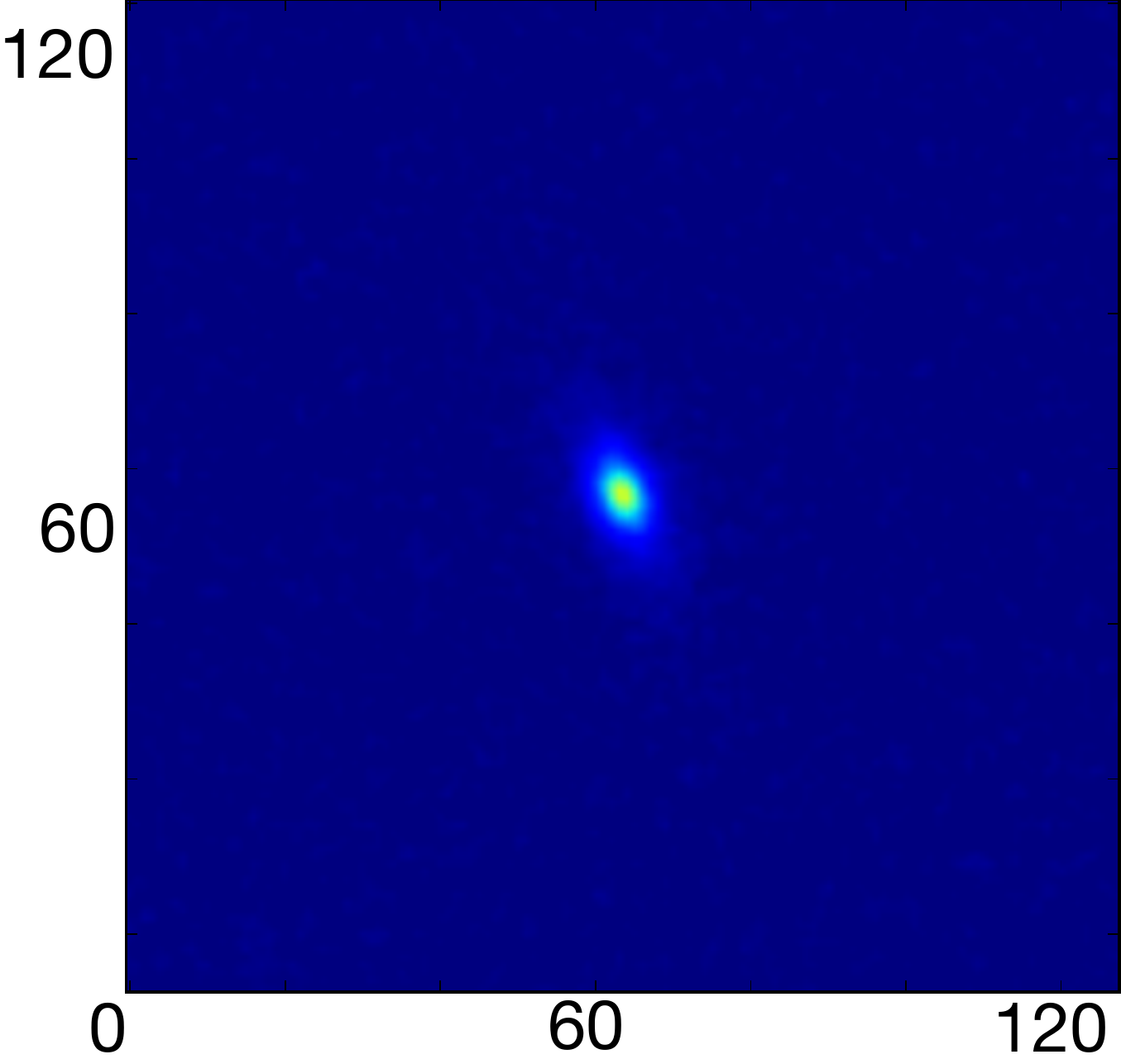}\\

\includegraphics[width=2.67cm, height=2.5cm]{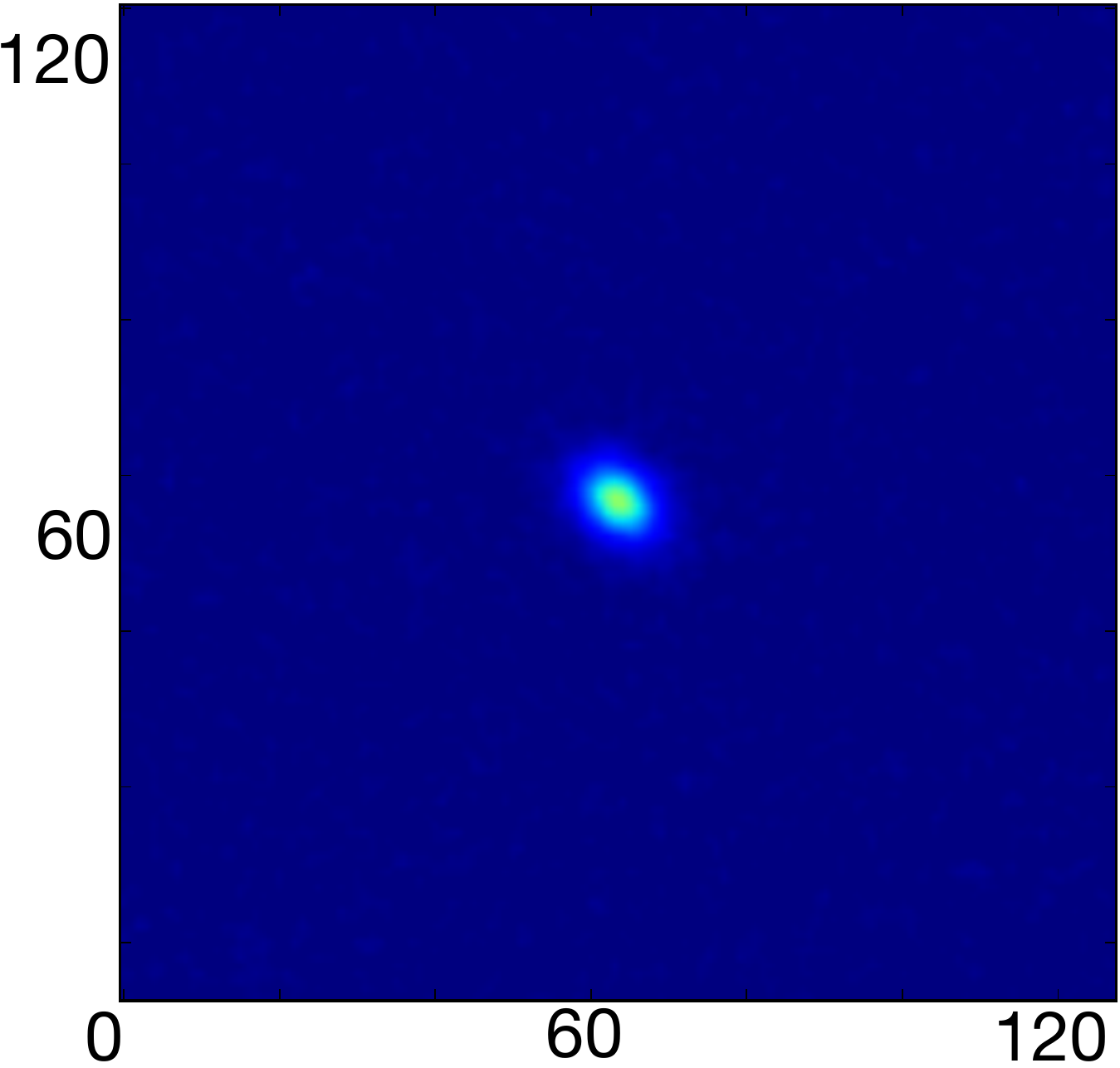} & \includegraphics[width=2.67cm, height=2.5cm]{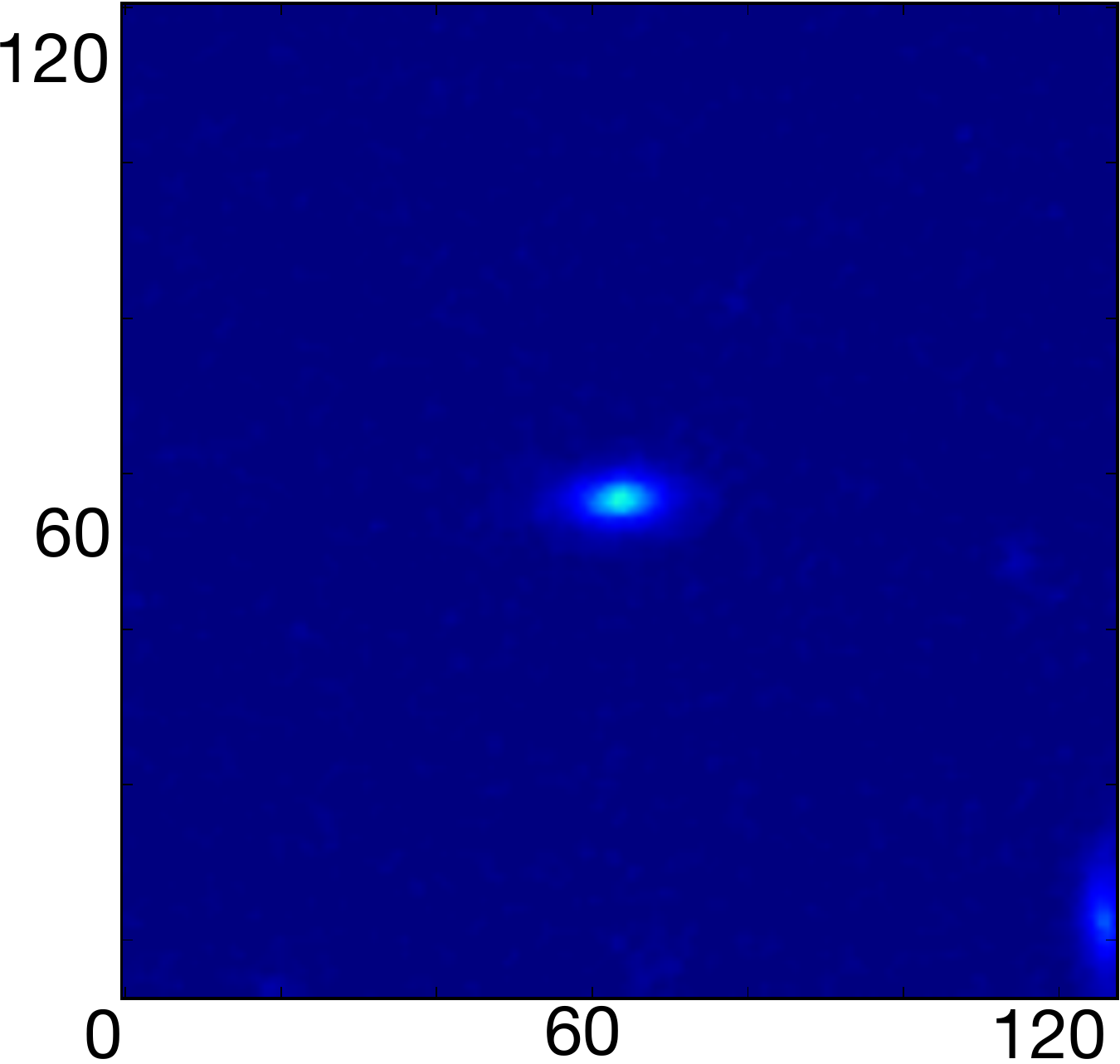} & \includegraphics[width=2.67cm, height=2.5cm]{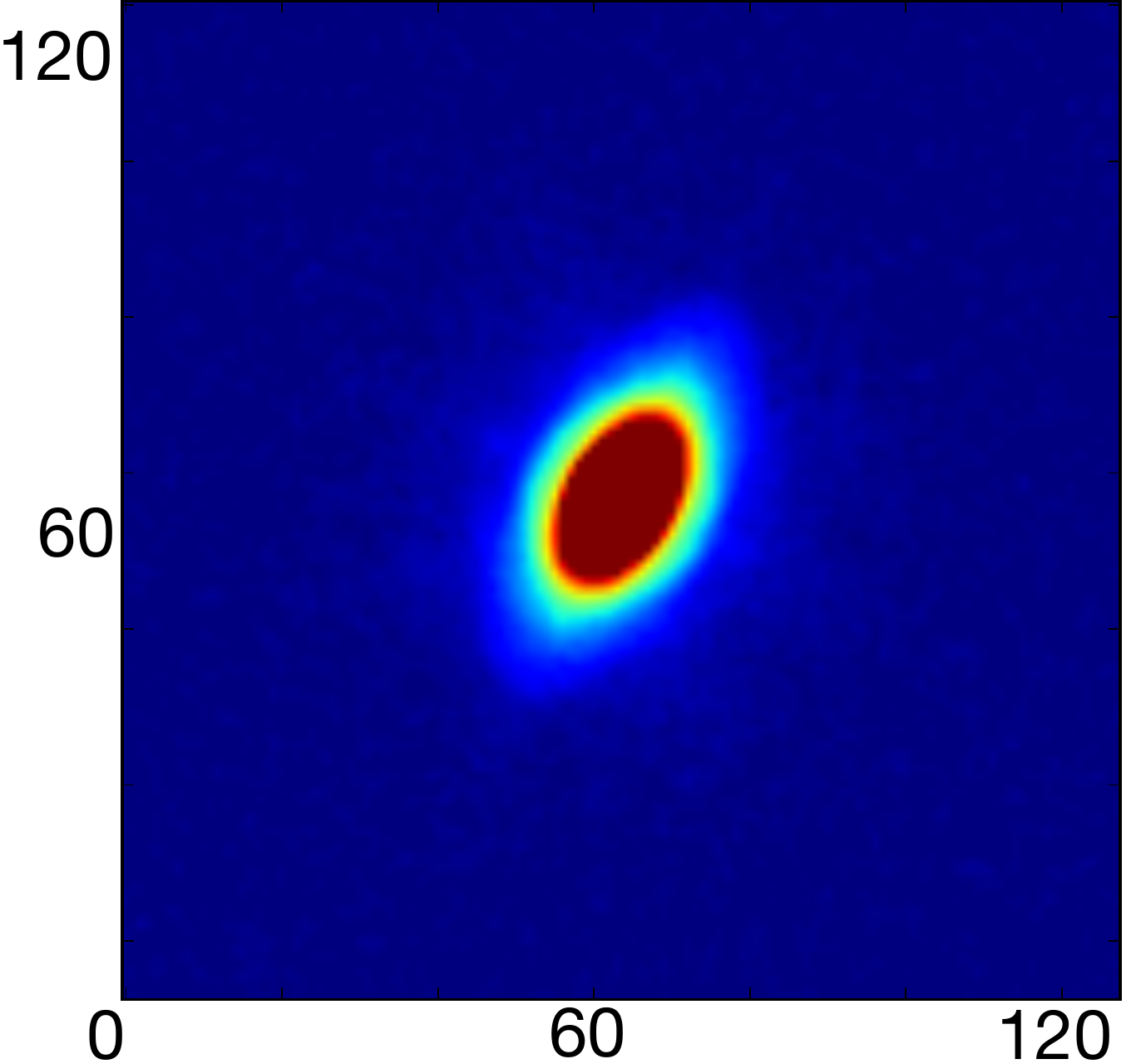} & \includegraphics[width=2.67cm, height=2.5cm]{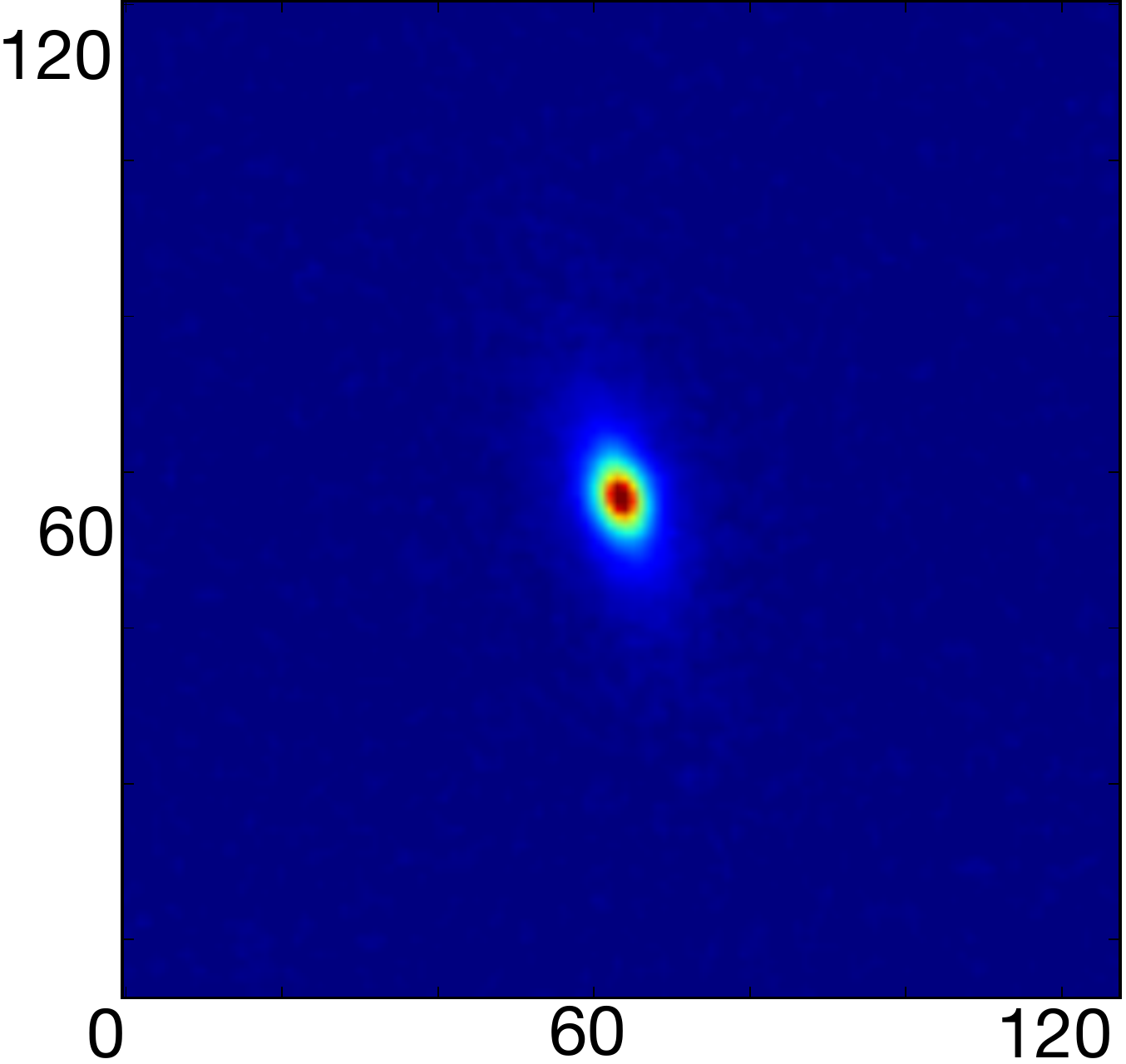} &
\includegraphics[width=2.67cm, height=2.5cm]{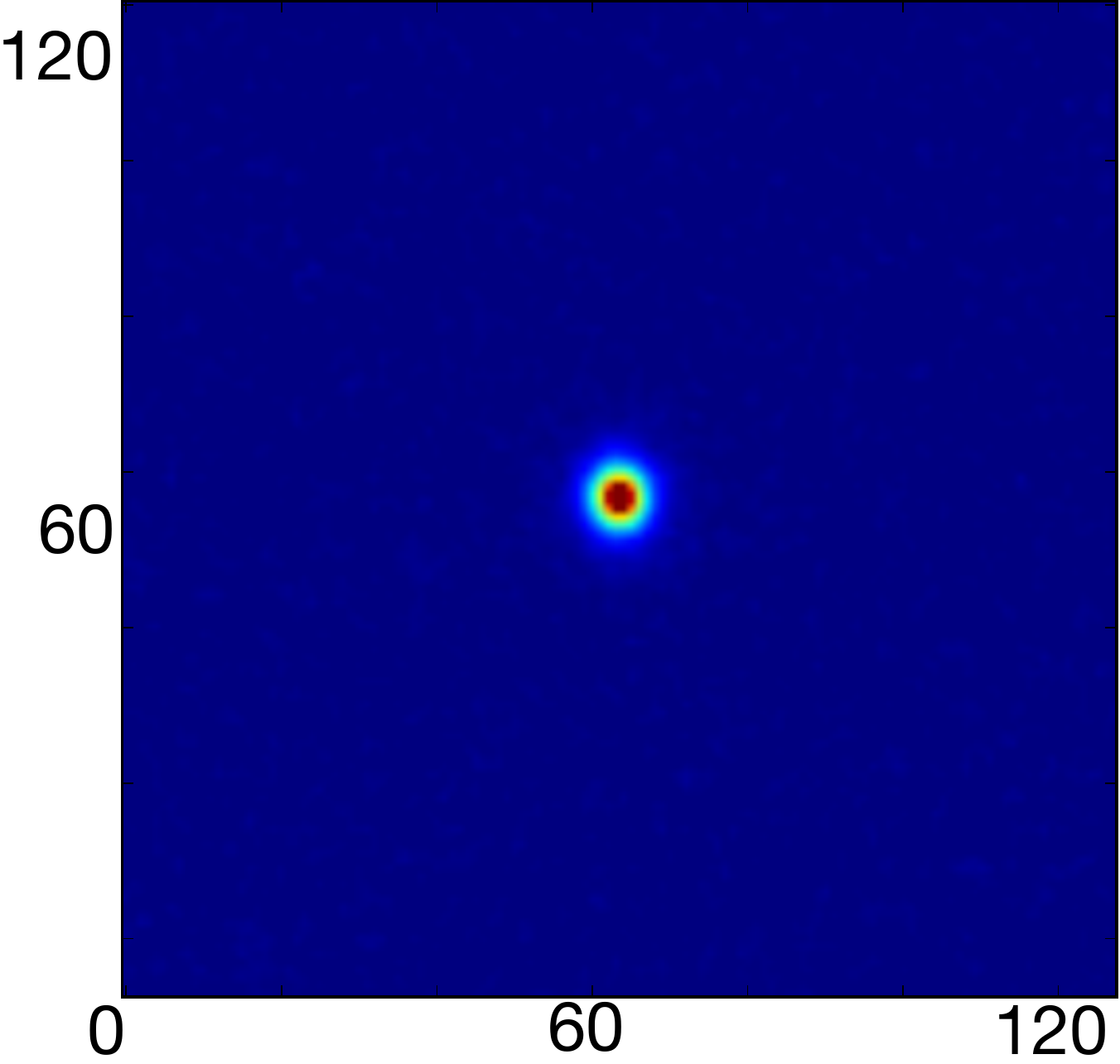} & \includegraphics[width=2.67cm, height=2.5cm]{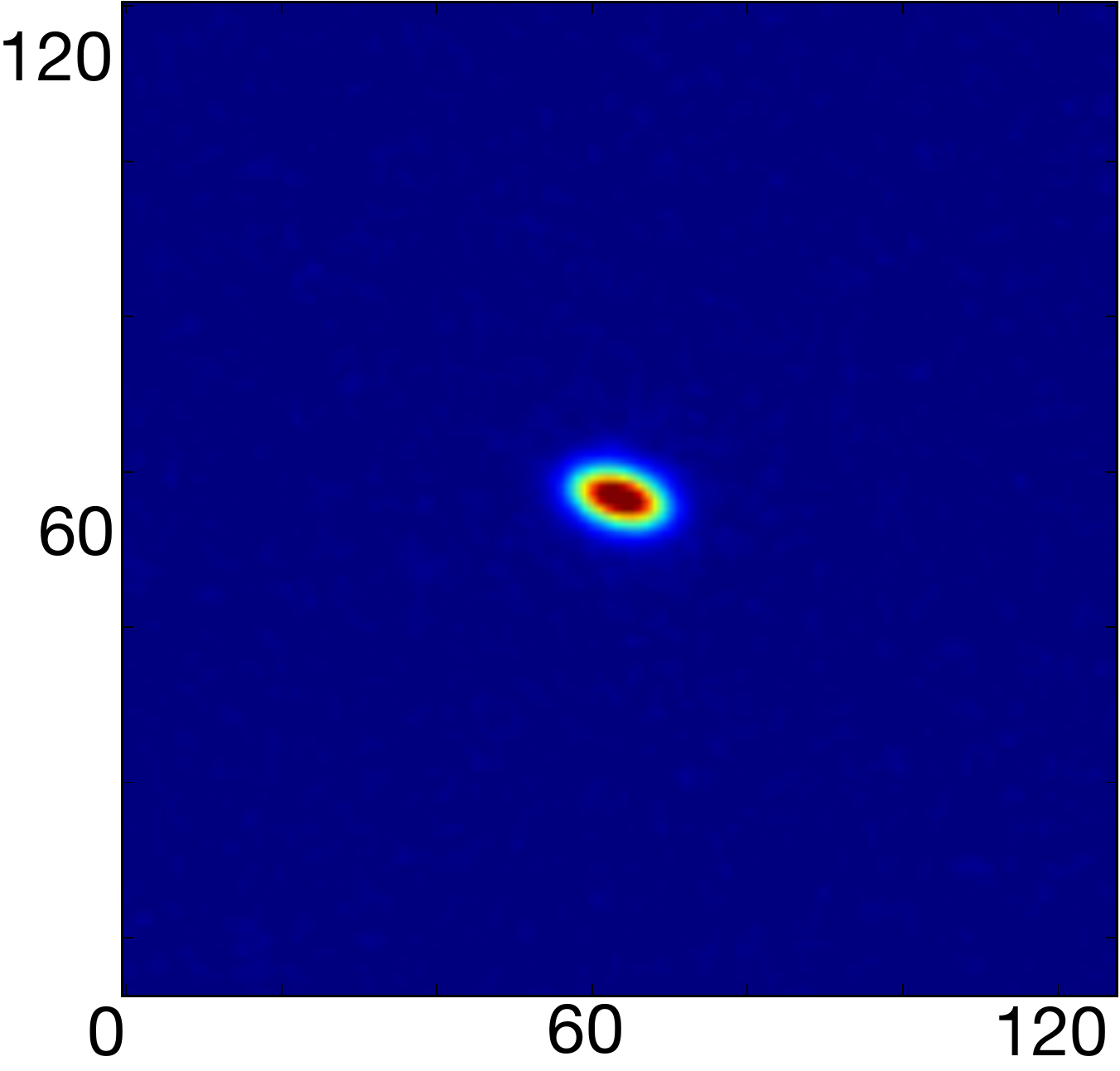}\\
\hline
\multicolumn{6}{c}{Stamps (128x128 pixels) of real galaxies well fitted by \textit{DeepLeGATo}}\\
\hline
\includegraphics[width=2.67cm, height=2.5cm]{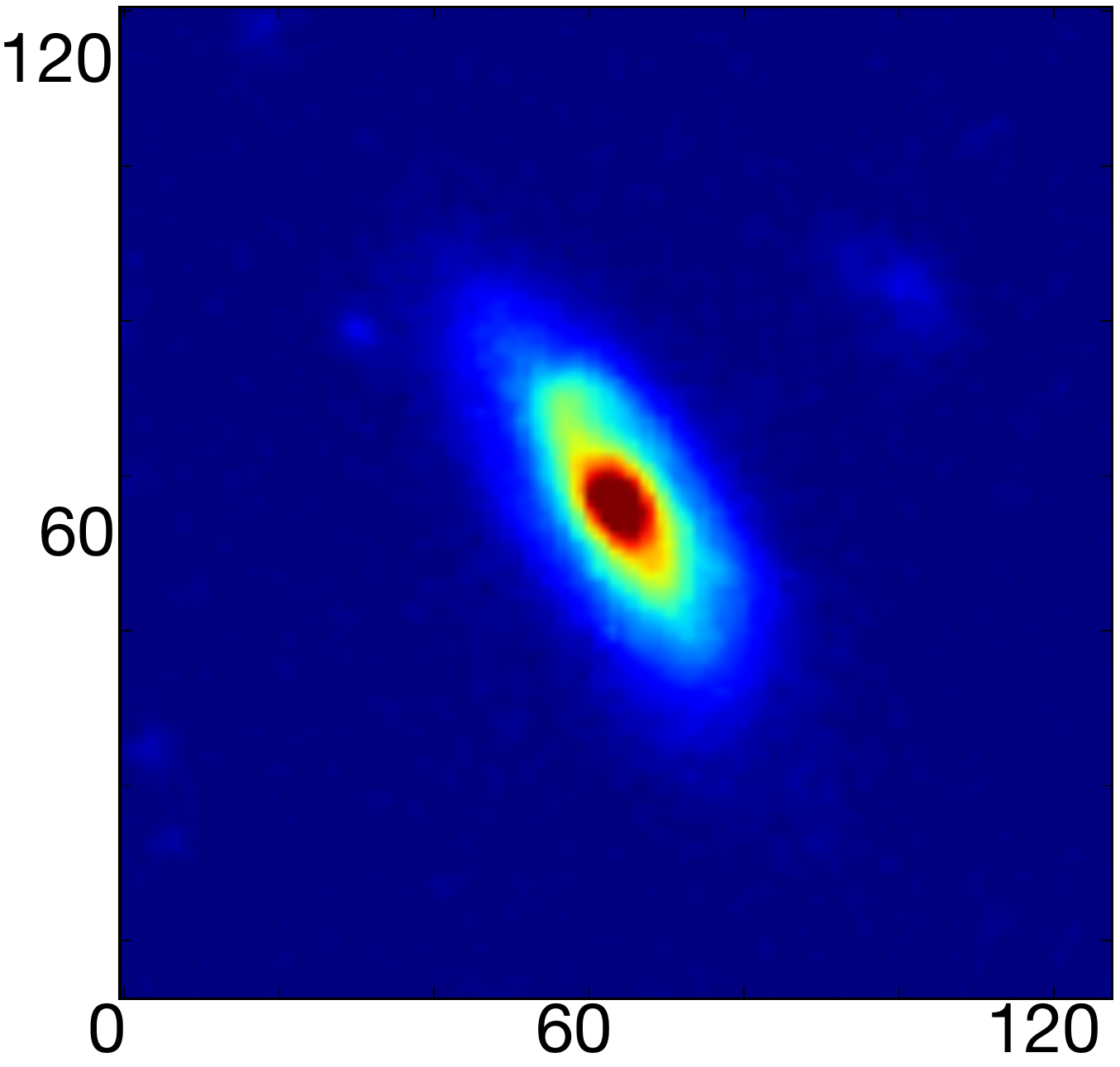} & \includegraphics[width=2.67cm, height=2.5cm]{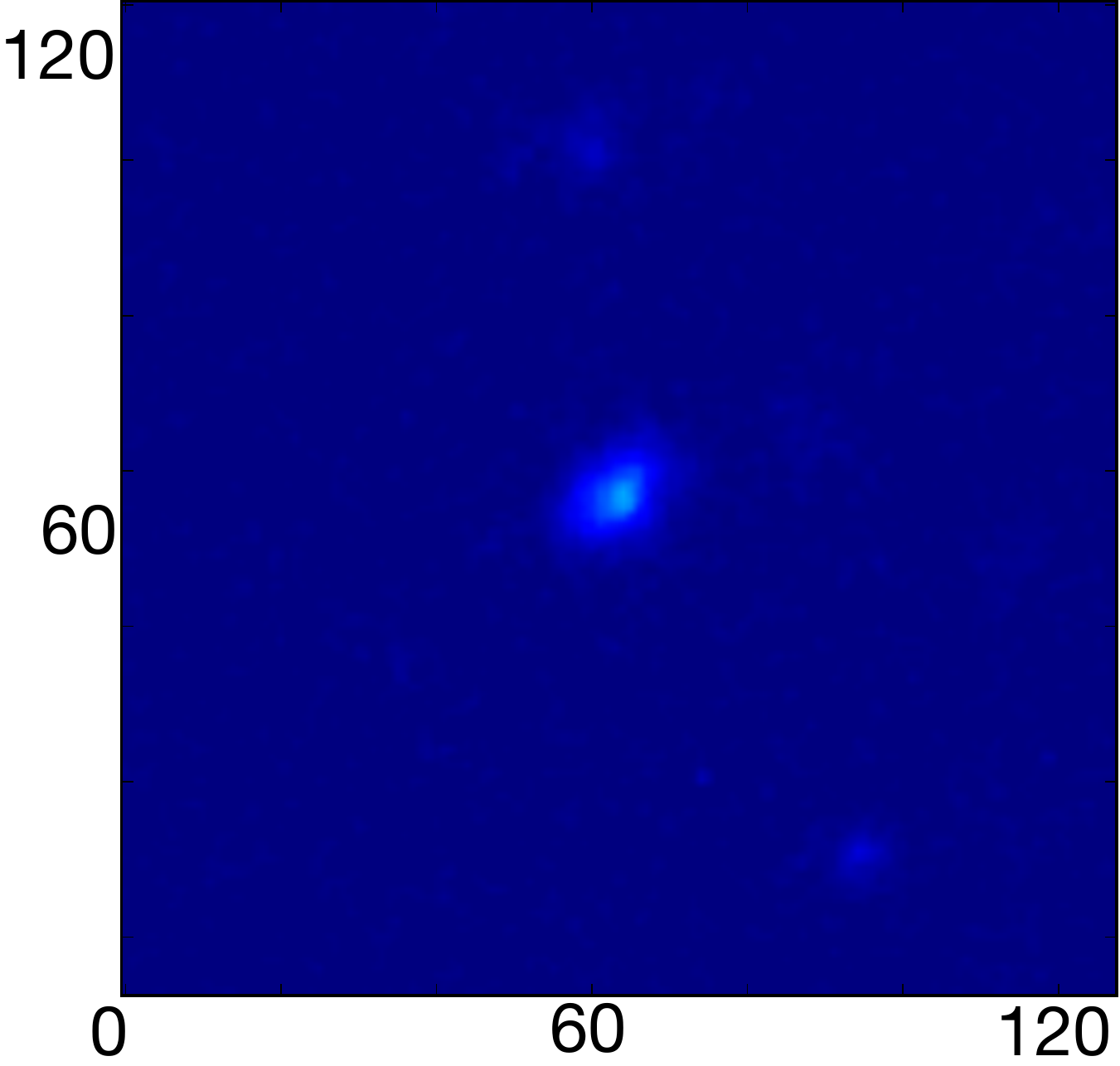} & \includegraphics[width=2.67cm, height=2.5cm]{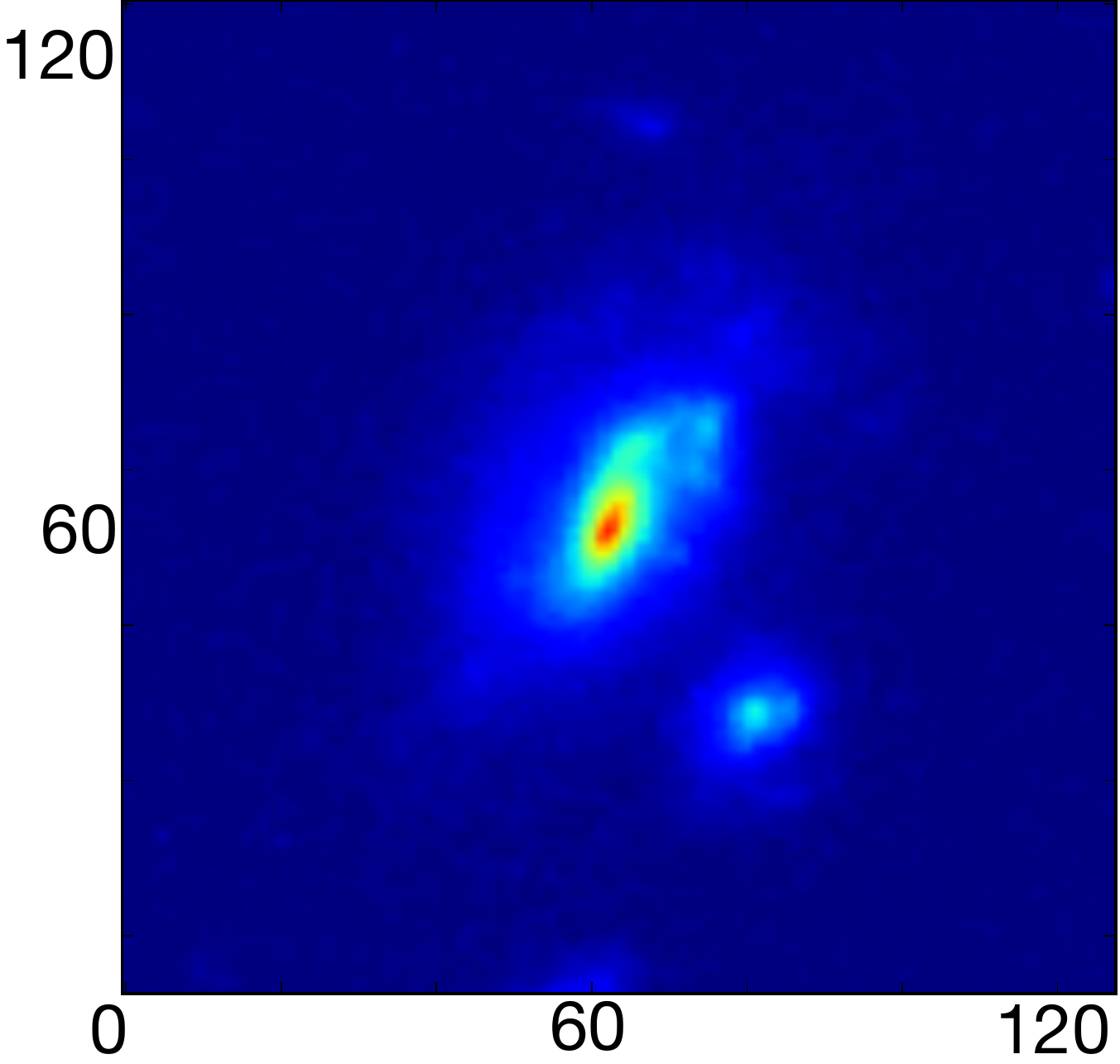} & \includegraphics[width=2.67cm, height=2.5cm]{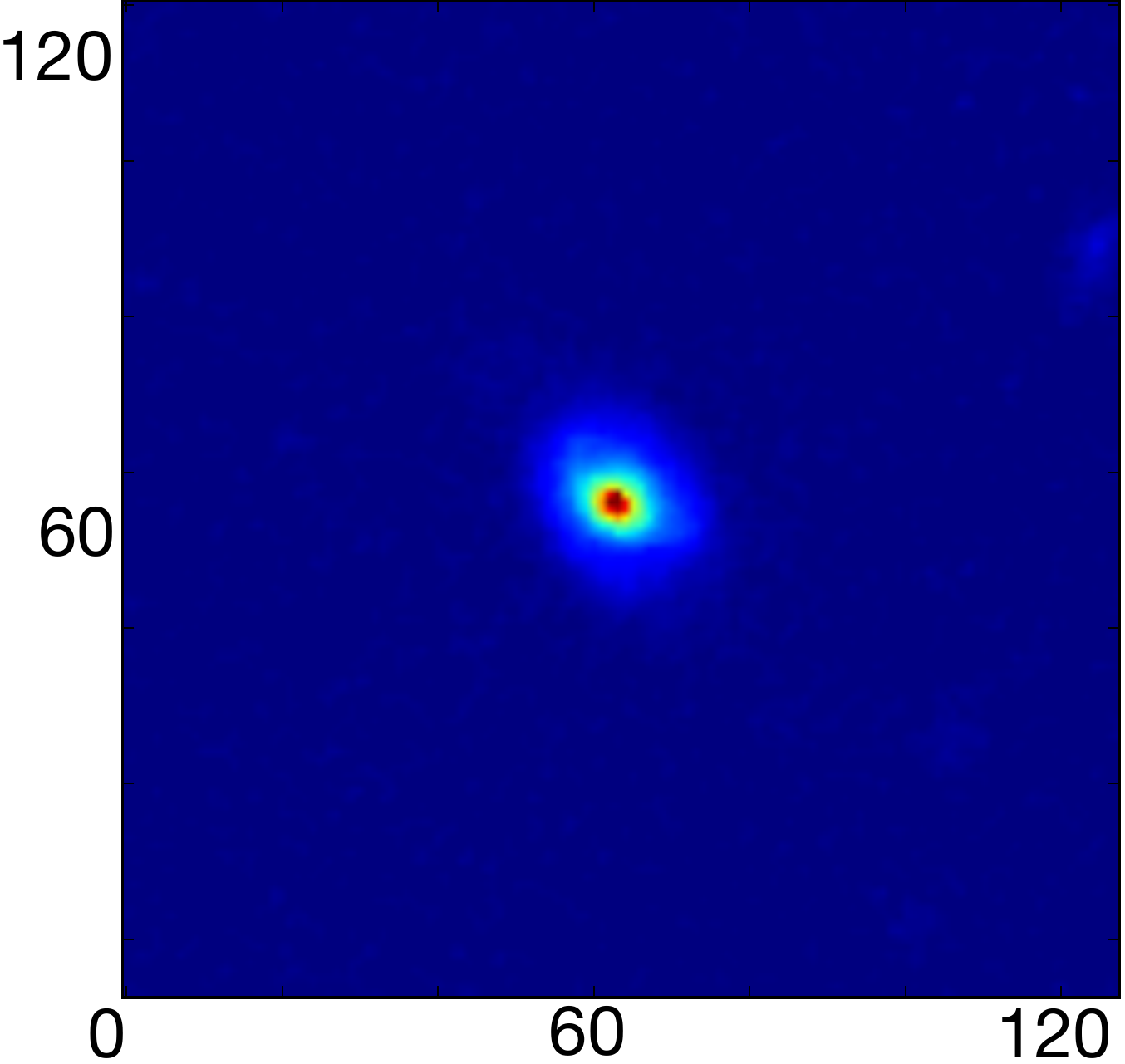} &
\includegraphics[width=2.67cm, height=2.5cm]{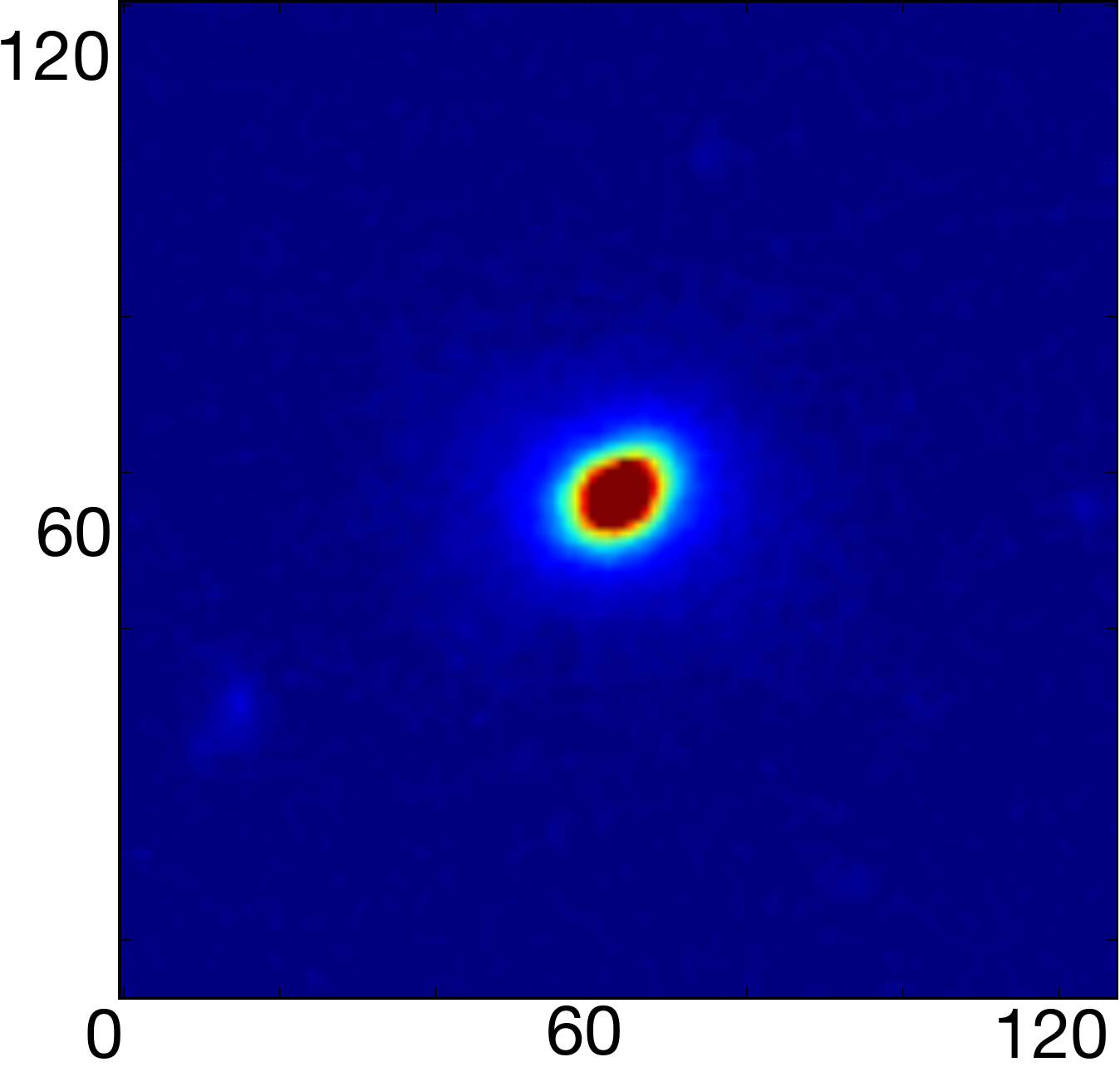} & \includegraphics[width=2.67cm, height=2.5cm]{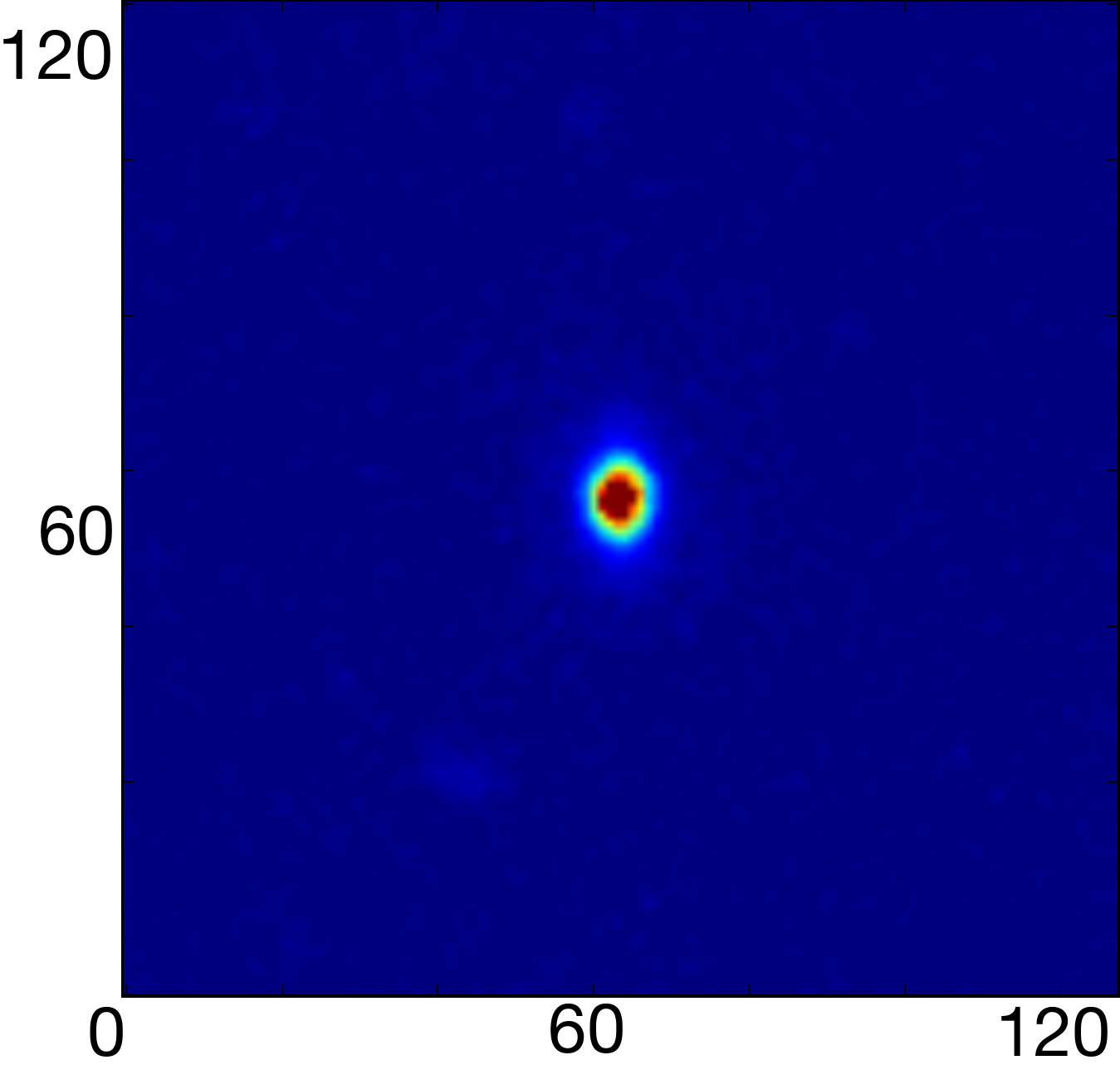}\\

\includegraphics[width=2.67cm, height=2.5cm]{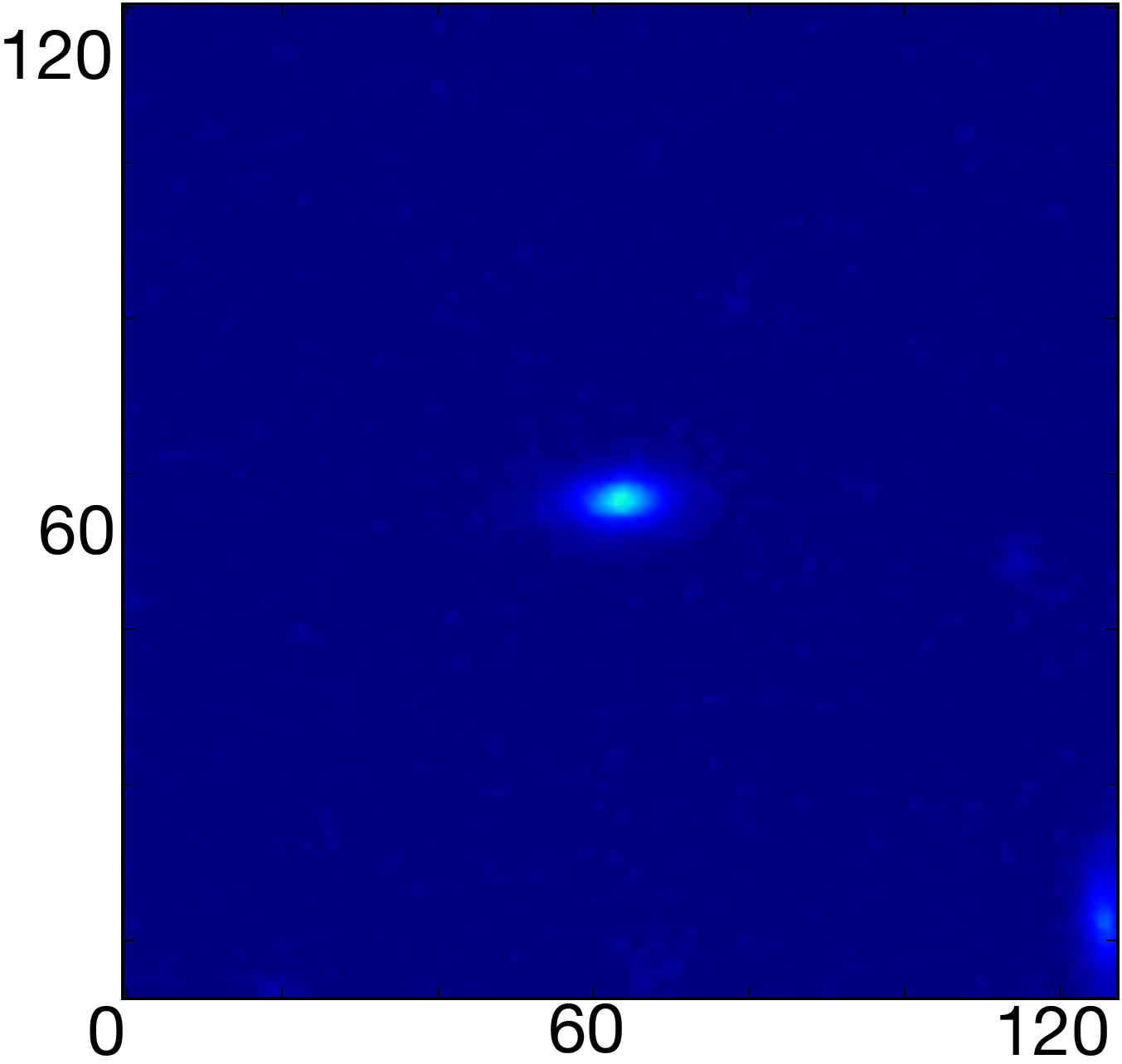} & \includegraphics[width=2.67cm, height=2.5cm]{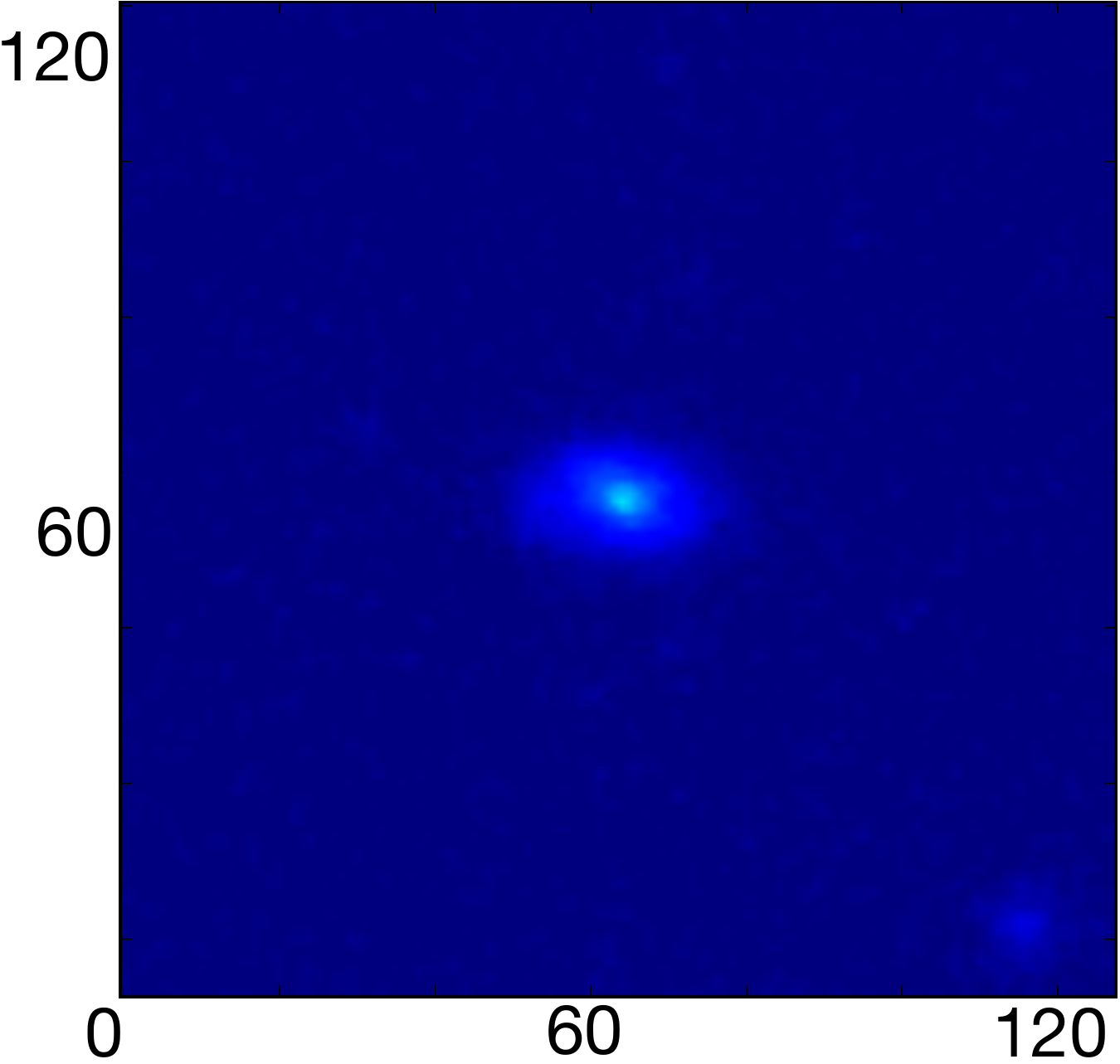} & \includegraphics[width=2.67cm, height=2.5cm]{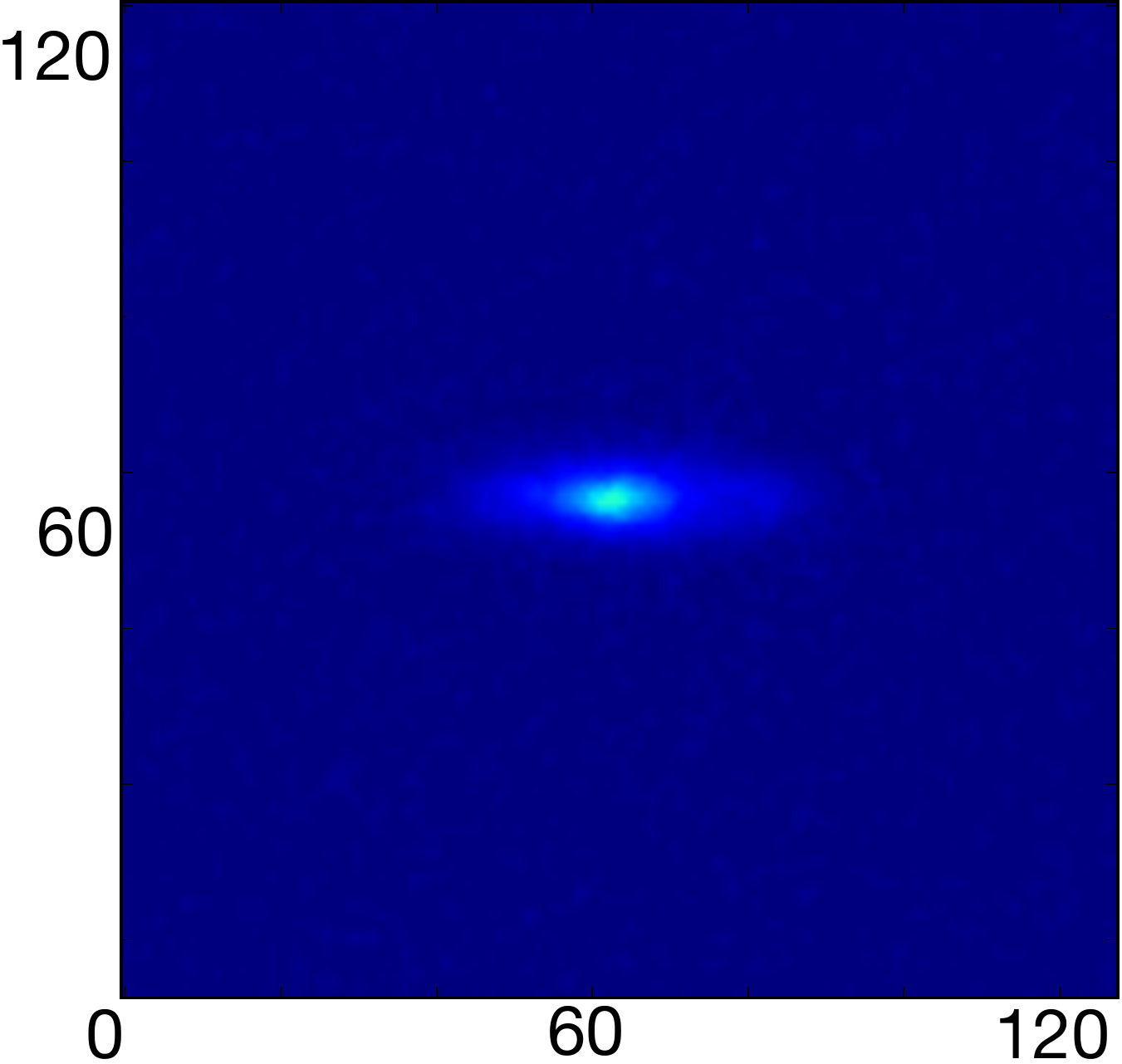} & \includegraphics[width=2.67cm, height=2.5cm]{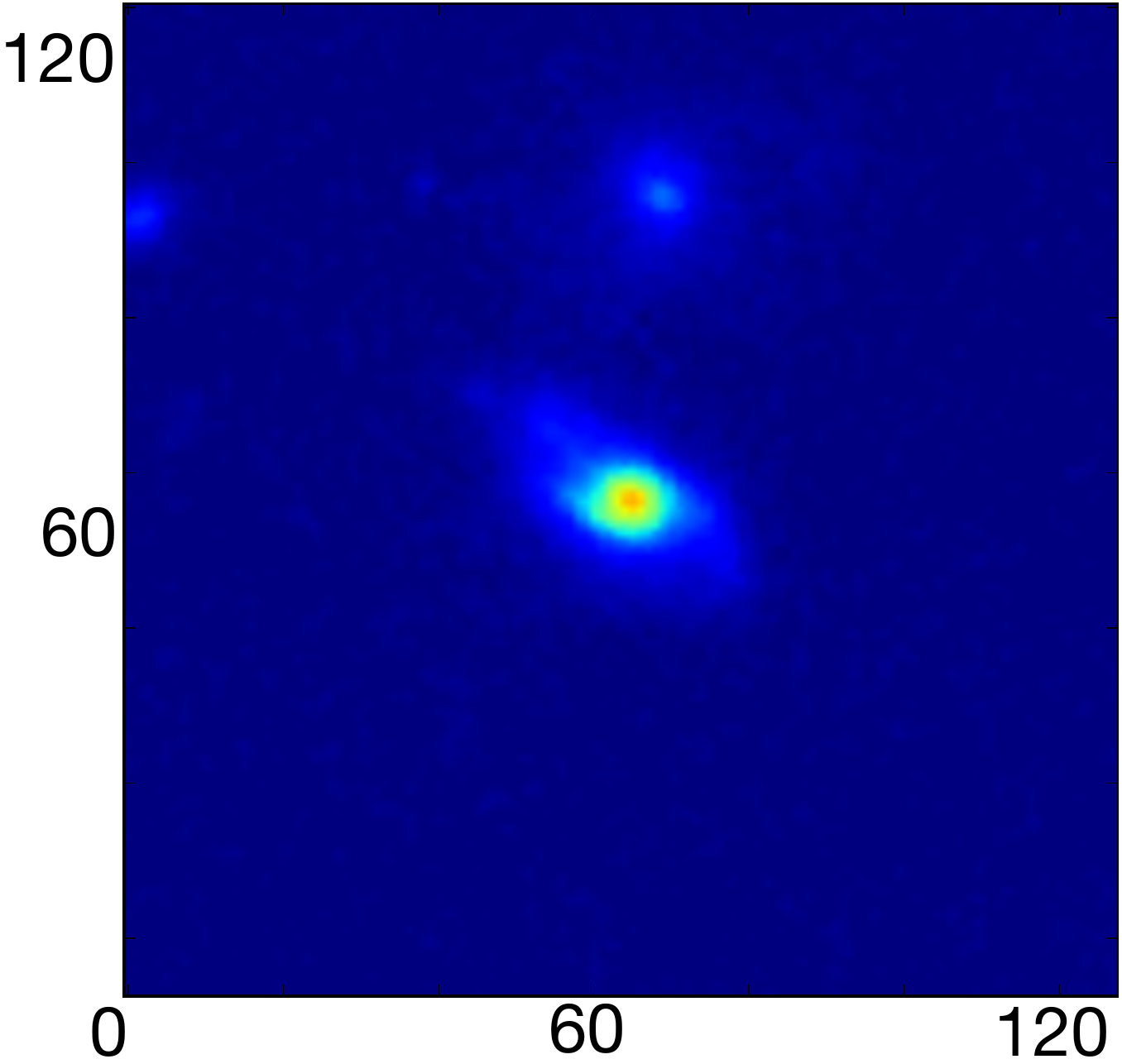} &
\includegraphics[width=2.67cm, height=2.5cm]{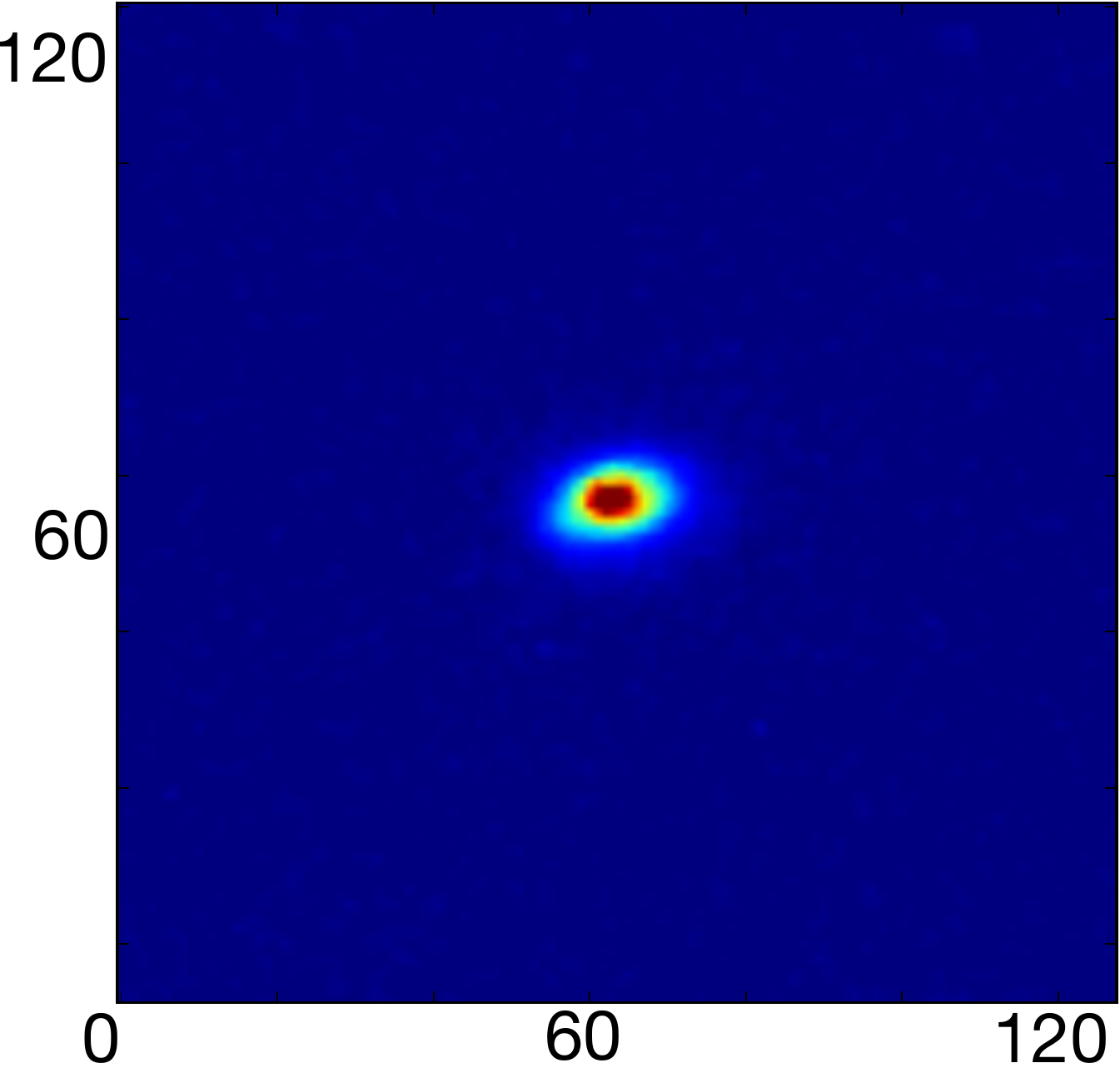} & \includegraphics[width=2.67cm, height=2.5cm]{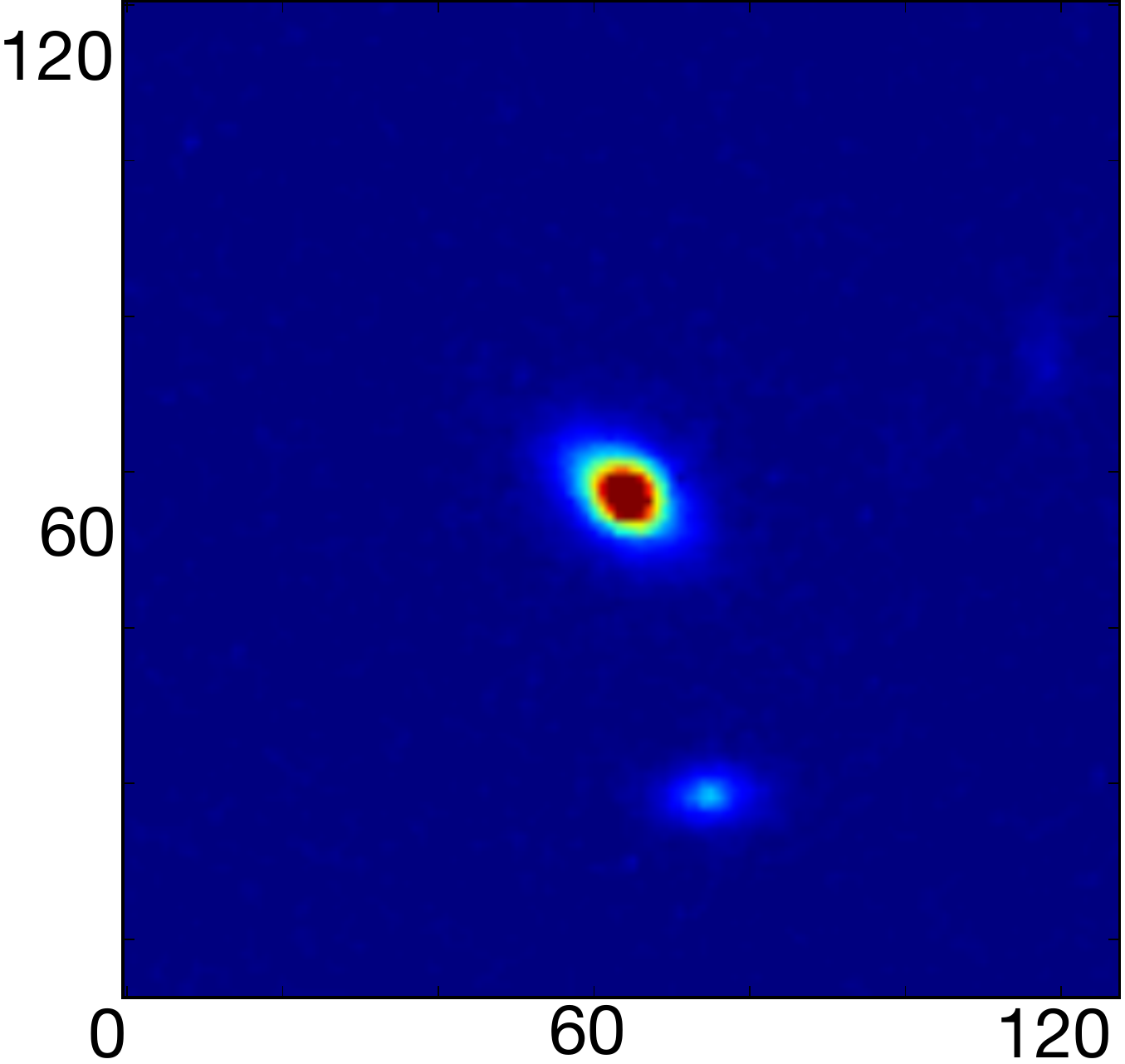}\\
\hline
\multicolumn{6}{c}{Stamps (128x128 pixels) of real galaxies poorly fitted by \textit{DeepLeGATo}}\\
\hline
\includegraphics[width=2.67cm, height=2.5cm]{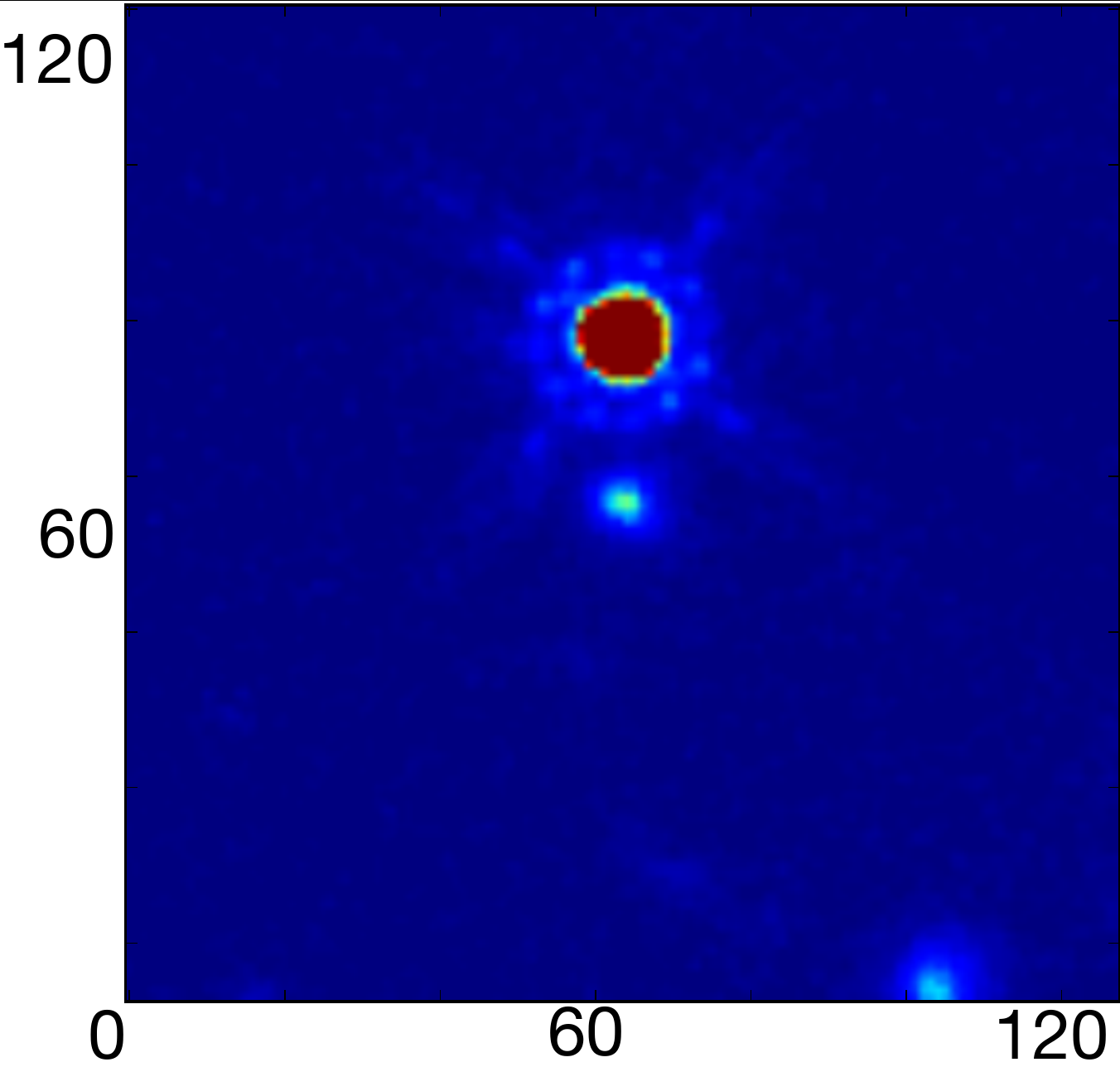} & \includegraphics[width=2.67cm, height=2.5cm]{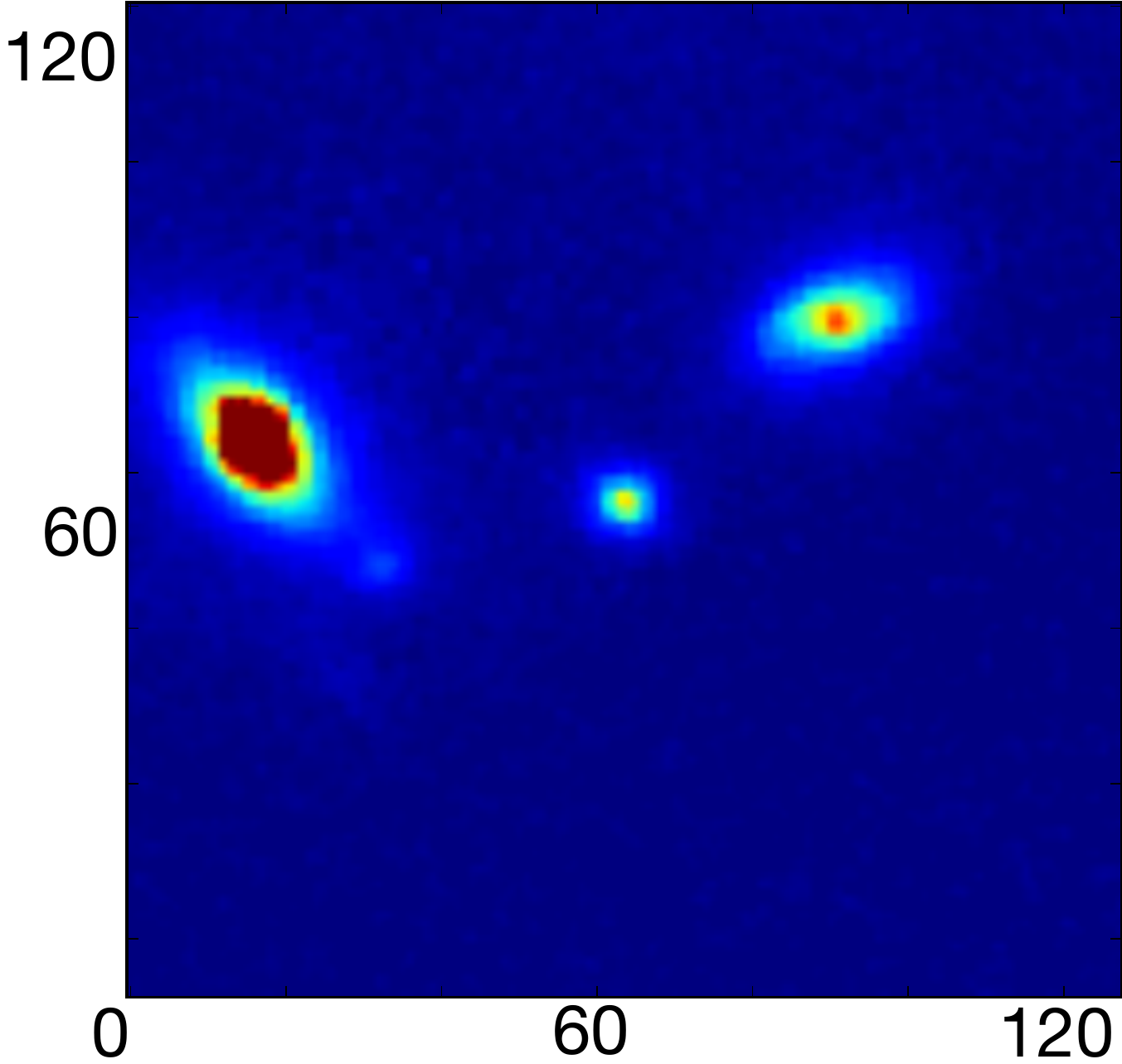} & \includegraphics[width=2.67cm, height=2.5cm]{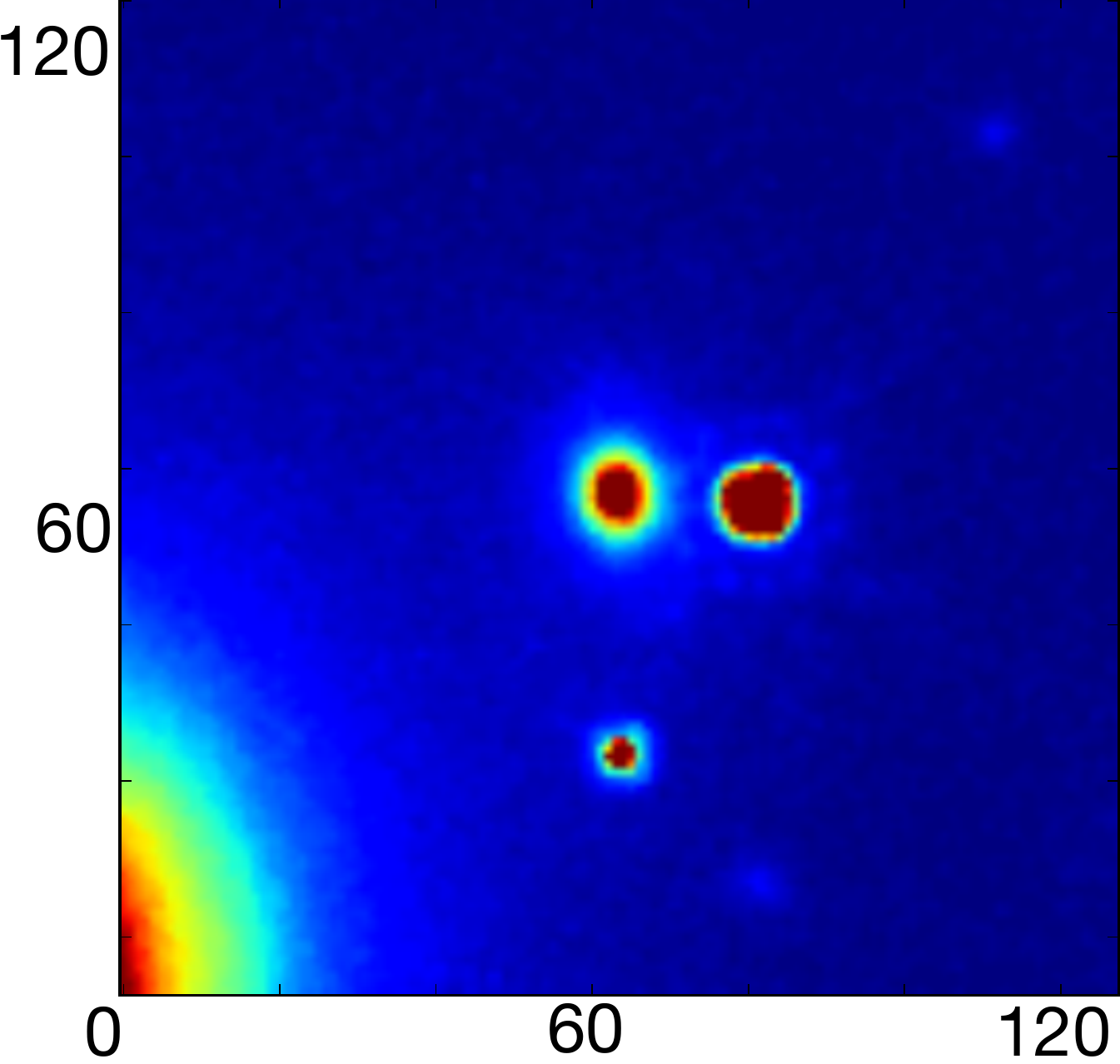} & \includegraphics[width=2.67cm, height=2.5cm]{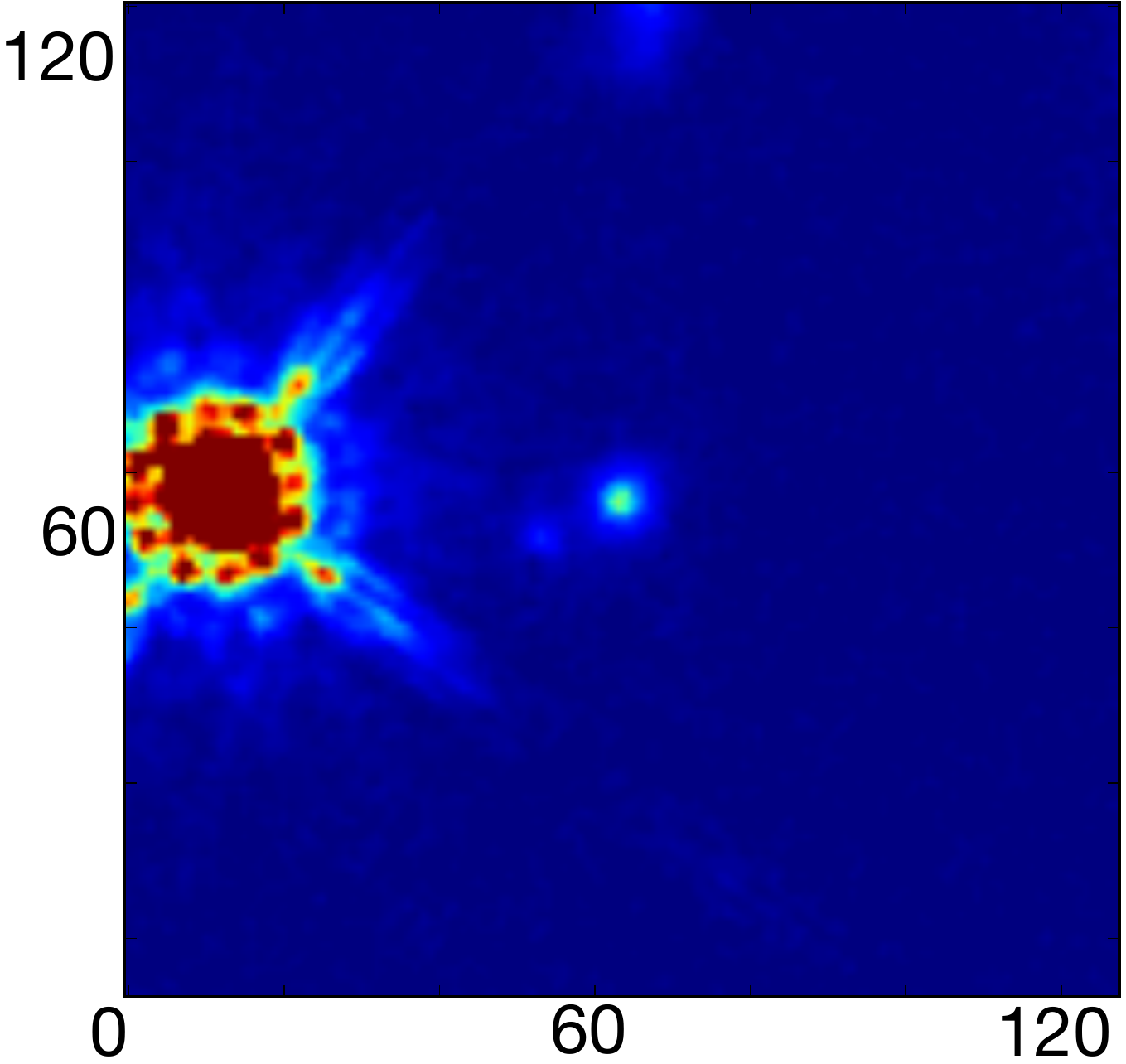} &
\includegraphics[width=2.67cm, height=2.5cm]{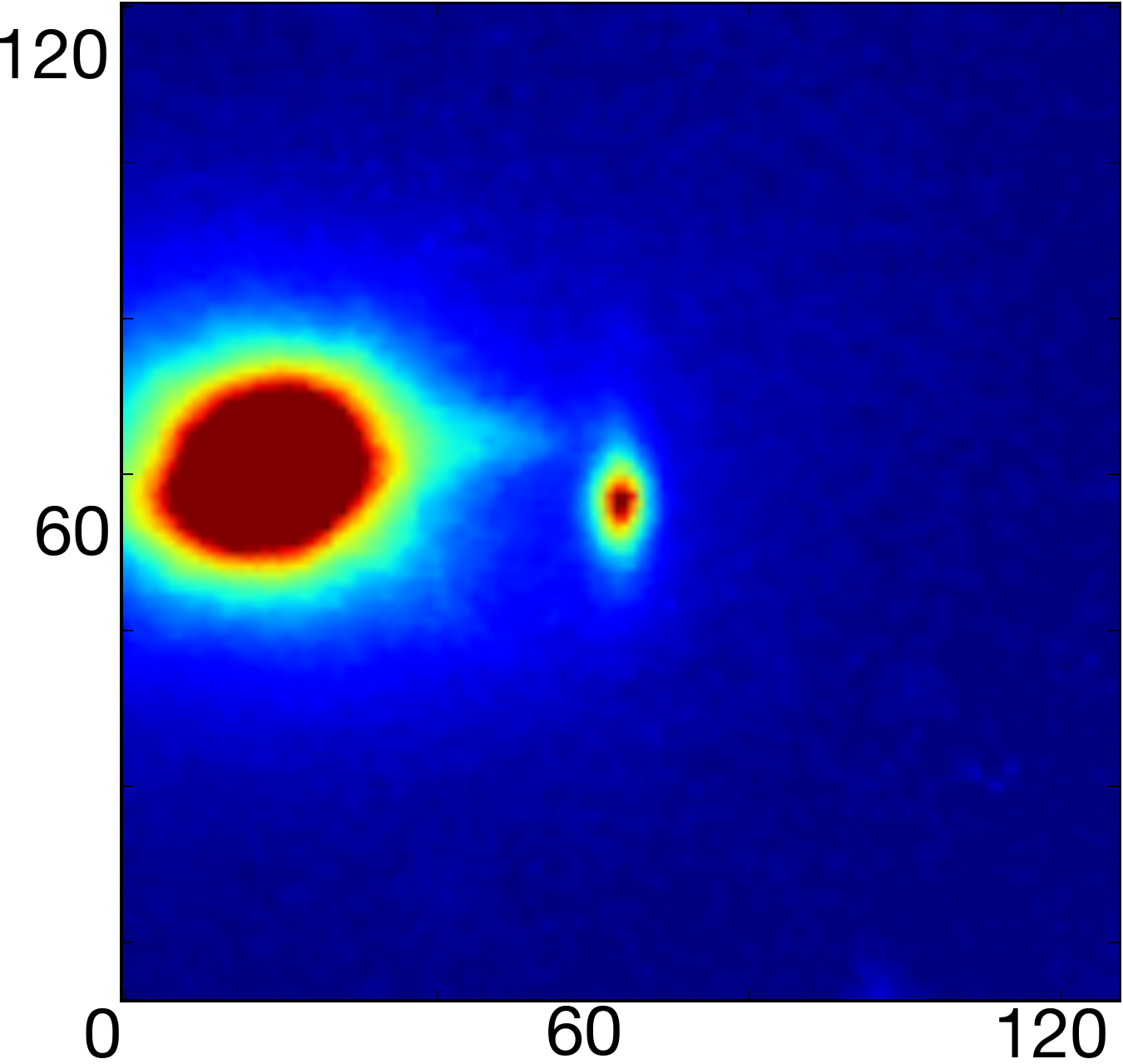} & \includegraphics[width=2.67cm, height=2.5cm]{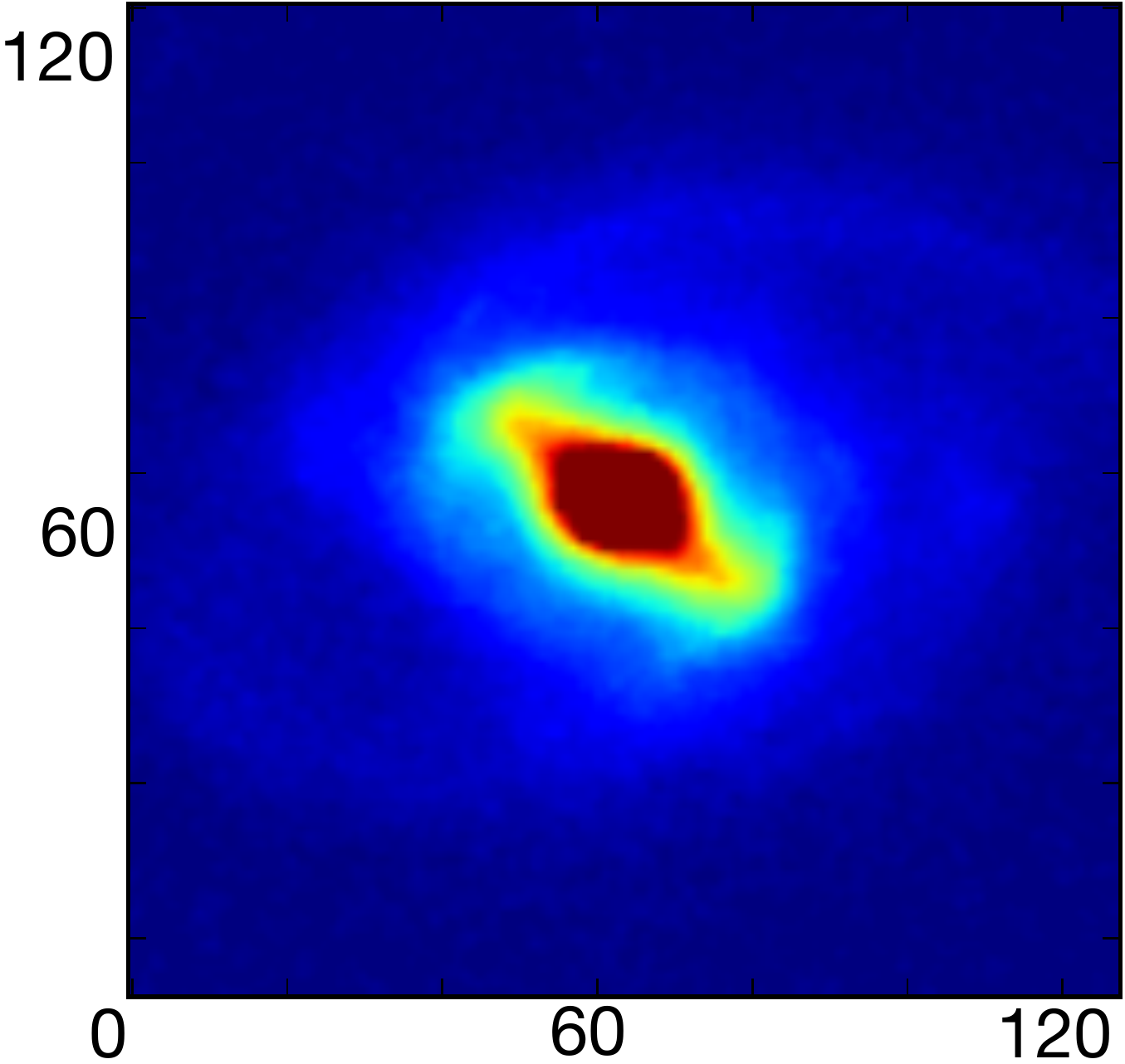}\\

\includegraphics[width=2.67cm, height=2.5cm]{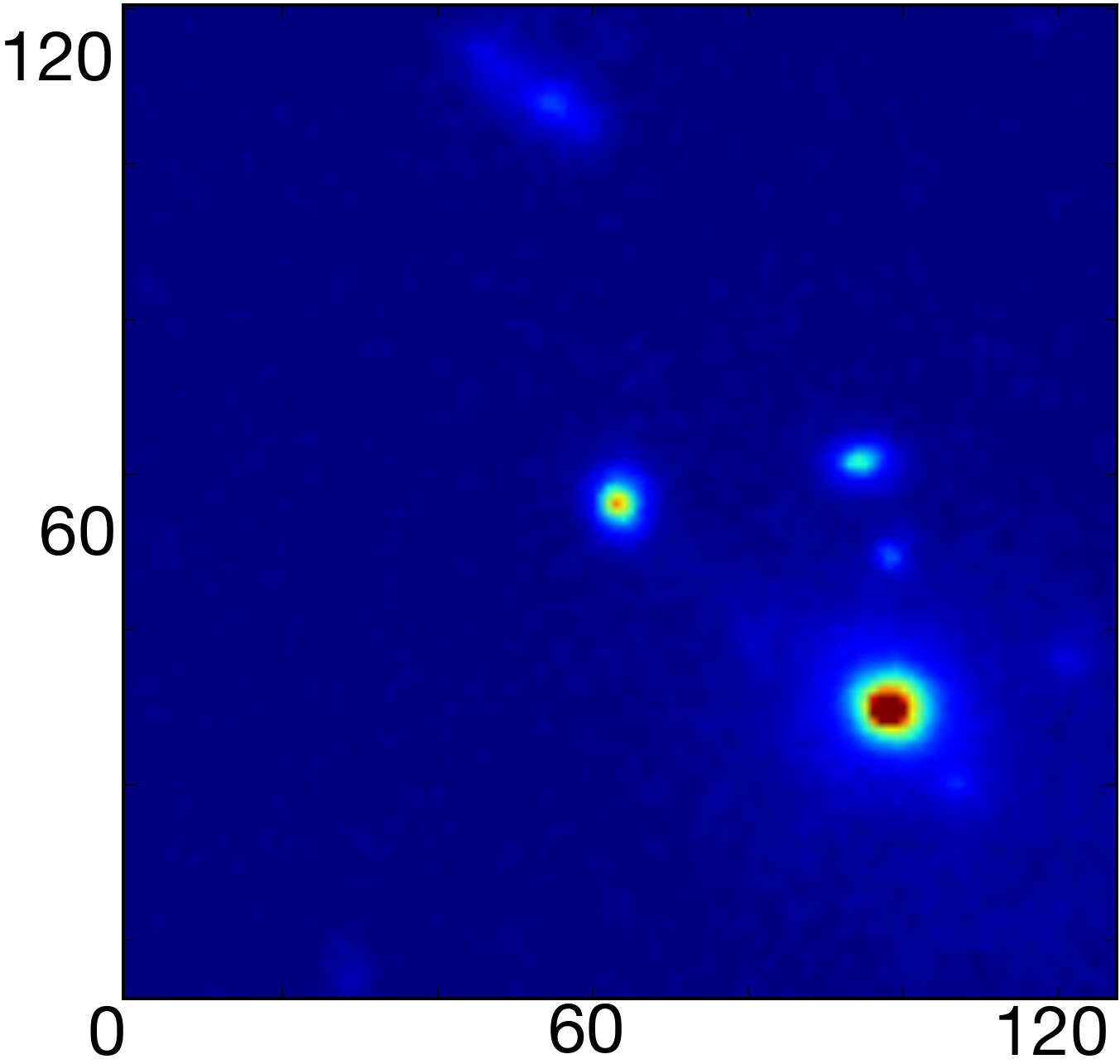} & \includegraphics[width=2.67cm, height=2.5cm]{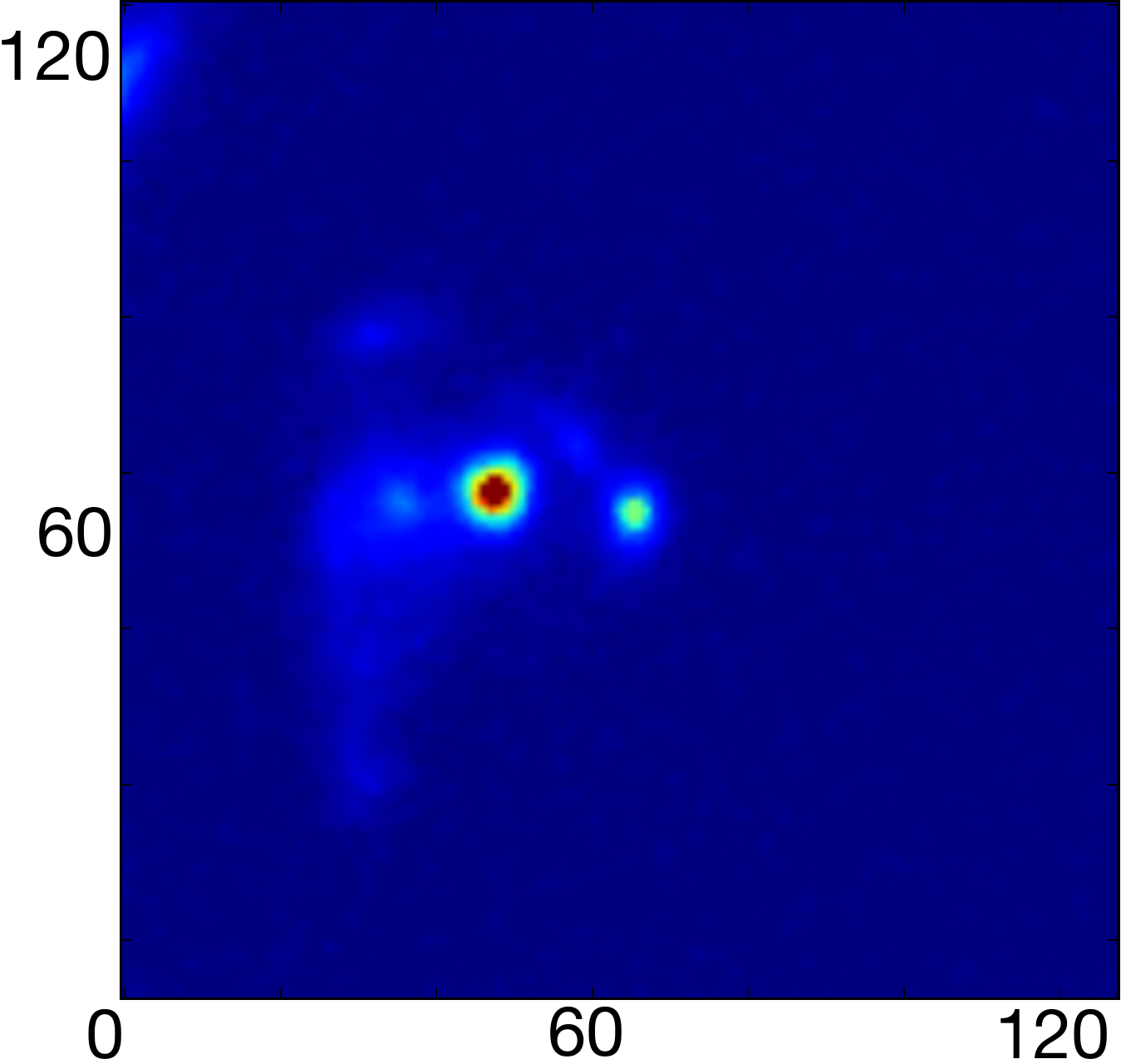} & \includegraphics[width=2.67cm, height=2.5cm]{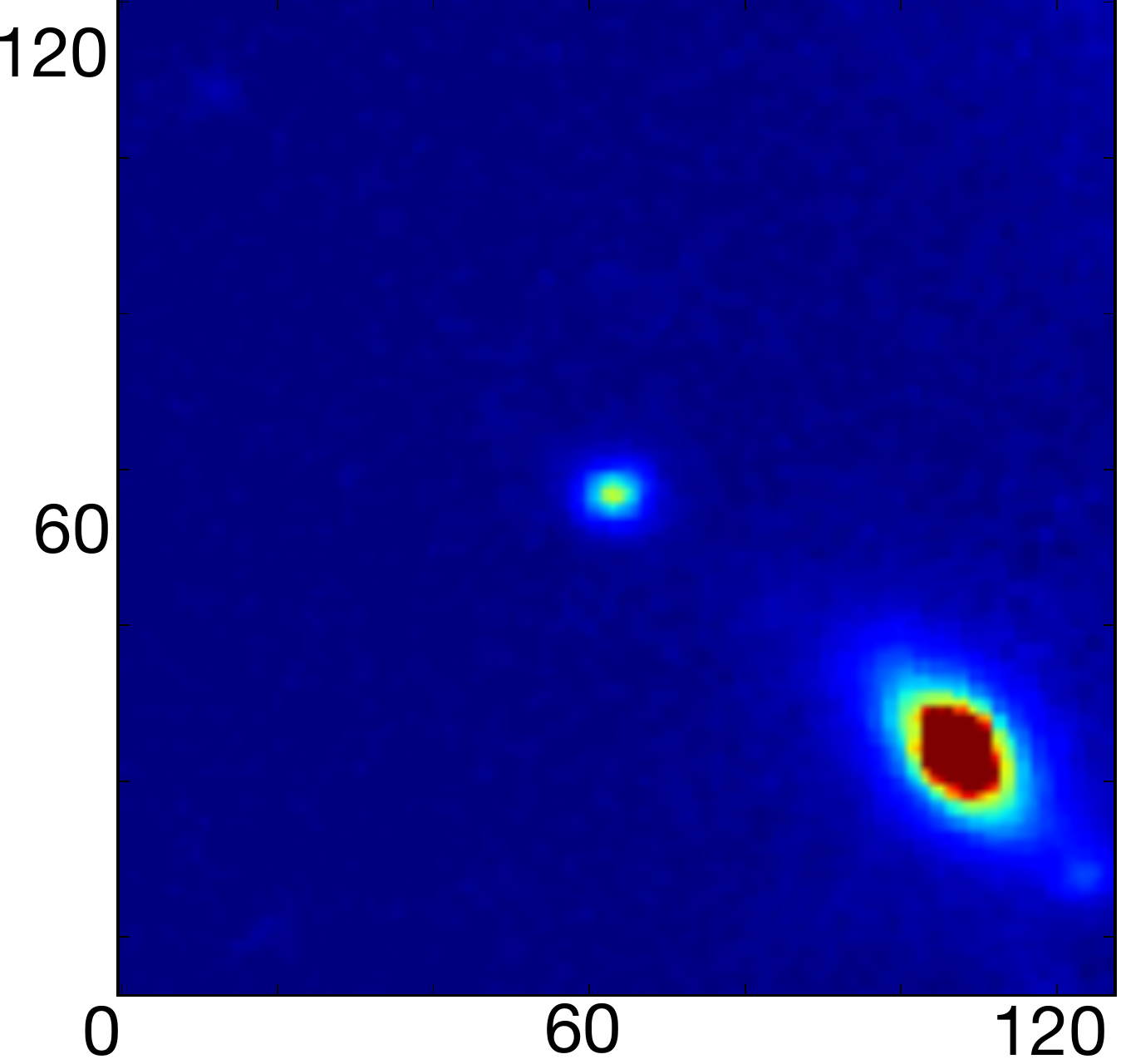} & \includegraphics[width=2.67cm, height=2.5cm]{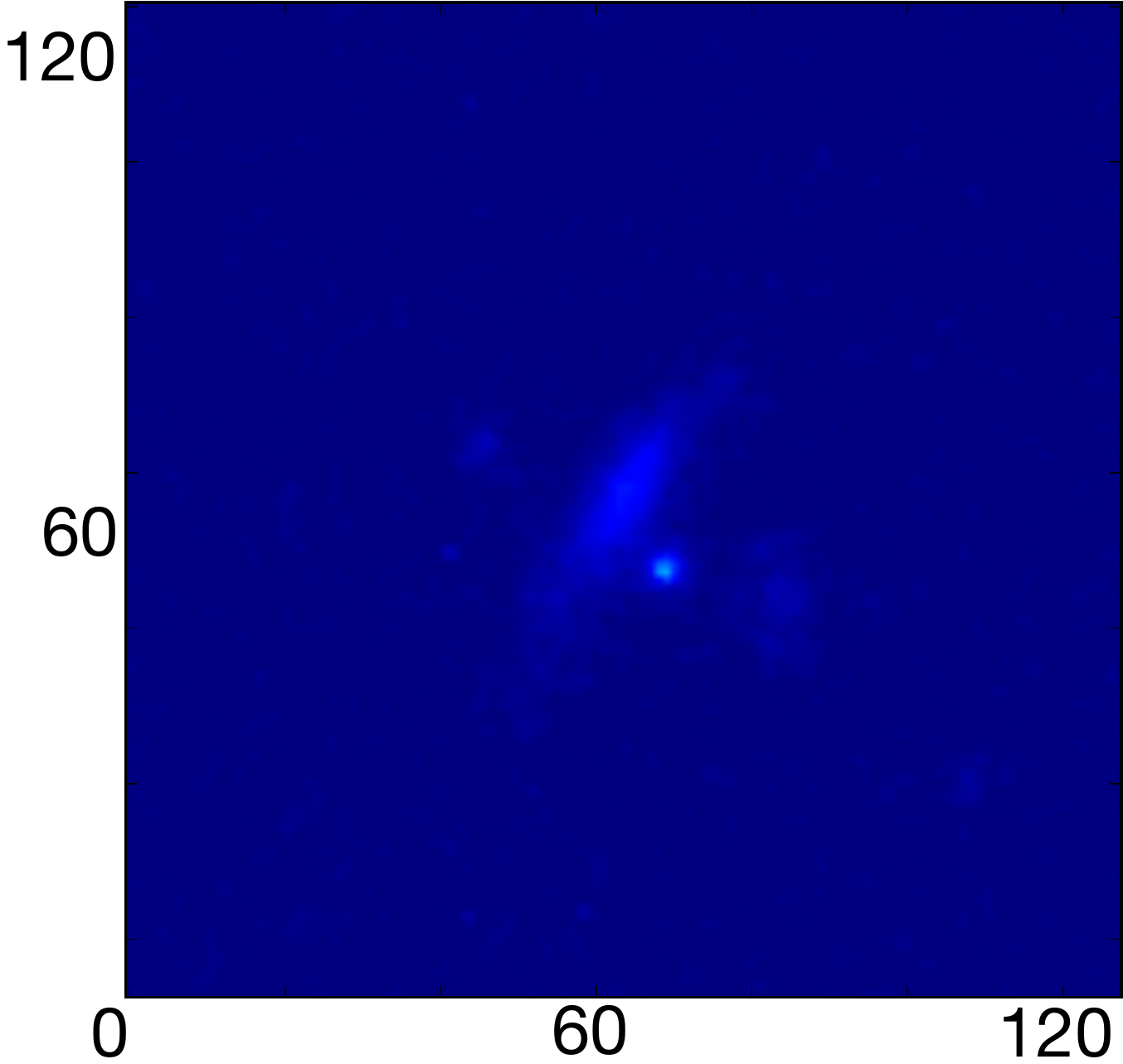} &
\includegraphics[width=2.67cm, height=2.5cm]{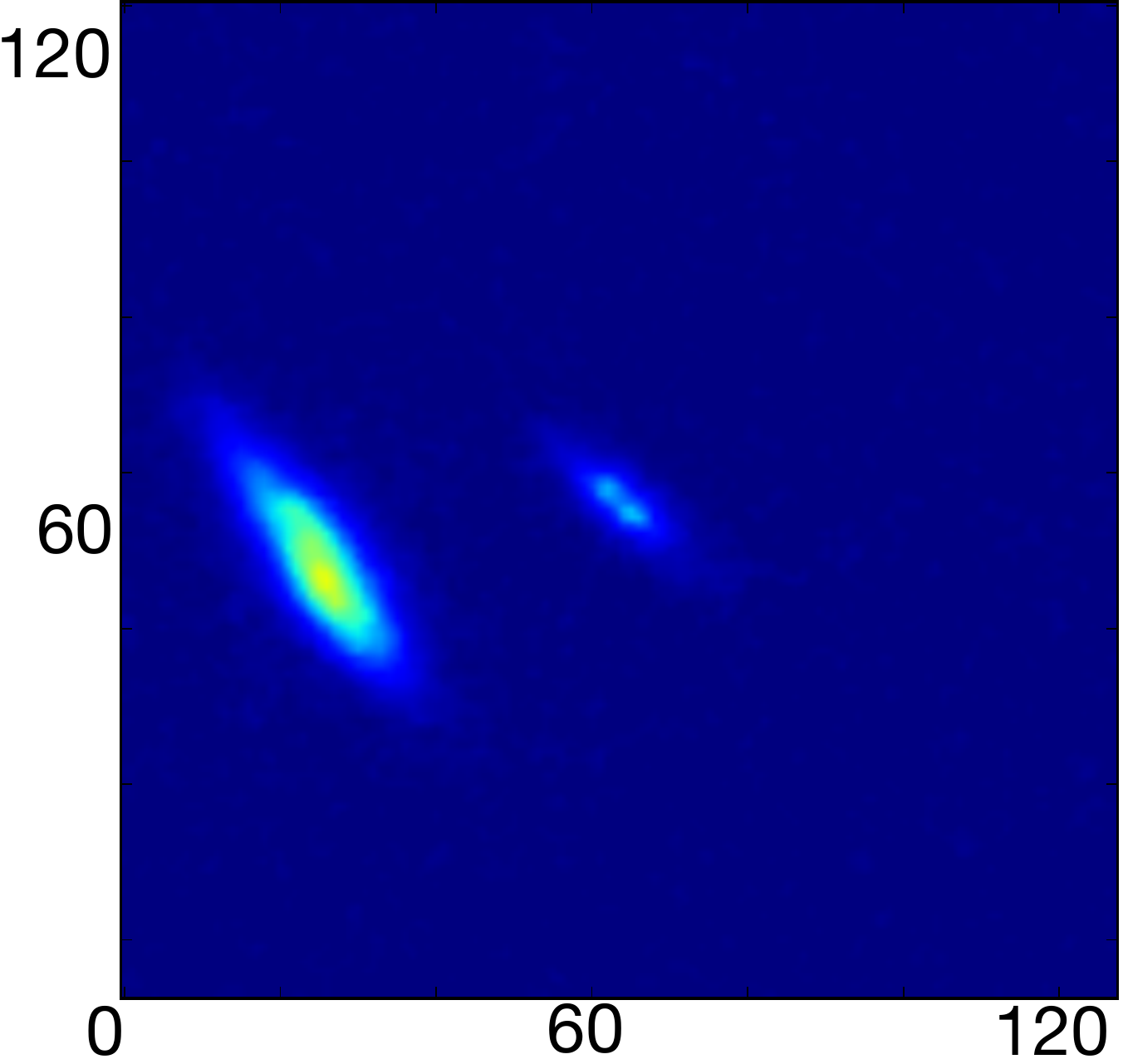} & \includegraphics[width=2.67cm, height=2.5cm]{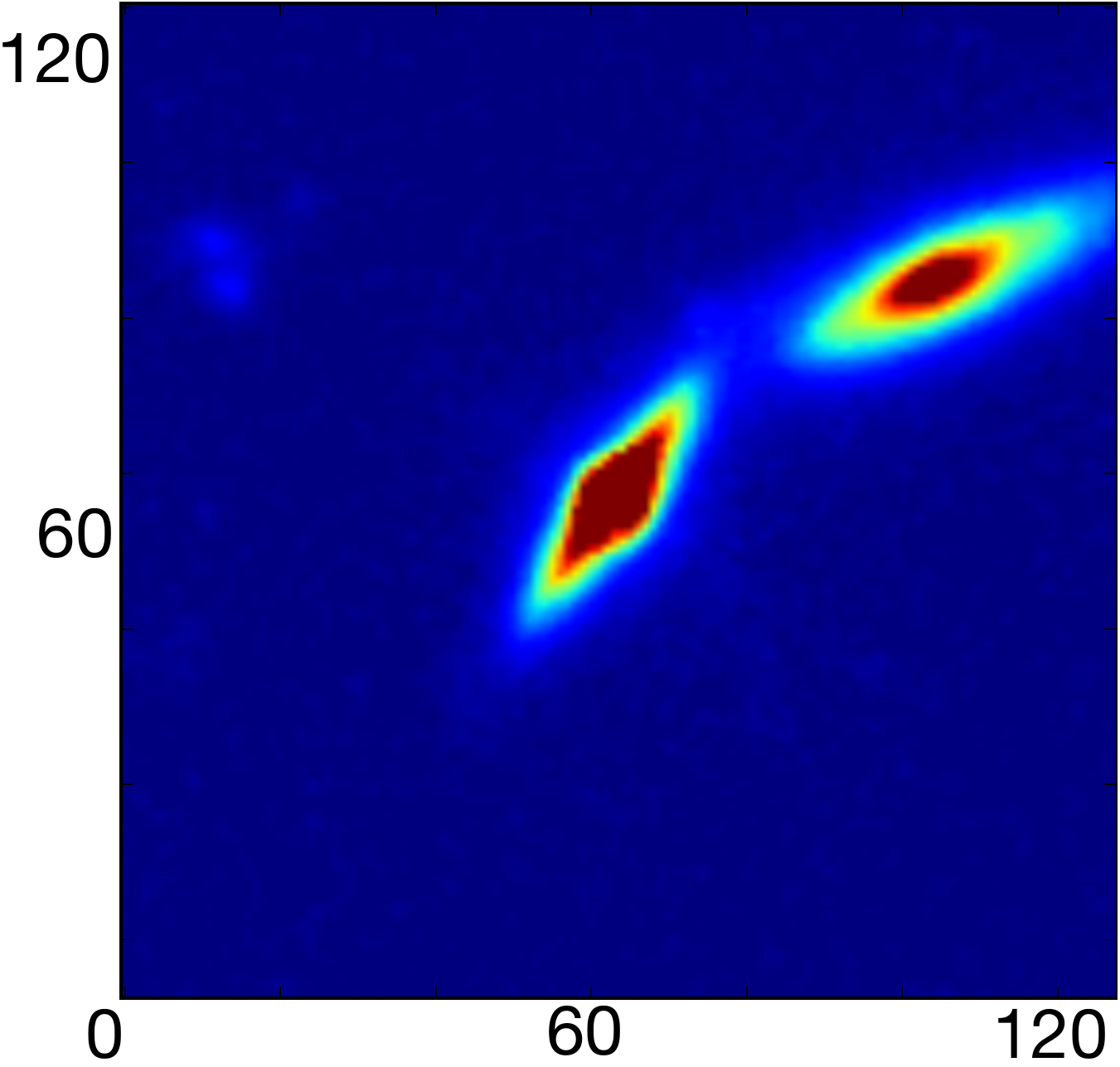}\\

\end{tabular}
\captionof{figure}{The first two lines we show twelve randomly selected stamps of simulated galaxies. The third and the fourth line show twelve stamps of real galaxies selected between the ones better predicted by our CNNs. Finally, the last two lines shows twelve stamps of real galaxies selected between the ones worse predicted by our CNNs. From these images we can see that the CNNs purely trained on simulated data are able to recover the parameters of real galaxies that are as regular and isolated as the simulated. The CNNs can still give accurate predictions when the real galaxies show smooth asymmetries and in the stamps there are other fainter and smaller galaxy companions. They give worse results when the stamps include several brighter and bigger galaxy companions.} 
\label{stamps}
\end{table*}

In order to verify the conclusions of this visual analysis, we calculated, for the whole test sample of real data, the number of galaxies having at least one companion close enough to fall within  the stamp (i.e., within $\sim 3\farcs8$). We obtain that almost all the galaxies (897 out of 1000) have at least one companion and that the mean number of companions for galaxy is $\sim 2.5$. We verified that, for all parameters, the accuracy of the CNNs predictions is clearly correlated with the presence of a bright companion. In the three bottom panels of Figure \ref{scatter1} we show this trend for the magnitude. We calculated the ratio between the flux of the fitted galaxy and the one of its brighter companion. In panel (c), we show the results obtained fitting the 142 galaxies whose companion has at least the 50\% of their flux. In panel (d) we show the results for the 450 galaxies whose companion have less than the 10\% of the flux of the galaxy. Finally, panel (e) shows the results for the 103 isolated galaxies of our test-sample, i.e. without companion within the stamp. For the latter sample, we also computed the $R^2$, reported in the second column of Table \ref{table_testRealData}. In Figures \ref{scatter2}, \ref{scatter3} and \ref{scatter4} of the Appendix A, and in the second column of Table \ref{table_testRealData}, we repeat the same analysis for the half-light radius, the S\'ersic index and the axis ratio. We conclude that on isolated galaxies, our machine trained on simulations is able to retrieve accurately the structural parameters.

\subsection{Domain adaptation}
\label{DomainAdap}

In the previous subsection, we show that a direct application on real data of the CNN models trained on the simulated data does not lead to predictions comparable to the ones obtained using GALFIT if galaxies have bright neighbors. This is because we did not include companions in our simulations. One simple possibility to overcome this problem is to produce more realistic simulations. In this section, we explore an alternative based on domain adaptation between networks. There is a vast literature in computer science dealing with this kind of problems, where methods are trained and evaluated on a certain kind of image distribution but then they are applied to changing visual domains. In general, visual domains could differ in some combination of (often unknown) factors, including viewing angle, resolution, intra-category variation, object location and pose. Studies have demonstrated a significant degradation in the performance of image methods due to these domain shifts. Therefore, methods of so-called  \textit{domain adaptation} have been developed to deal with these situations, where the task of the machine learning method remains the same between each set, but the input distribution is slightly different. 

Adopting a \textit{domain adaptation} strategy, we saved our best models trained and validated on simulated data, and we repeated their training and validation using a small sample of real data. The core idea of this strategy is that the same representation learned on simulated data may be useful to adapt the learning system on the second setting of real data. This way, we exploit what the CNNs have learned on one setting to improve the generalization in another one.

In section \ref{realData} we described a sample of 5000 real galaxies  included in the \cite{vanderWel_2012} catalogue. We divided that sample into a test sample $A$ of 1000 galaxies, and a training sample $B$ including the remaining 4000 galaxies. We trained and validate our CNNs with different subsets $(B_1, B_2, B_3, ...)$ of $B$, having size variating between 50 and 4000. In agreement with the procedure that we followed in this work, each of the $B_i$  samples were divided in the proportion of 4/5  for the training and 1/5 for the validation. After the training and validation of the \textit{domain adapted} CNN, we applied the result of the new model on the sample $A$ and we calculated the  $R^2$ of the prediction for each galaxy structural parameter retrieved.  Through all this step we considered the structural parameters estimations given in the \cite{vanderWel_2012} catalogue as the \textit{ground truth} of our models. 

\begin{figure*} 
\centering
\subfloat[Total magnitude (mag)]{
  \includegraphics[width=86mm]{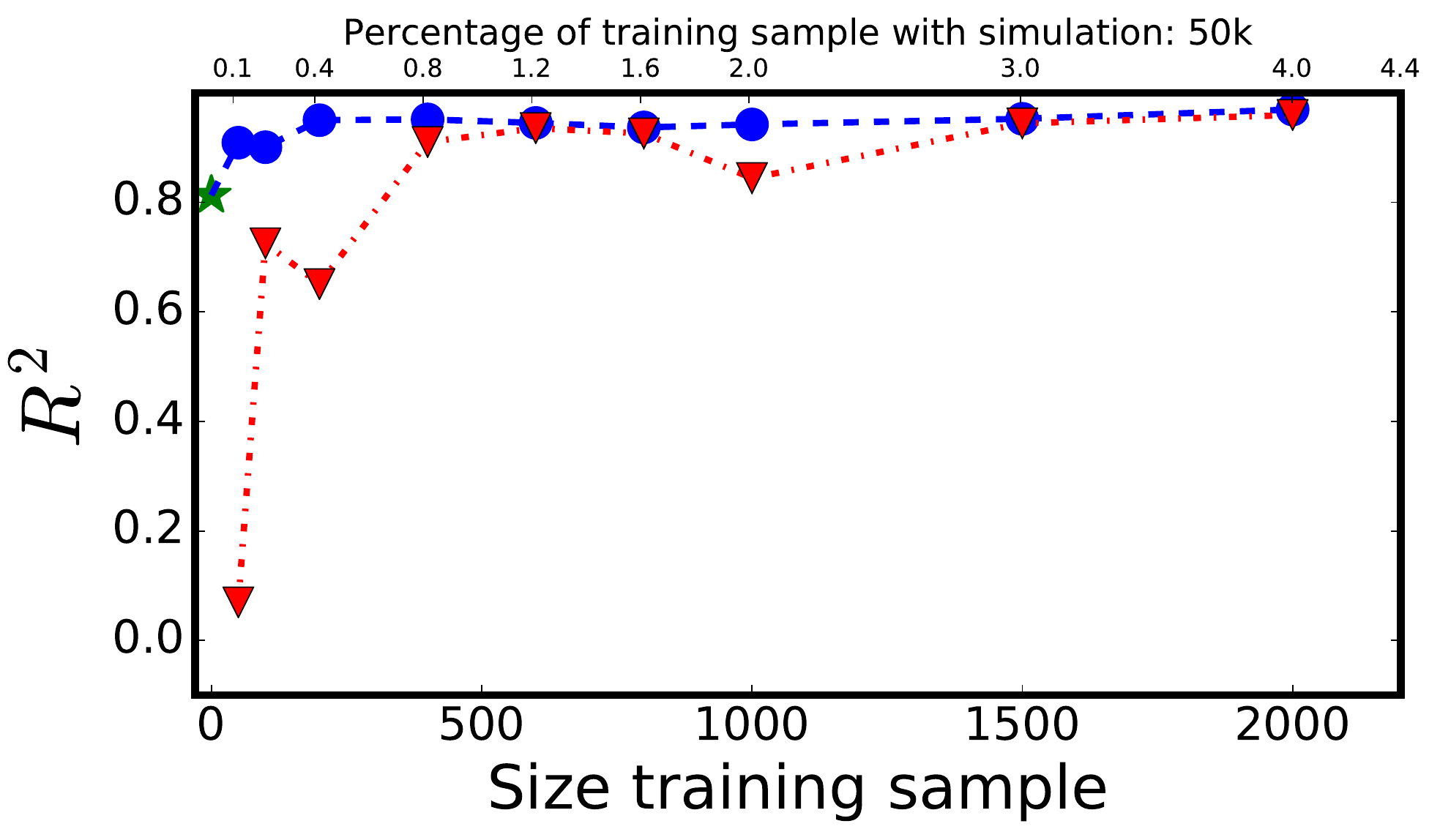}
}
\subfloat[Half-light radius $(R_e)$ ]{
  \includegraphics[width=86mm]{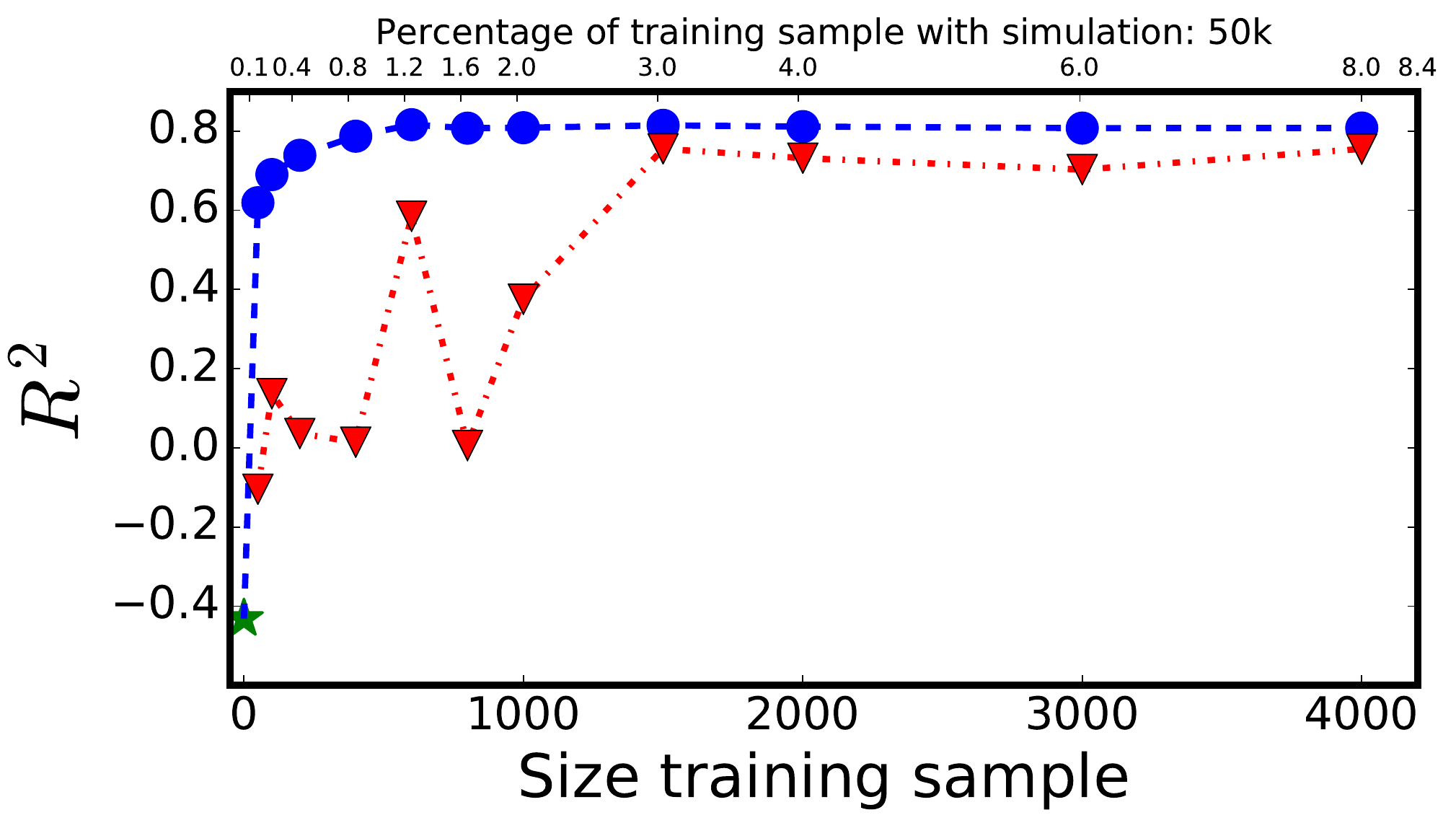}
}
\hspace{0mm}
\subfloat[S\'ersic  index ($n$)]{
  \includegraphics[width=86mm]{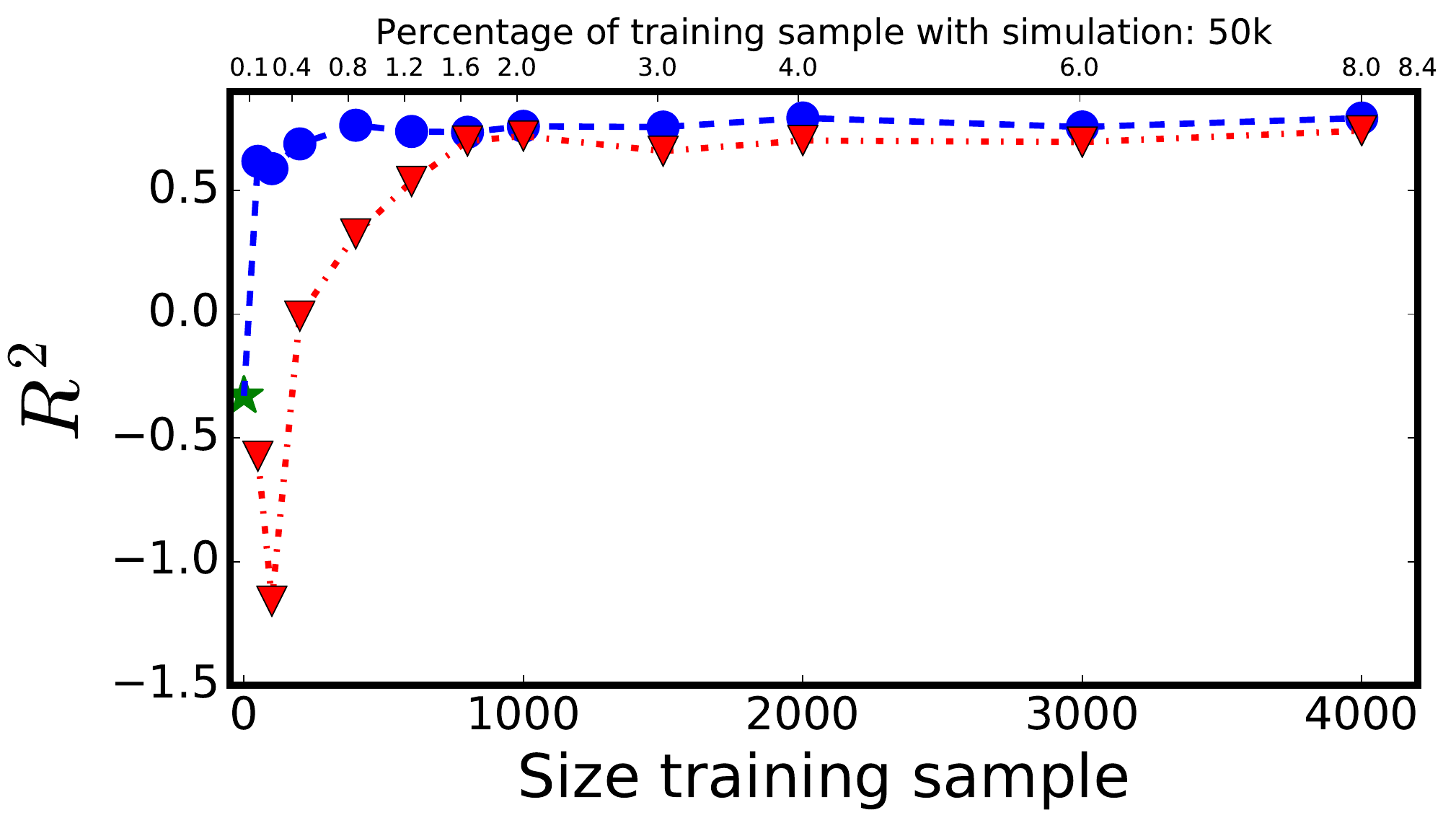}
}
\subfloat[Axis ratio (q)]{
  \includegraphics[width=86mm]{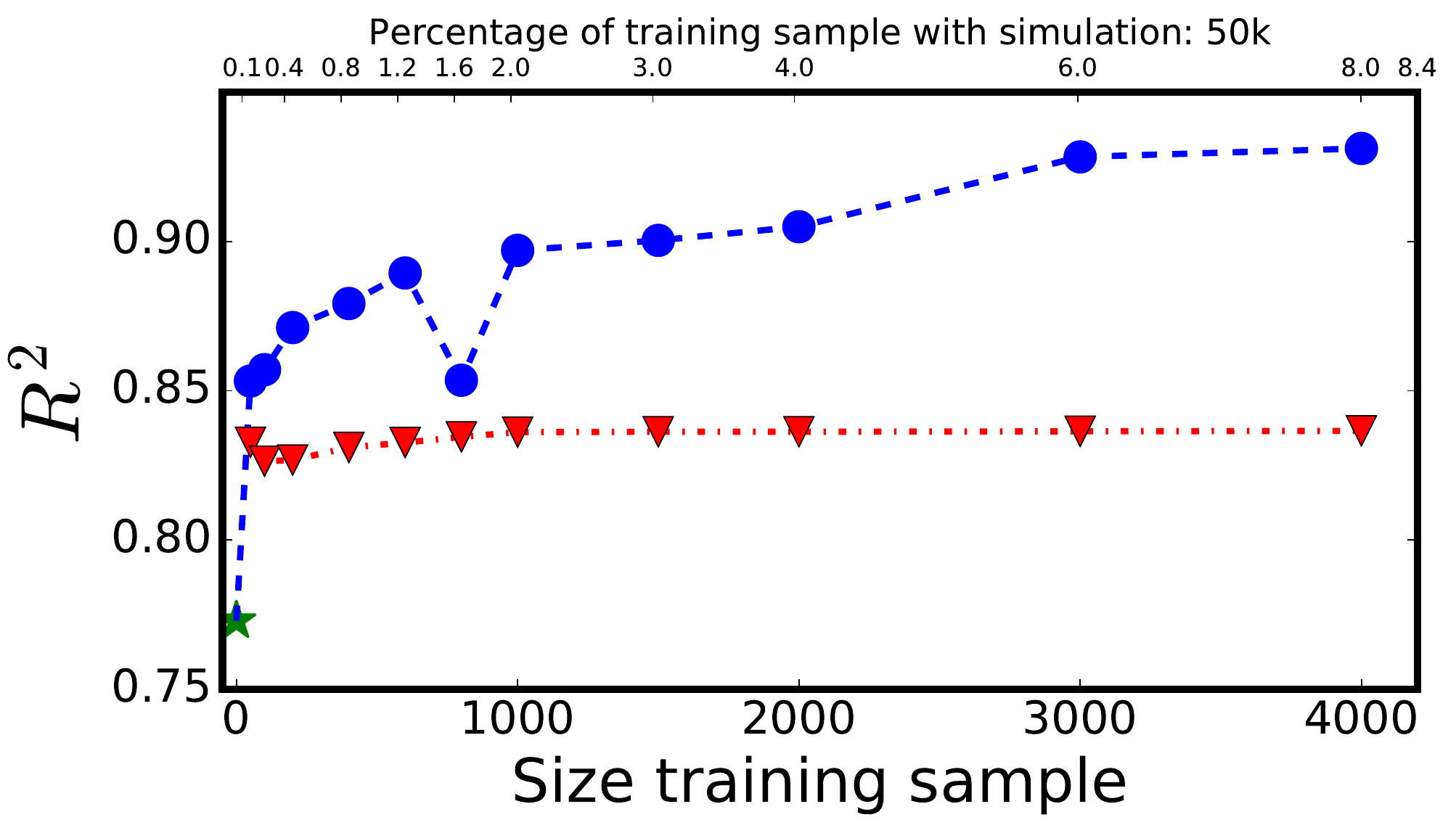}
  }
\caption{We show the $R^2$ for the profile fitting of 1000 real CANDELS/HST galaxies, obtained using our CNN code. The values of $R^2$ are plotted as a function of the size of real data sample used to train and validate the method. In particular, the blue line represents the behavior of our CNN previously trained on simulated data and then \textit{domain adaptated} using real galaxies. The staring point of the blue line, the green star, represent the $R^2$ value obtained using the CNN purely trained on simulated data (see section \ref{directApp}). The red line represents the behavior of our CNN directly trained on real data (i.e. without \textit{domain adaptation}). The points on the x-axis, i.e. the size of the training/validation sample are: 50, 100, 200,400, 600, 800, 1000, 1500, 2000, 3000, 4000. See section \ref{DomainAdap}}
\label{transferLear}
\end{figure*}

In Figure \ref{transferLear} we show how the $R^2$ of the prediction varies as a function of the size of the real data used to train and validate the CNN. The blue lines represent the behavior of our CNNs previously trained on simulated data, to which we then applied \textit{domain adaptation}; the green star at the beginning of the blue lines represents the $R^2$ of the prediction before the \textit{domain adaptation}. The red lines represent the behavior of our CNNs directly trained on the $B_i$ samples of real data (i.e. without \textit{domain adaptation}). Each parameter is estimated independently by the CNN that showed the best accuracy for the profiling of the simulated data (see Table \ref{Architectures}). 
For the magnitude, with only 50 real galaxies used as \textit{domain adaptation}, we obtain an $R^2 \sim 0.9$ and with 200 real galaxies an $R^2 \sim 0.97$ (i.e a 0.4 per cent of the 50k simulated galaxies formerly used to train and validate the CNN). In the case of the magnitude, without \textit{domain adaptation} (red line), with 800 real galaxies we already reach an $R^2 \sim 0.96$. 
For the half-light radius, the $R^2$ has a value of $\sim -0.43$ when trained only with simulated data (the green star), and it improves to $\sim 0.62$ with \textit{domain adaptation} made with only 50 real galaxies. It reaches an $R^2 \sim 0.81$ with 600 real galaxies, while without \textit{domain adaptation}, it needs $\sim 1500$ galaxies to reach an $R^2 \sim 0.75$
In the case of the S\'ersic index, the best $R^2 \sim 0.79$ is obtained with \textit{domain adaptation} made with 400 real galaxies, while the best performance for the CNN trained without \textit{domain adaptation} is  $R^2 \sim 0.76$ obtained with 1000 real galaxies. 
The trend for the axis ratio is smoother, with the blue line (\textit{domain adaptation curve}) improving in the whole range of sizes and obtaining $R^2 \sim 0.93$ when trained/validated with 4000 real galaxies. On the contrary, the red line (trend without \textit{domain adaptation}) reaches a plateau of $R^2 \sim 0.70$ with 800 real galaxies. 
We notice that for all parameters the blue line is always above the red line, i.e. when using \textit{domain adaptation curve} the CNNs need fewer real galaxies examples to reach their best performance on real galaxies and they are always more accurate than CNNs not previously trained with simulated data. 

In Figure \ref{scatter1} and in appendix \ref{appendix_a} we compare the result of the CNNs profiling of the one thousand galaxies before and after the \textit{domain adaptation}.

\subsection{Comparison with GALFIT}
\label{GalfitCompReal}

The results of the GALFIT predictions greatly depend on the initial parametrization used. Different parametrization may easily lead to different results. To take this effect into account, in the third column of Table \ref{table_testRealData} we compare, for our test set of real data, the predictions of \cite{vanderWel_2012} catalogue (used as ground truth) with the ones of another catalogue using GALFIT (Dimauro et al. in prep). As we can see, the $R^2$ of our models after the  \textit{domain adaptation} are very similar to the results obtained for the same set of data from these two catalogues using GALFIT.

In Figure \ref{PerformanceReal} we detail the comparison between our CNNs predictions and the two GALFIT catalogues, as a function of the magnitude.

\begin{table*}
\begin{tabular}{c c}
Results using \textit{DeepLeGATo} & Results using GALFIT\\
\includegraphics[width=8.5cm]{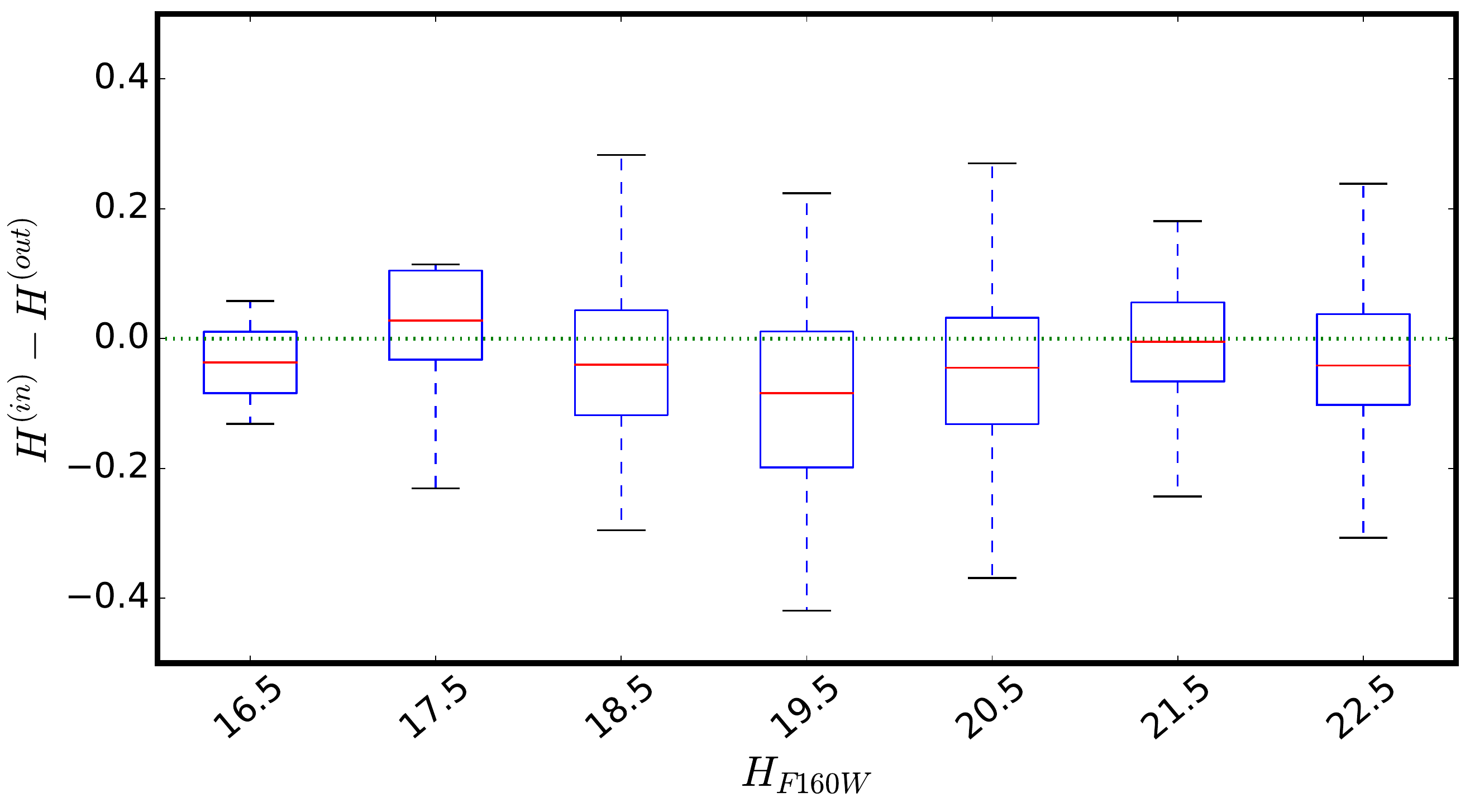} & \includegraphics[width=8.5cm]{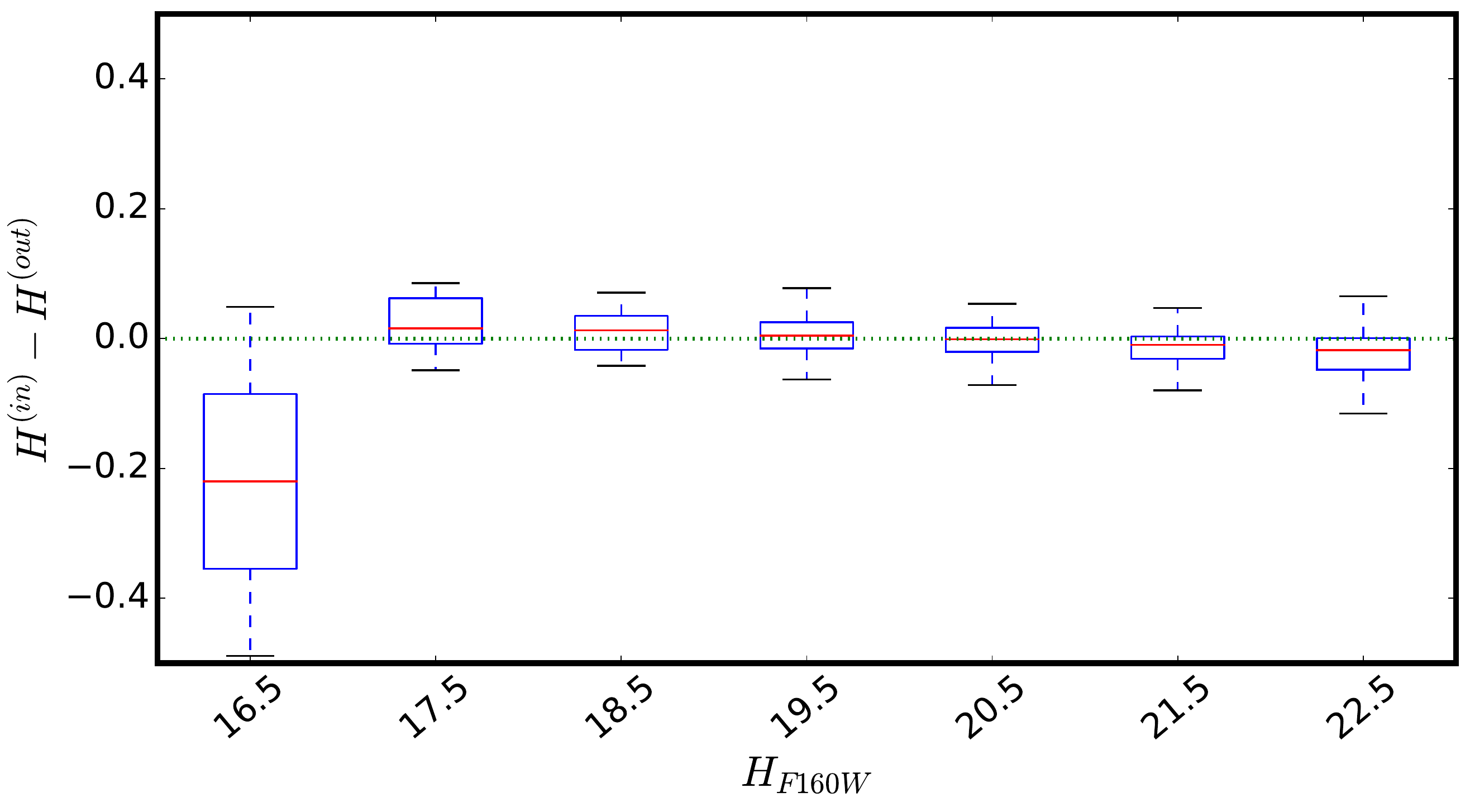} \\

\includegraphics[width=8.5cm]{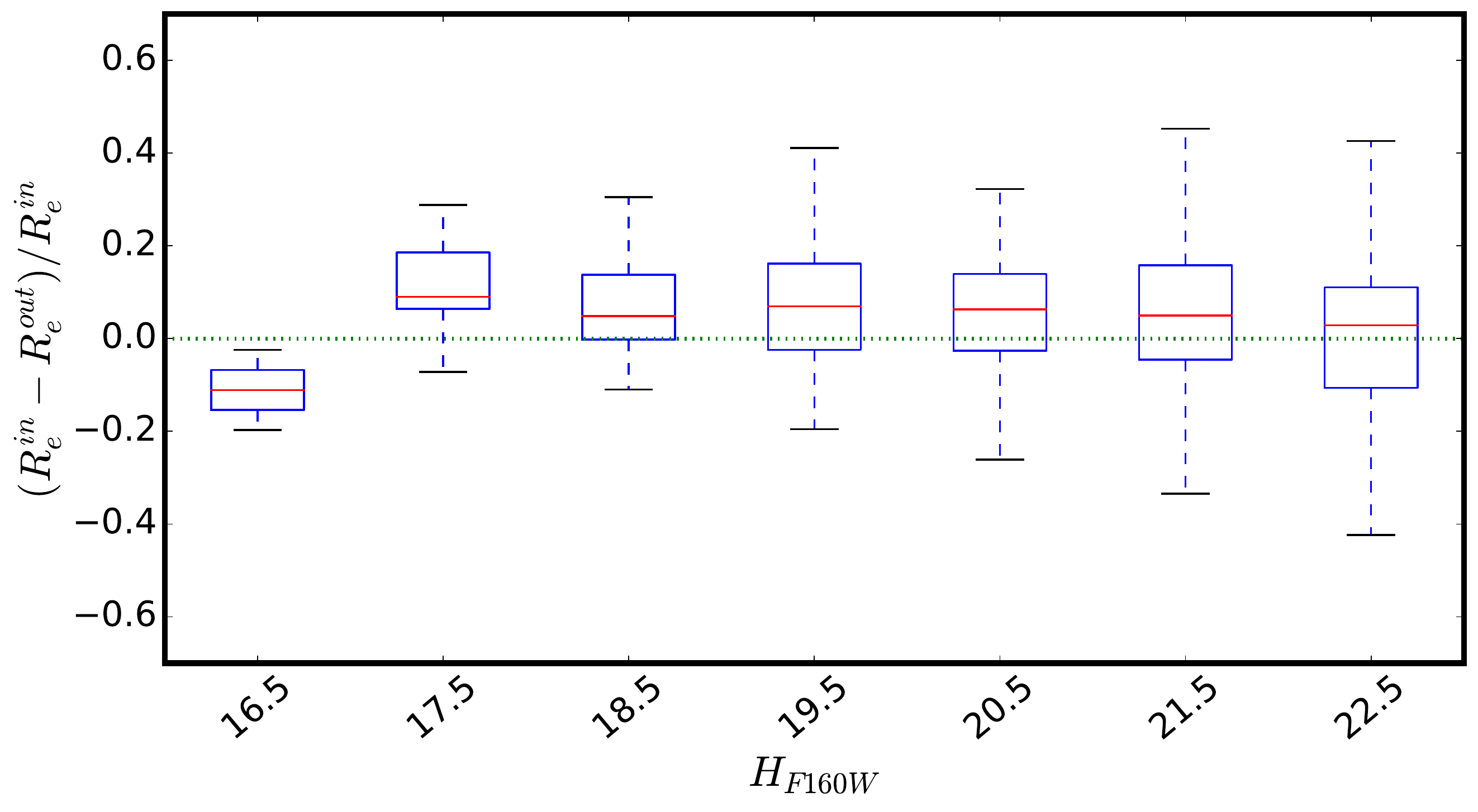} & \includegraphics[width=8.5cm]{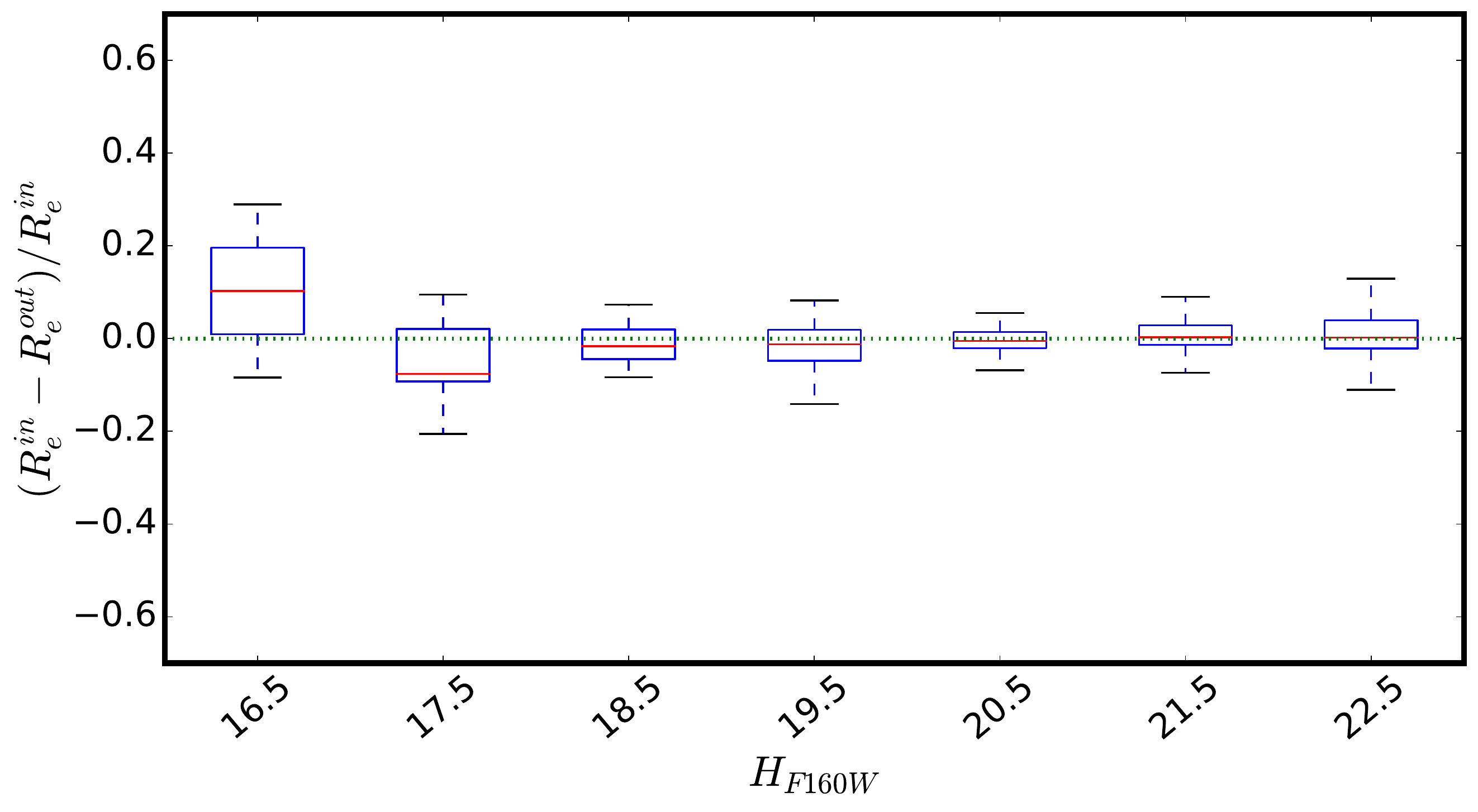} \\

\includegraphics[width=8.5cm]{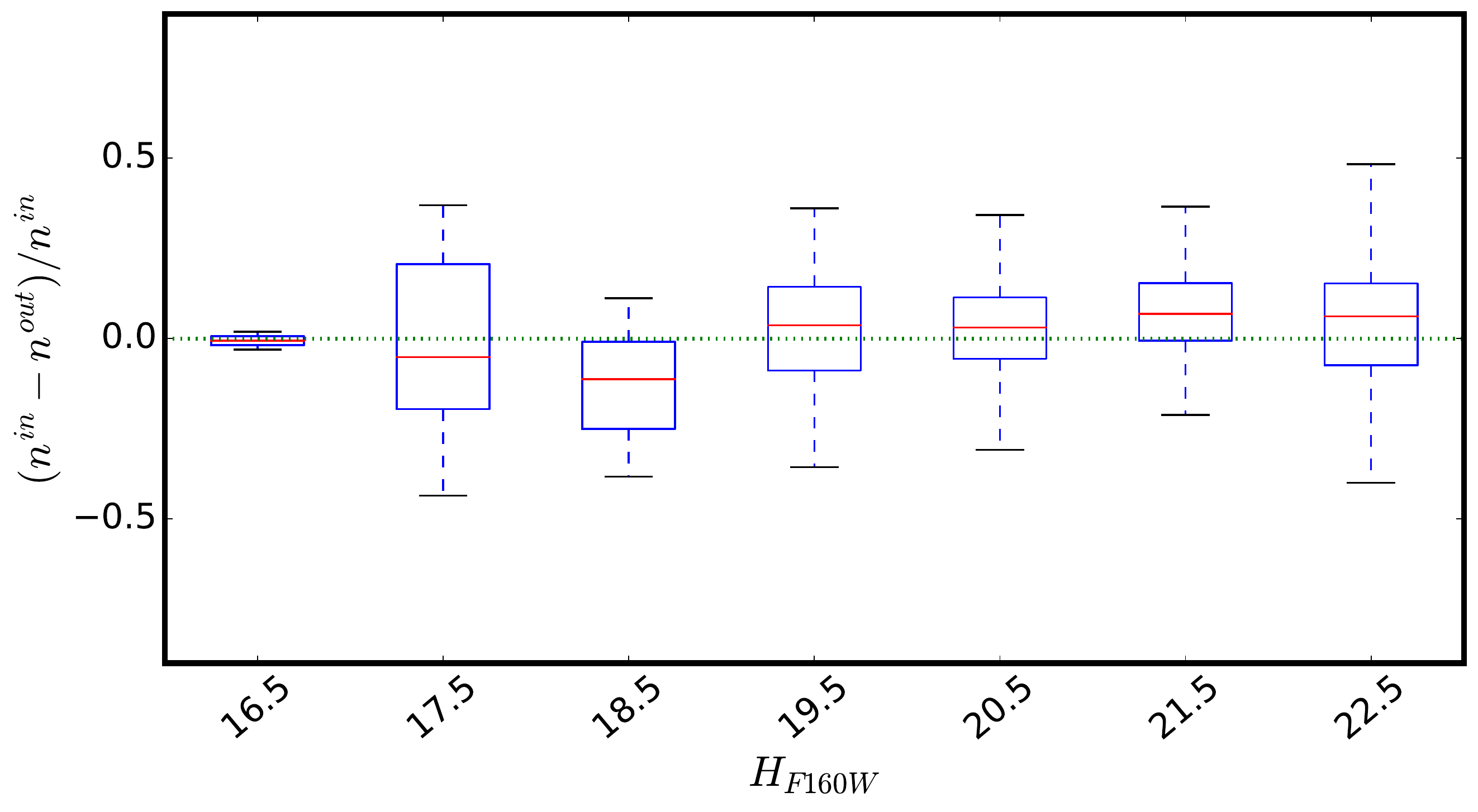} & \includegraphics[width=8.5cm]{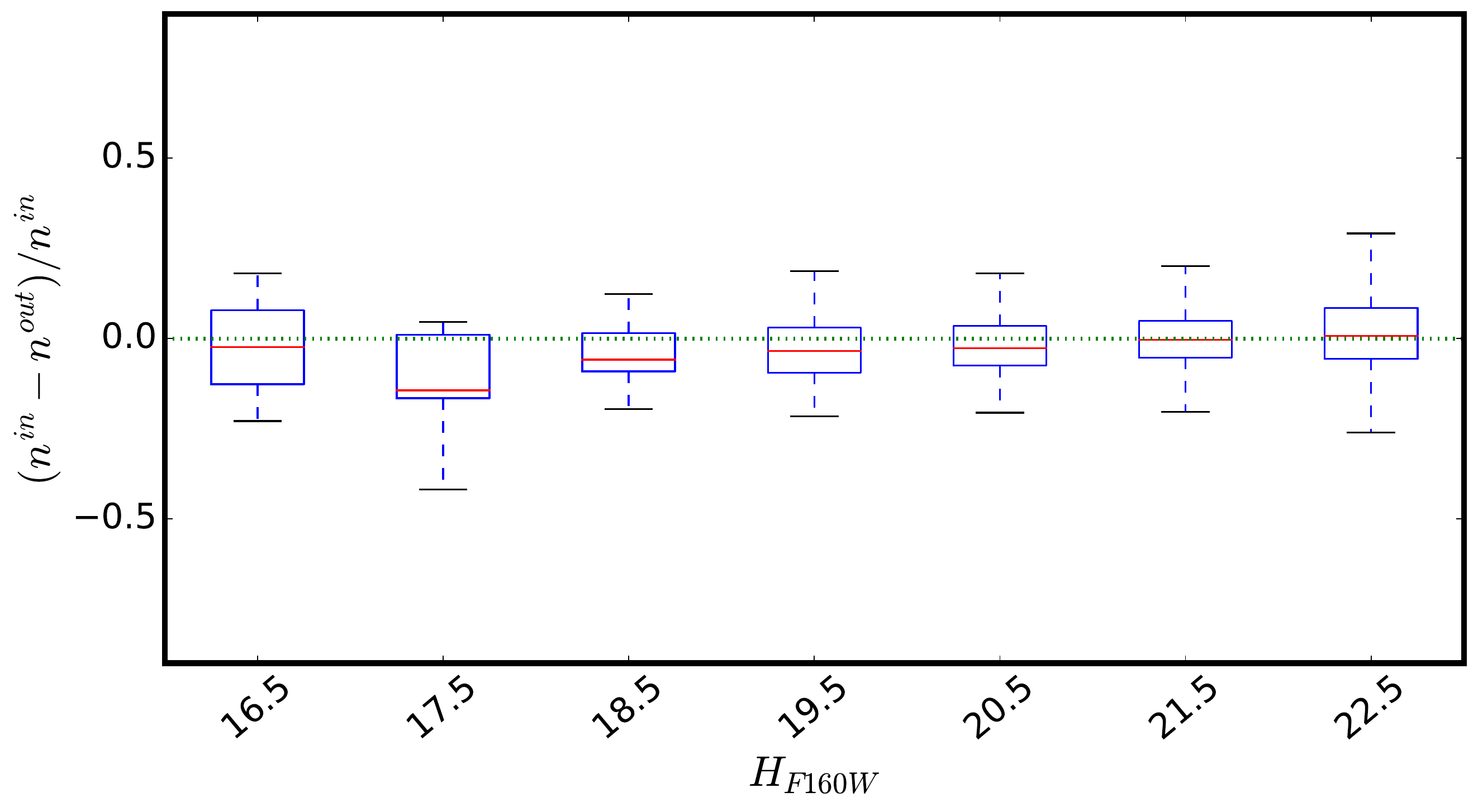} \\
\includegraphics[width=8.5cm]{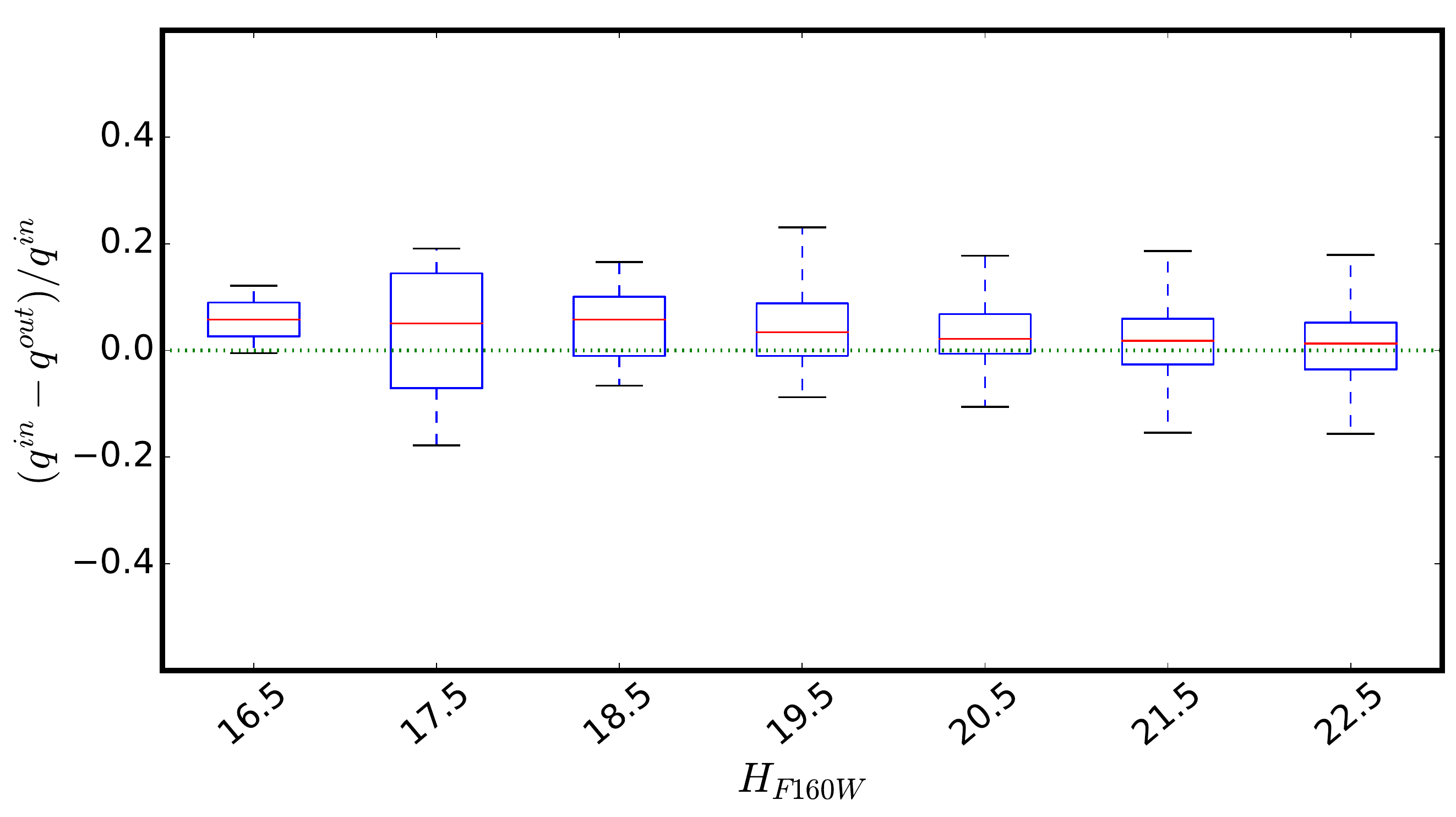} & \includegraphics[width=8.5cm]{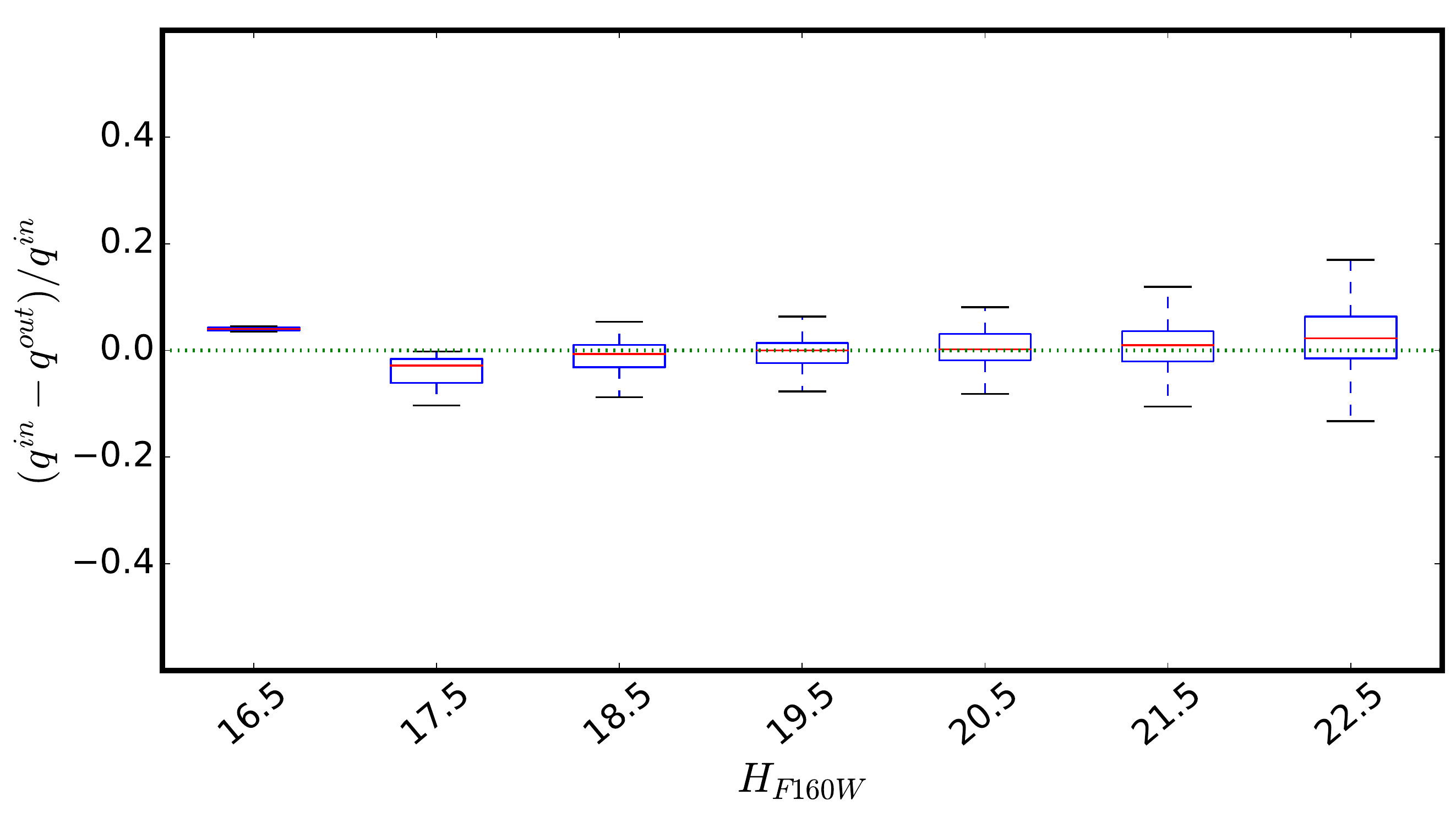} \\

\end{tabular}
\captionof{figure}{Results of profile fitting for the images of 1000 real HST/CANDELS galaxies. The left column shows the results obtained using \textit{DeepLeGATo} after domain adaptation, the right column the results obtained using two different catalogues based on the use of GALFIT. For each galaxy and for each parameter $X$ we calculated the difference between the parameter estimation given in  Van der Wel et al 2012 ($X^{in}$) and, on the left column the estimation ($X^{out}$) obtained from  \textit{DeepLeGATo}, and on the right the estimation published in Dimauro et al. 2017. The results are shown in bins of magnitude (bin width = 1 mag) and non parametrically in the form of box plots. As usually, the boxes are delimitated by the first and third interquartile of the data, while the whisker indicates the range between $\pm1.5$ the interquartile range (IQR). The red line in the box indicates the median of the data.} 
\label{PerformanceReal}
\end{table*}

\section{Conclusions and perspectives}


In this work, we present \textit{DeepLeGATo}, a convolutional neural network method designed for two-dimensional light profile fitting of one-component galaxies. We trained, validated and tested  \textit{DeepLeGATo} on an extensive set of simulations of HST/CANDELs galaxies images and we find it to be robust in terms of parameter recoverability and consistent with the results obtained with GALFIT. We find that the use of different architectures depending on the output parameter helps to improve the performance of the models, therefore \textit{DeepLeGATo} uses two different architectures. On simulated data, our method obtains more accurate results than GALFIT for the parameters (magnitude, half-light radius, S\'ersic index and axis ratio) of the structural decomposition. Moreover, it is $\sim 3000$ times faster when running on GPU, and $\sim 50$ times when run on the same CPU machine. On real data, \textit{DeepLeGATo} trained on simulations behaves similarly to GALFIT on isolated and fairly regular galaxies. Moreover, after a \textit{domain adaptation} step made with the $0.1-0.8$ per cent of the size of the training set,  \textit{DeepLeGATo} obtains results consistent with the ones presented in the \cite{vanderWel_2012} catalogue even for galaxies having several bright companions. Considering this test, we conclude that our method is able to obtain reliable results either using more complex and realistic simulations for its training, either with a \textit{domain adaptation} step made with a sample of reliable estimations.

Overall, in this work we prove that deep neural networks represent an exciting prospect for conducting large scale galaxy-decomposition, as they are capable of automated feature extraction and do not need an hand-made user-defined parameter set-up. While the accuracy of other methods greatly depends on the choice of the input set-up of parameters, initial conditions and centering. Deep learning and CNNs in particular,  have the potential to significantly cut down on the need for human visual inspection and make the galaxy decomposition a powerful tool in the era of new and future wide-field surveys such as LSST, Euclid and WFIRST.

\textit{DeepLeGATo} aims to be a first step towards the systematic use of deep learning methods for fast, accurate and precise measurements of galaxy structural parameters. Future steps are needed before this method can be implemented effectively on large photometric datasets. A fundamental issue that we plan to address is the measurement of uncertainties. Realistic error bars are crucial to confirm or rule out models when compared to data. Several solutions to this problem may be viable outside the CNN framework. For instance, calculating the residuals after subtracting away from the galaxy image a model generated with the fitted parameters. More elegant solutions having interest beyond our specific task would implement the estimation of the uncertainties within the CNN framework although this is still an unsolved issue. We tested one possible approach, consisting in fitting several slightly different images of the same galaxy. We test the self consistence of the CNN regression using different observations (or artificially disturbed images) of the galaxies, thus obtaining an empirical sample of the predictive distribution for each one of them. From the latter we infer an empirical estimator for the predictive variance (our uncertainty in the estimations). Another possible solution involves the use of Bayesian neural networks (BNNs), that are currently considered the state-of-the-art for estimating predictive uncertainty (\citealt{Gal_2016}). BNNs learn a-posterior distribution over the parameters of the neural network and use this approximate distribution as a prior to provide a complete probability distribution of the estimations. Classical deep neural networks provide only a single estimation.

As shown in the paper, in this initial test we excluded all extreme cases from our analysis, including very round or flat galaxies. Future and ongoing work includes the extension of the parameter space, and the application of CNNs on more complex cases consisting of two-component bulge-disc galaxies, implementing a variable PSF and making the algorithm more robust for dense and noisy regions of the sky. Last, we plan to render \textit{DeepLeGATo} freely available with a full set of instructions to adapt the method to different surveys and galaxy samples.


\section*{Acknowledgements}

This work is funded by the project GAMOCLASS (P.I. MHC) as part as the \emph{appel structuration de la recherche 
lanc\'e par PSL au titre de l'ann\'ee 2015}. The authors are also thankful to Google for the unrestricted gift given at the University of Santa Cruz for the project \emph{Deep learning for galaxies} which  contributed to improve this work. MHC would like to thank Sandy Faber, David Koo and Joel Primack for very useful discussions. The authors gratefully thank the referee, Arjen van der Wel, for the constructive and positive comments which help to improve the paper.

\bibliography{mybib}
\bibliographystyle{mn2e}

\appendix

\section{CNNs prediction on real data}
\label{appendix_a}

\begin{figure*}
\centering
\subfloat[Before domain adaptation (BDA)]{
  \includegraphics[width=85mm]{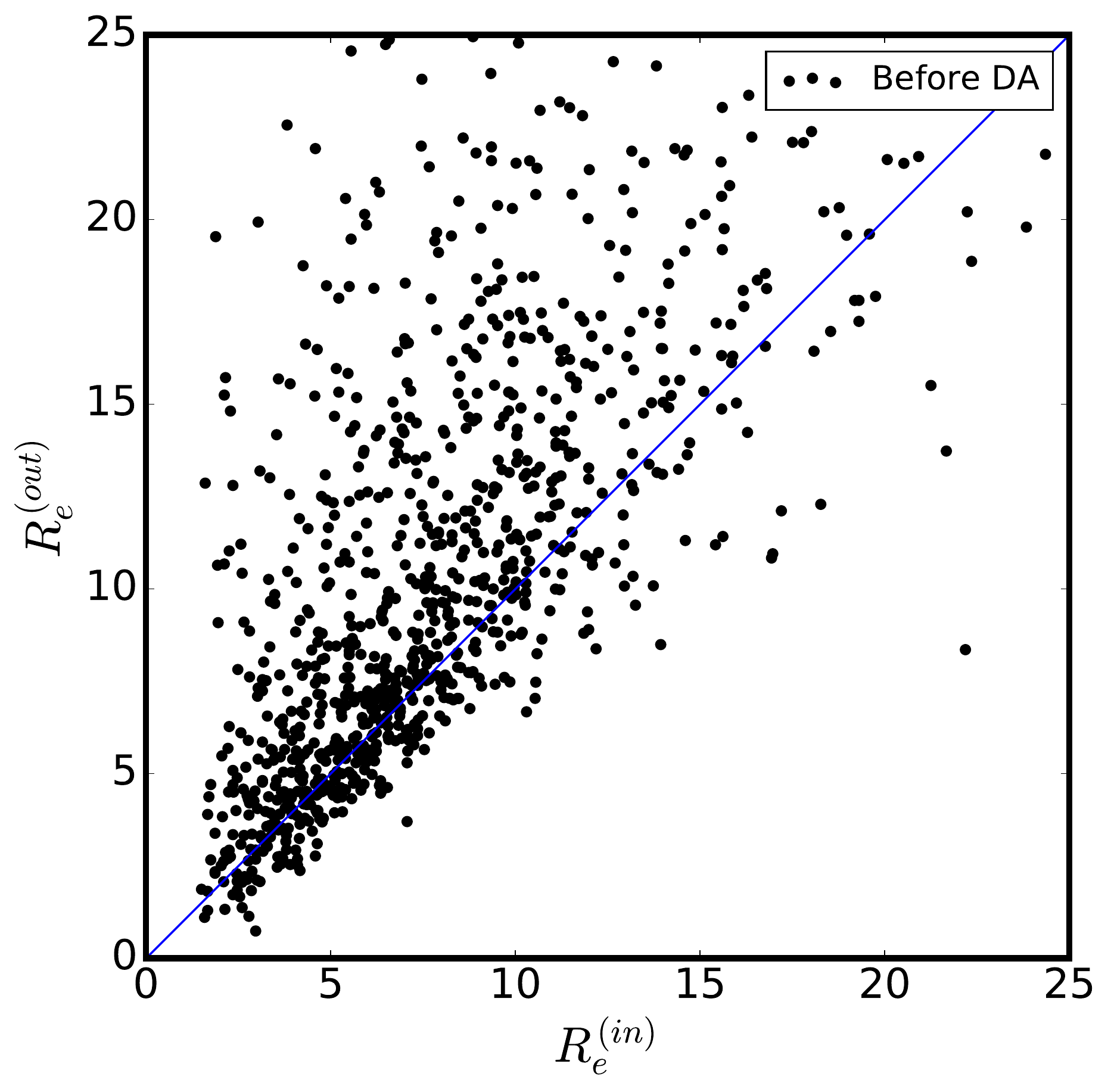}
}
\subfloat[After domain adaptation ]{
  \includegraphics[width=85mm]{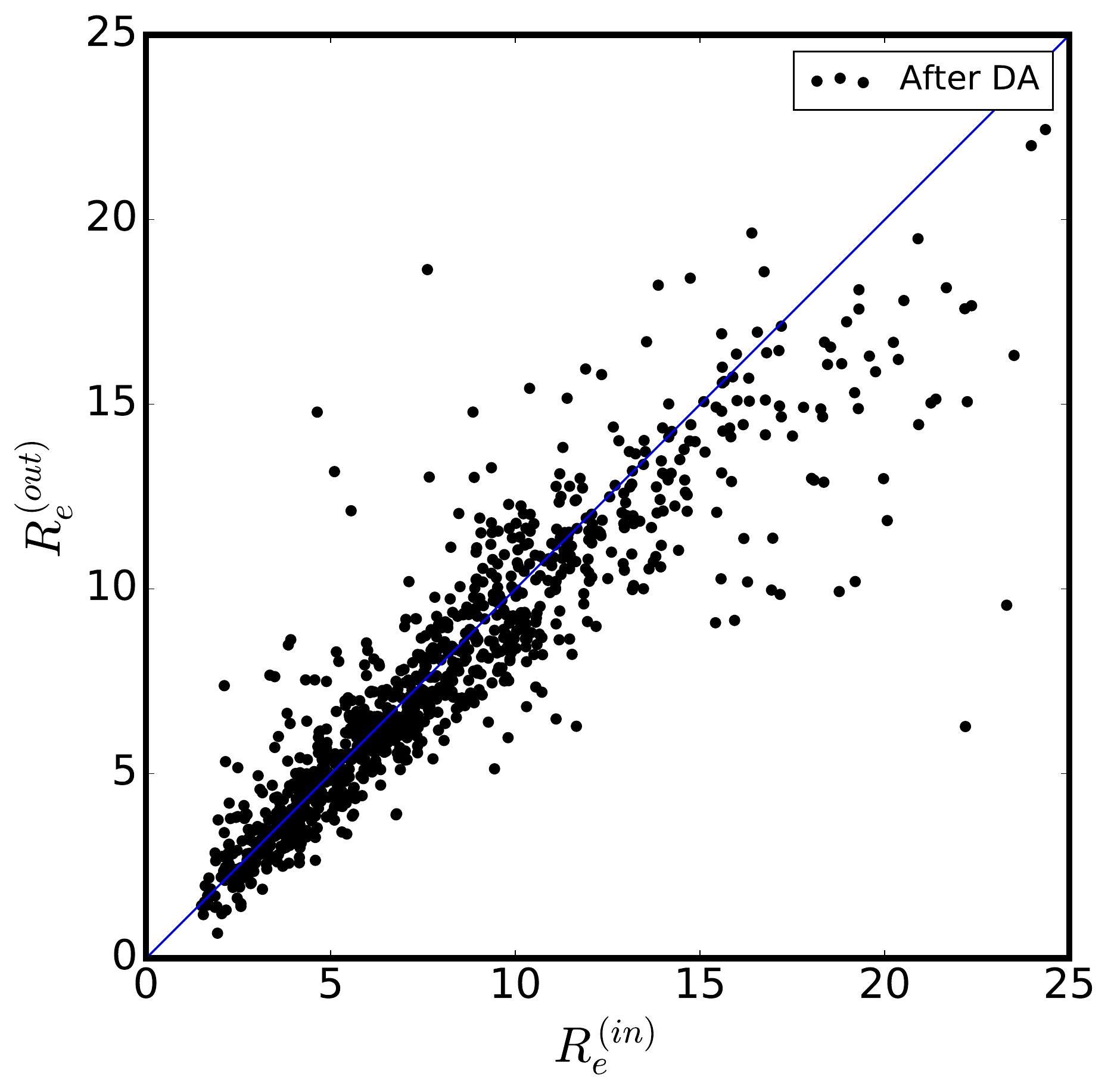}
}
\hspace{0mm}
\subfloat[BDA bright neigbours]{
  \includegraphics[width=60mm]{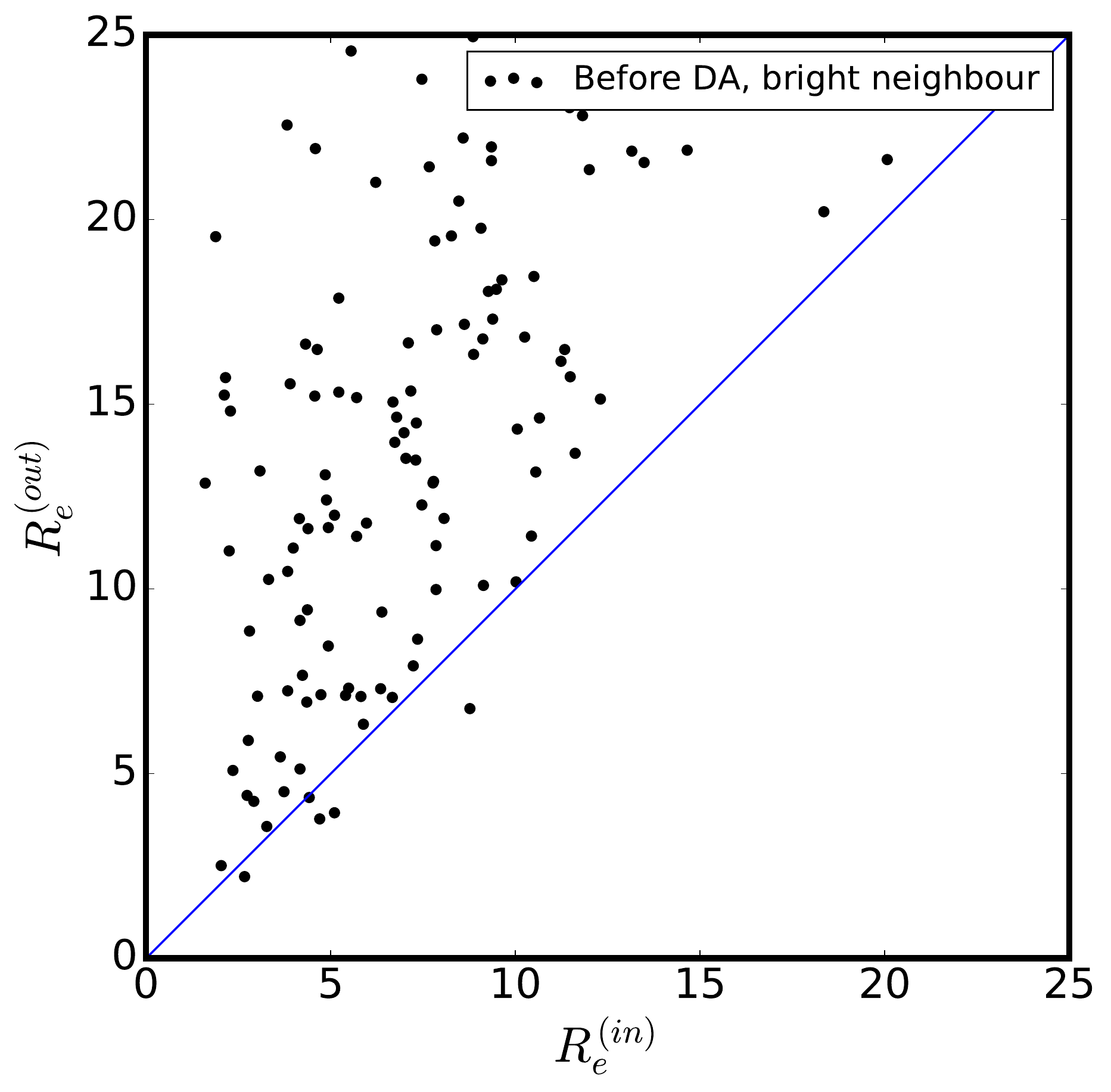}
}
\subfloat[BDA faint neigbours]{
  \includegraphics[width=60mm]{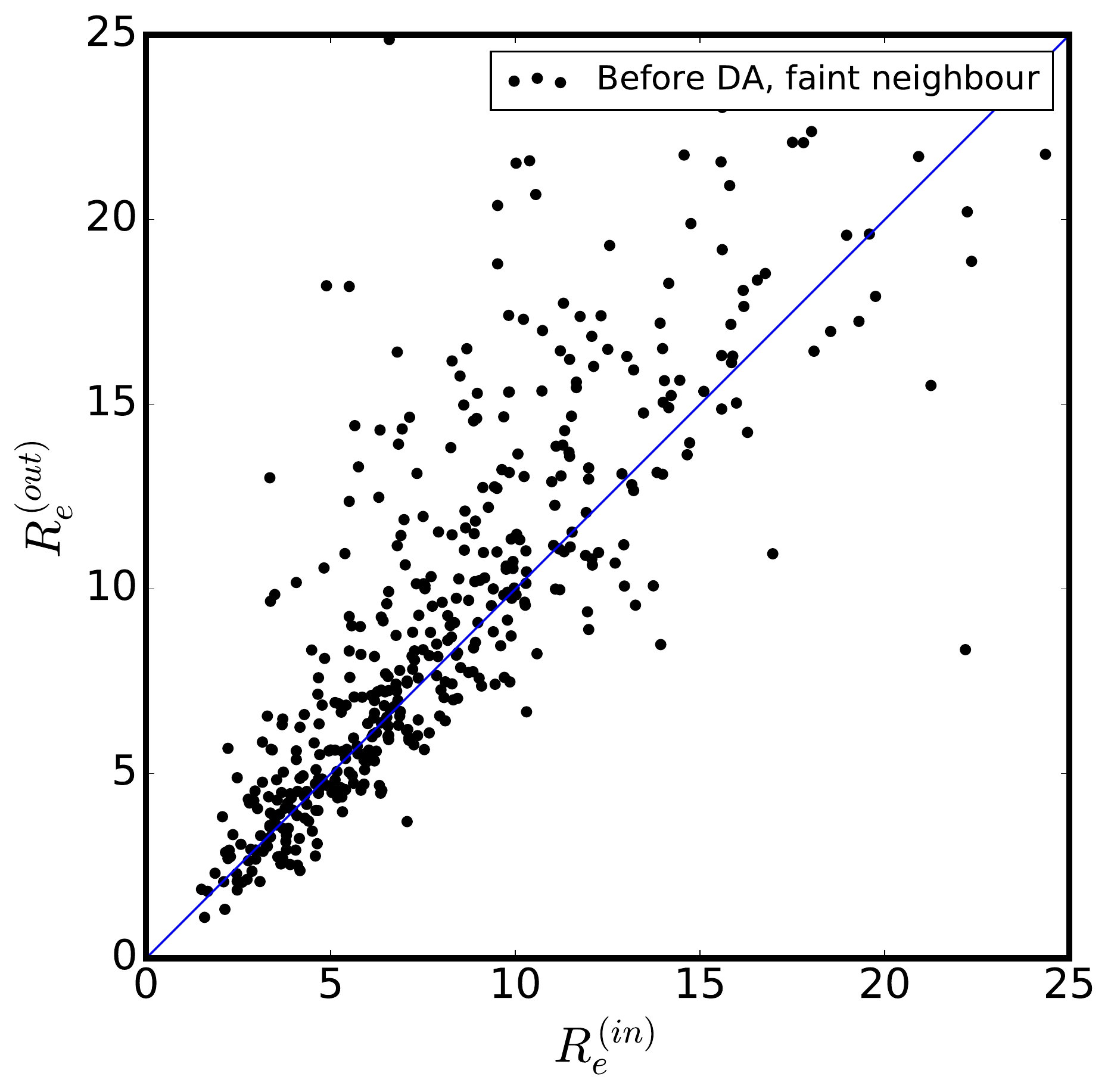}
}  
\subfloat[BDA isolated galaxies]{
  \includegraphics[width=60mm]{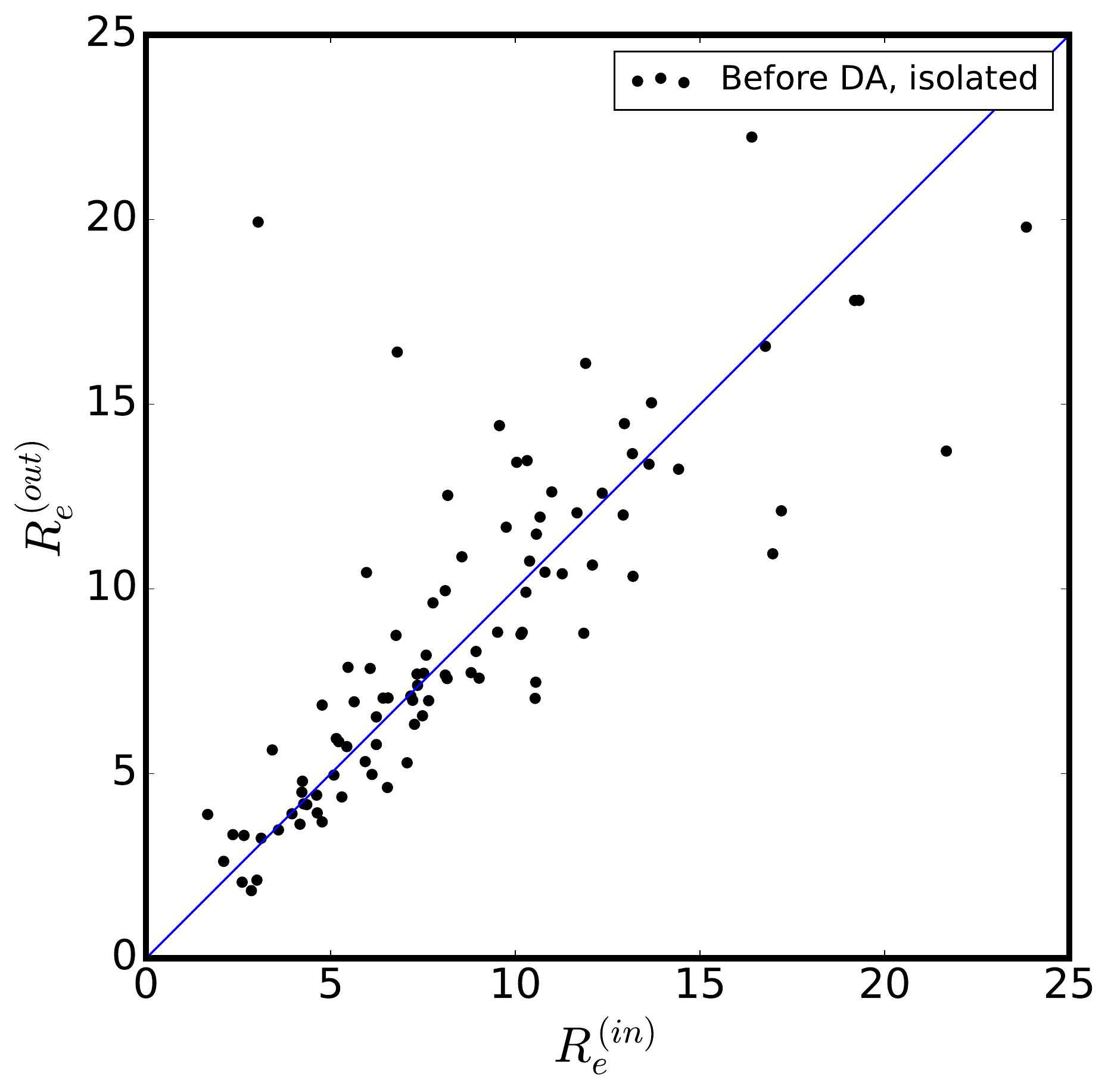}
}  

\caption{We show the results of our CNNs in fitting the half-light radius of one thousand real galaxies. On the x-axis we plot the parameter estimation given from the vanderWel et al. (2012) catalogue, while on the y-axis we give the parameter values estimated with DeepLeGATo. The upper panels show the results obtained for the whole sample before \textit{domain adaptation} step (a), and after the \textit{domain adaptation} (b). The three bottom panels show: c)  the results obtained on the 142 galaxies whose companion has at least the 50\% of their flux; c) the results for the 450 galaxies whose companion have less than the 10\% of the flux of the galaxy; e) results for the 103 isolated galaxies of our test-sample, i.e. without companion within the stamp}
\label{scatter2}
\end{figure*}

\begin{figure*}
\centering
\subfloat[Before domain adaptation (BDA)]{
  \includegraphics[width=85mm]{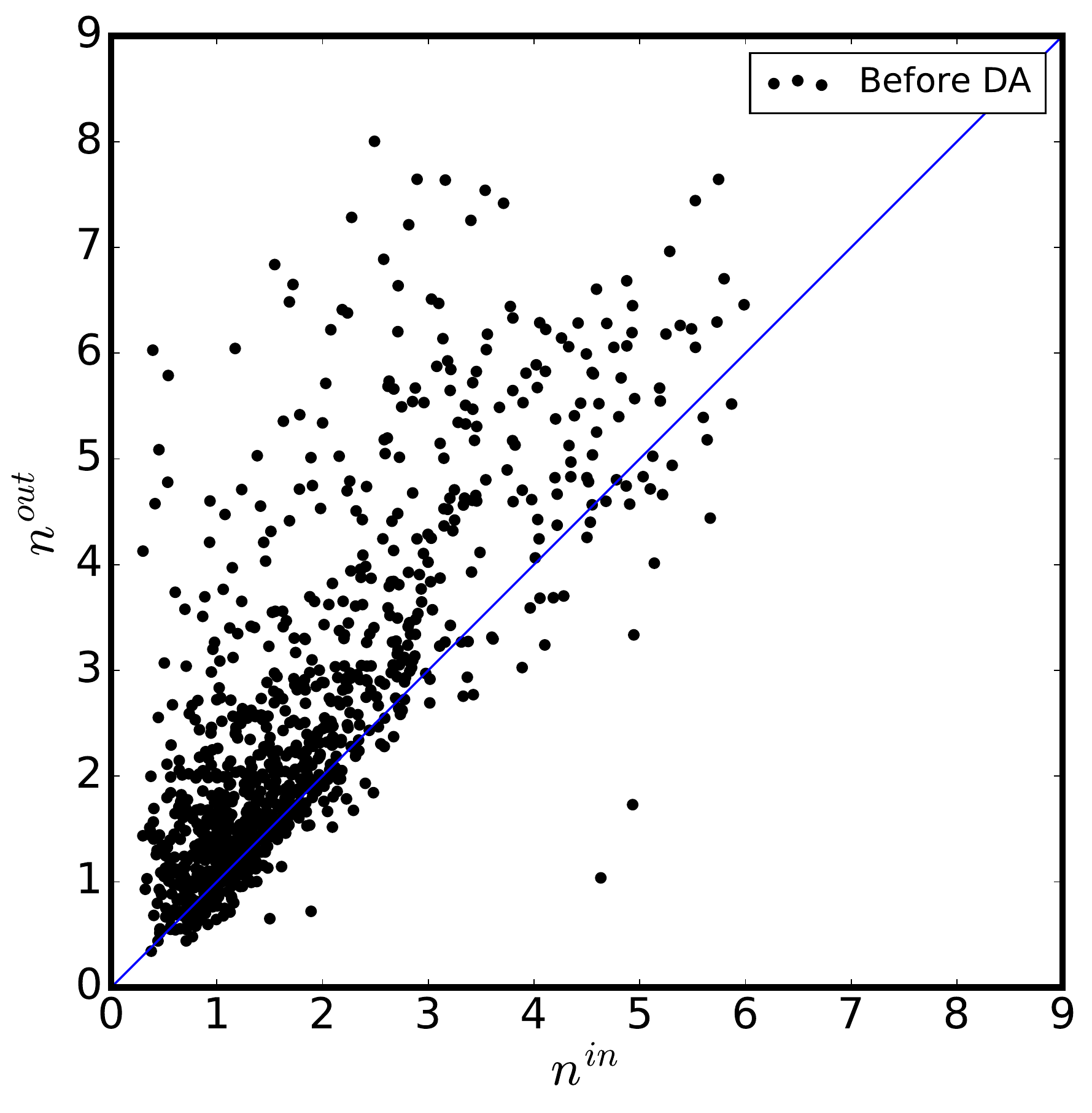}
}
\subfloat[After domain adaptation ]{
  \includegraphics[width=85mm]{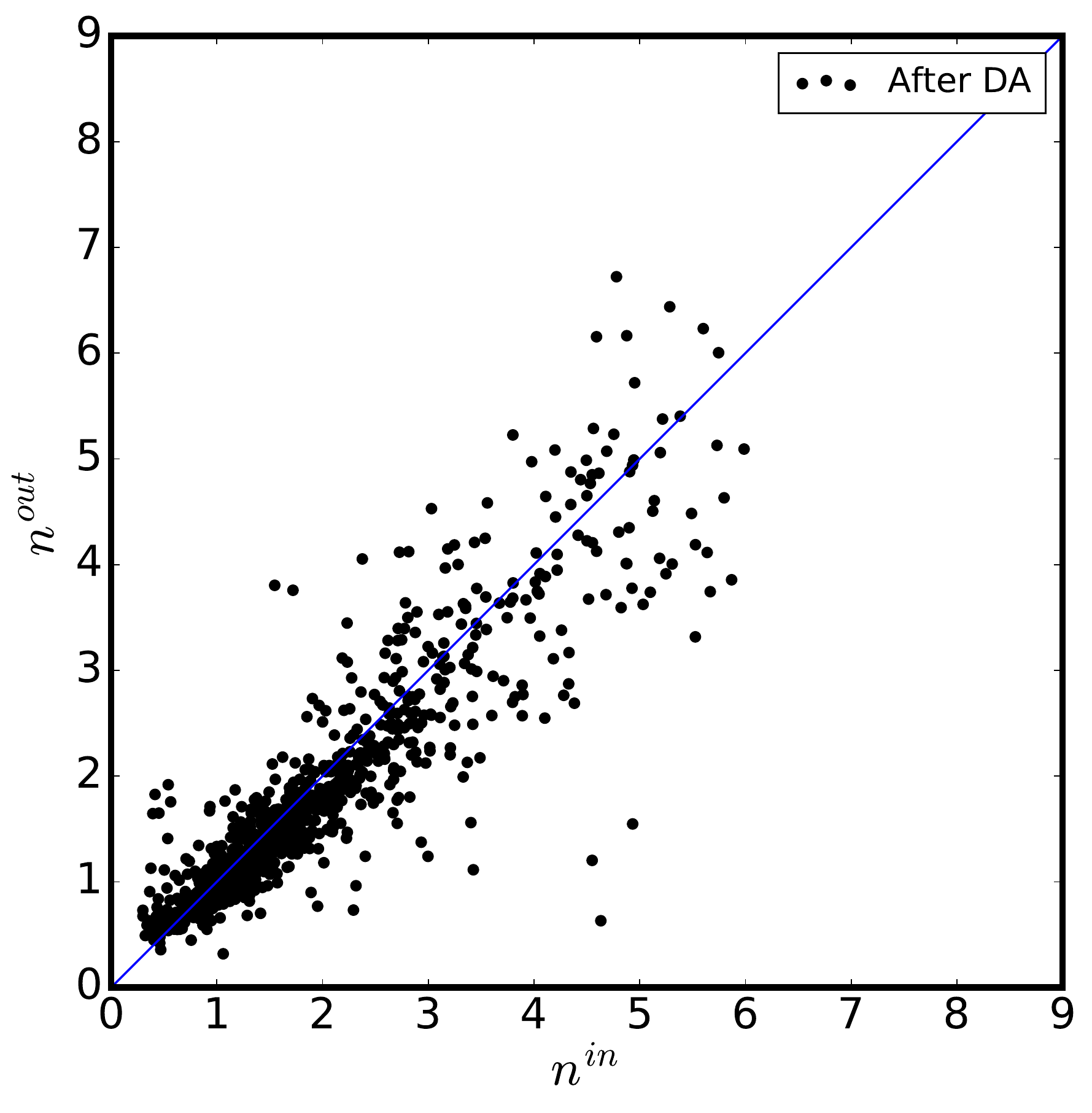}
}
\hspace{0mm}
\subfloat[BDA bright neigbours]{
  \includegraphics[width=60mm]{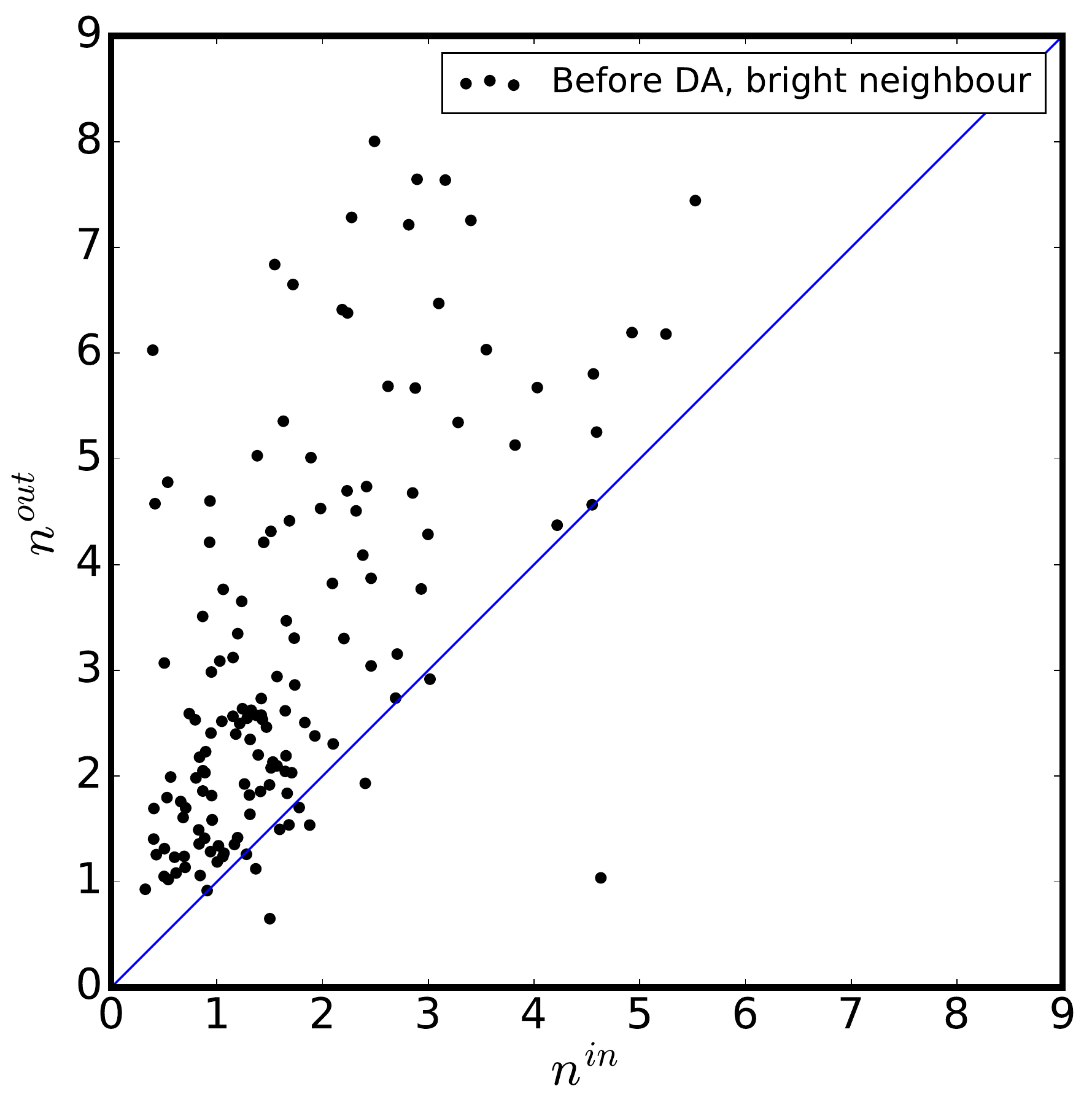}
}
\subfloat[BDA faint neigbours]{
  \includegraphics[width=60mm]{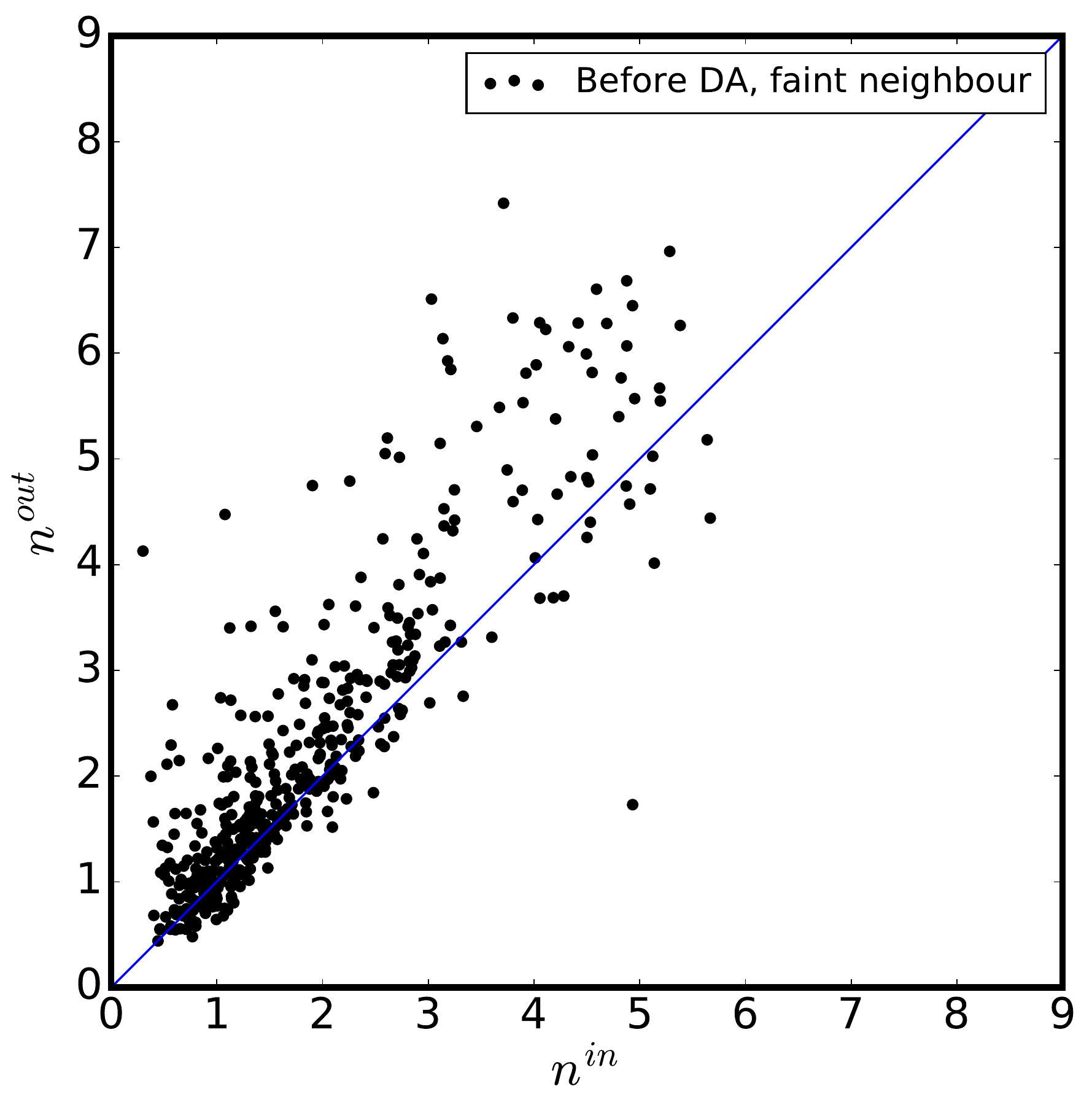}
}  
\subfloat[BDA isolated galaxies]{
  \includegraphics[width=60mm]{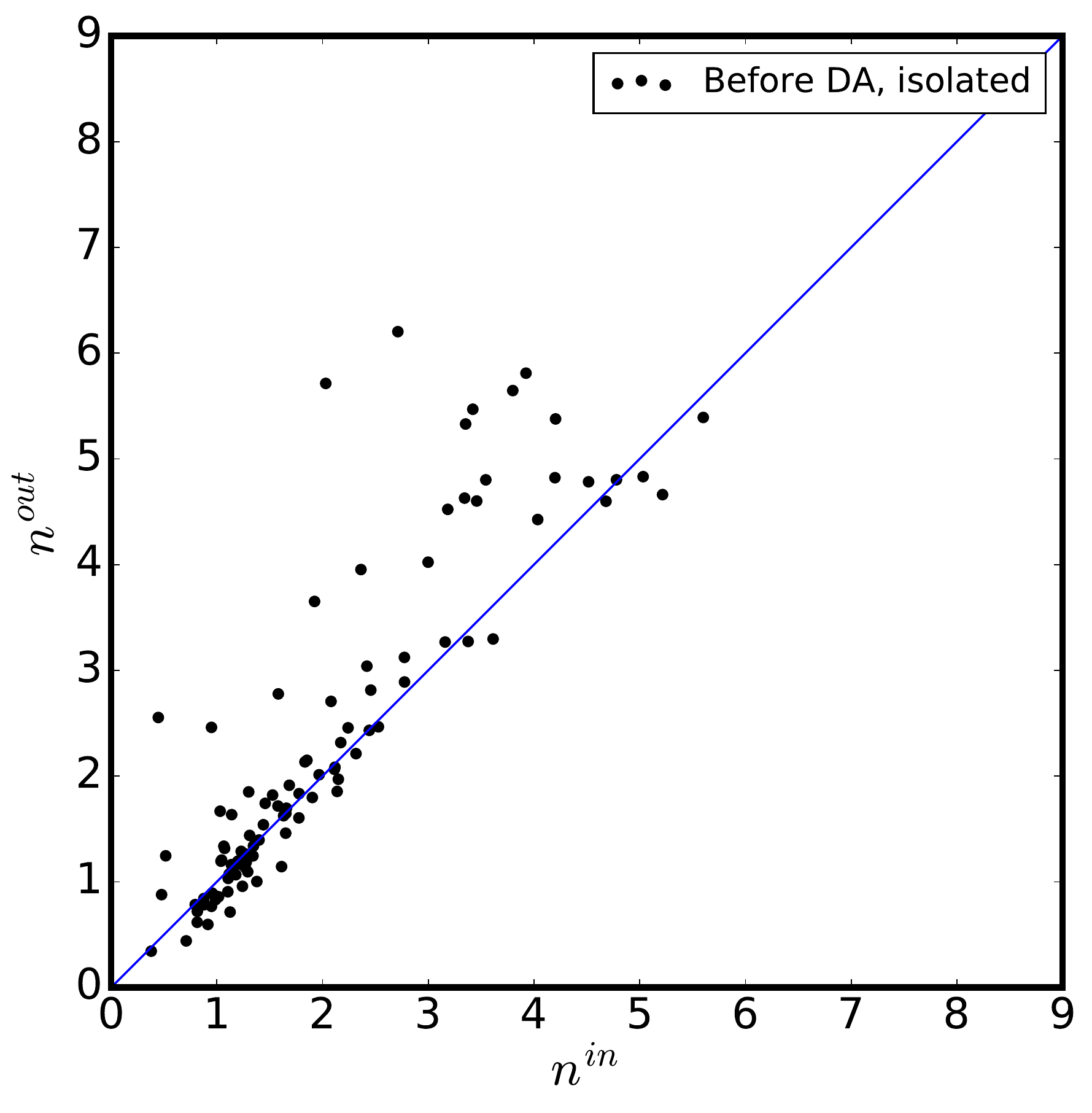}
}  

\caption{We show the results of our CNNs in fitting the S\'ersic index of one thousand real galaxies. On the x-axis we plot the parameter estimation given from the vanderWel et al. (2012) catalogue, while on the y-axis we give the parameter values estimated with DeepLeGATo. The upper panels show the results obtained for the whole sample before \textit{domain adaptation} step (a) and after the \textit{domain adaptation} (b). The three bottom panels show: c) the results obtained on the 142 galaxies whose companion has at least the 50\% of their flux; d)  the results for the 450 galaxies whose companion have less than the 10\% of the flux of the galaxy; e) the results for the 103 isolated galaxies of our test-sample, i.e. without companion within the stamp}
\label{scatter3} 
\end{figure*}

\begin{figure*}
\centering
\subfloat[Before domain adaptation (BDA)]{
  \includegraphics[width=85mm]{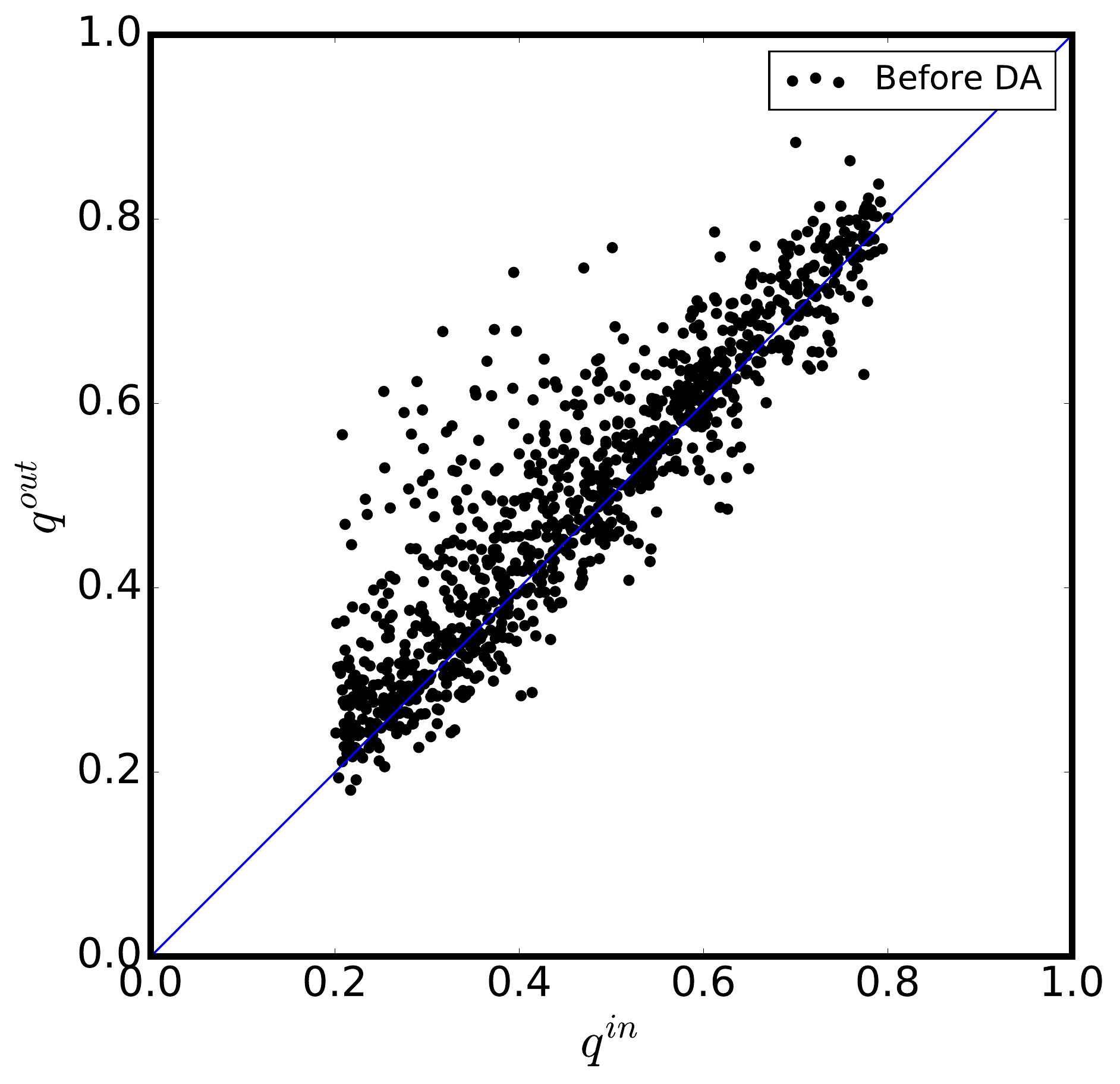}
}
\subfloat[After domain adaptation ]{
  \includegraphics[width=85mm]{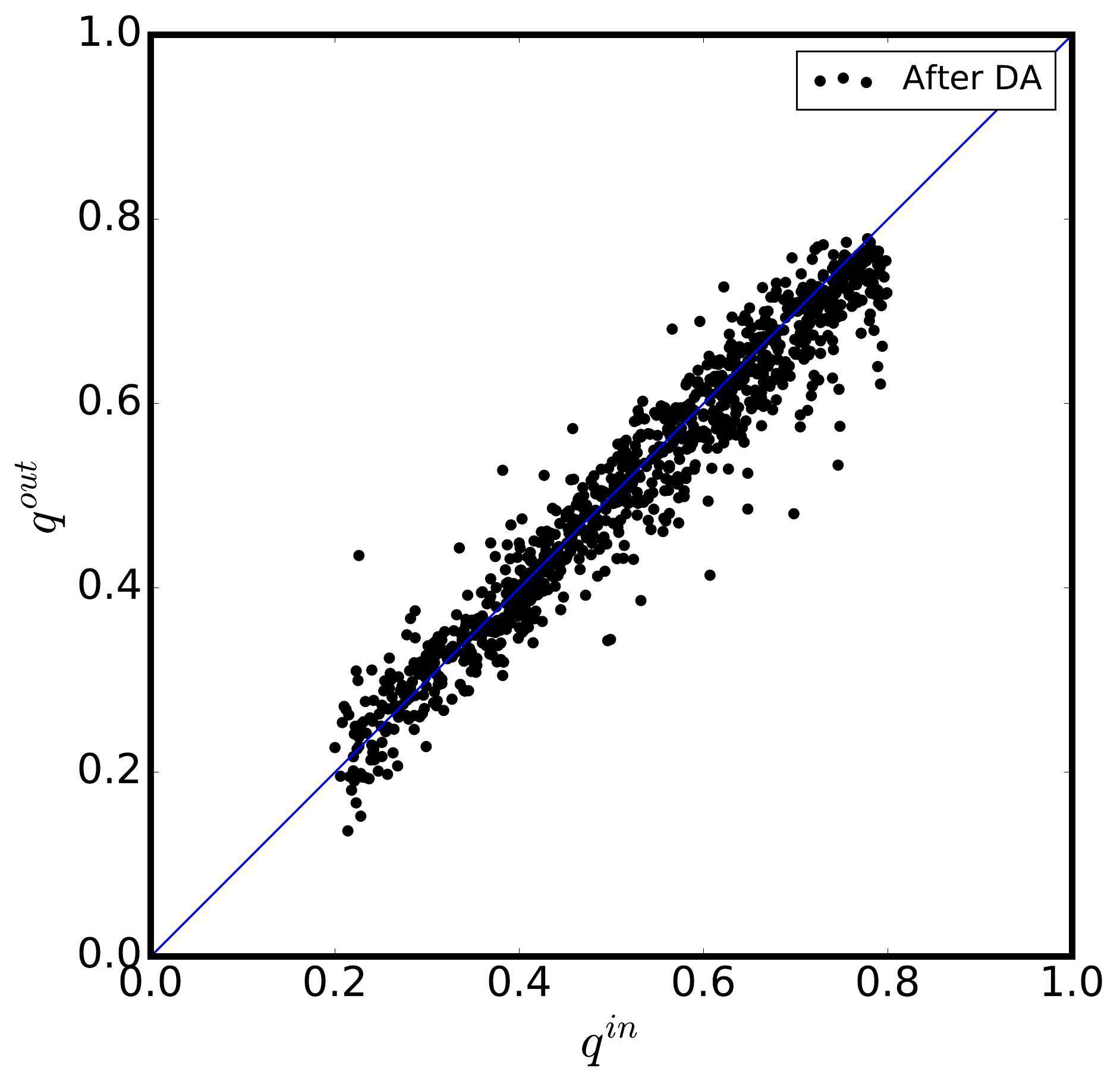}
}
\hspace{0mm}
\subfloat[BDA bright neigbours]{
  \includegraphics[width=60mm]{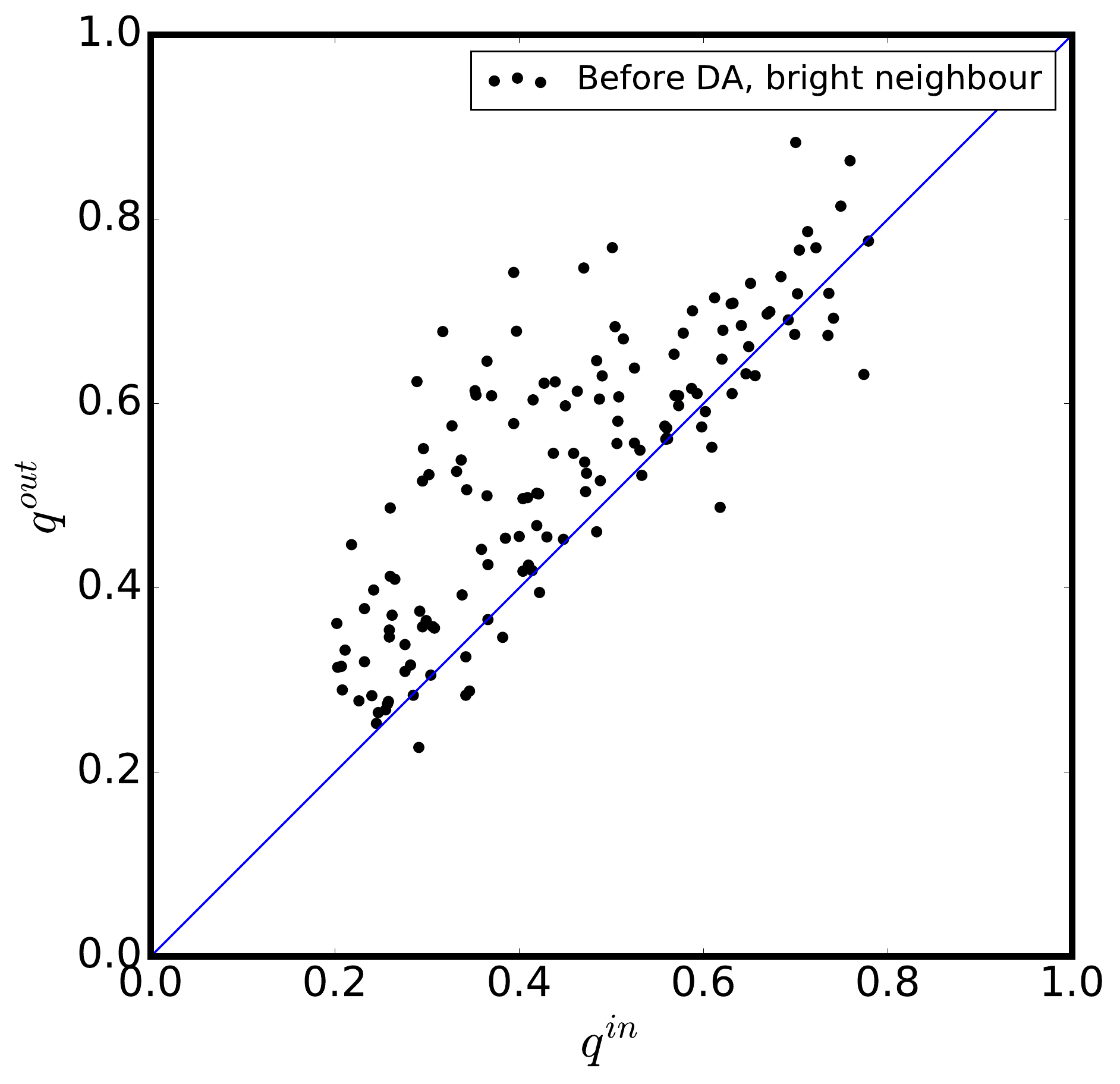}
}
\subfloat[BDA faint neigbours]{
  \includegraphics[width=60mm]{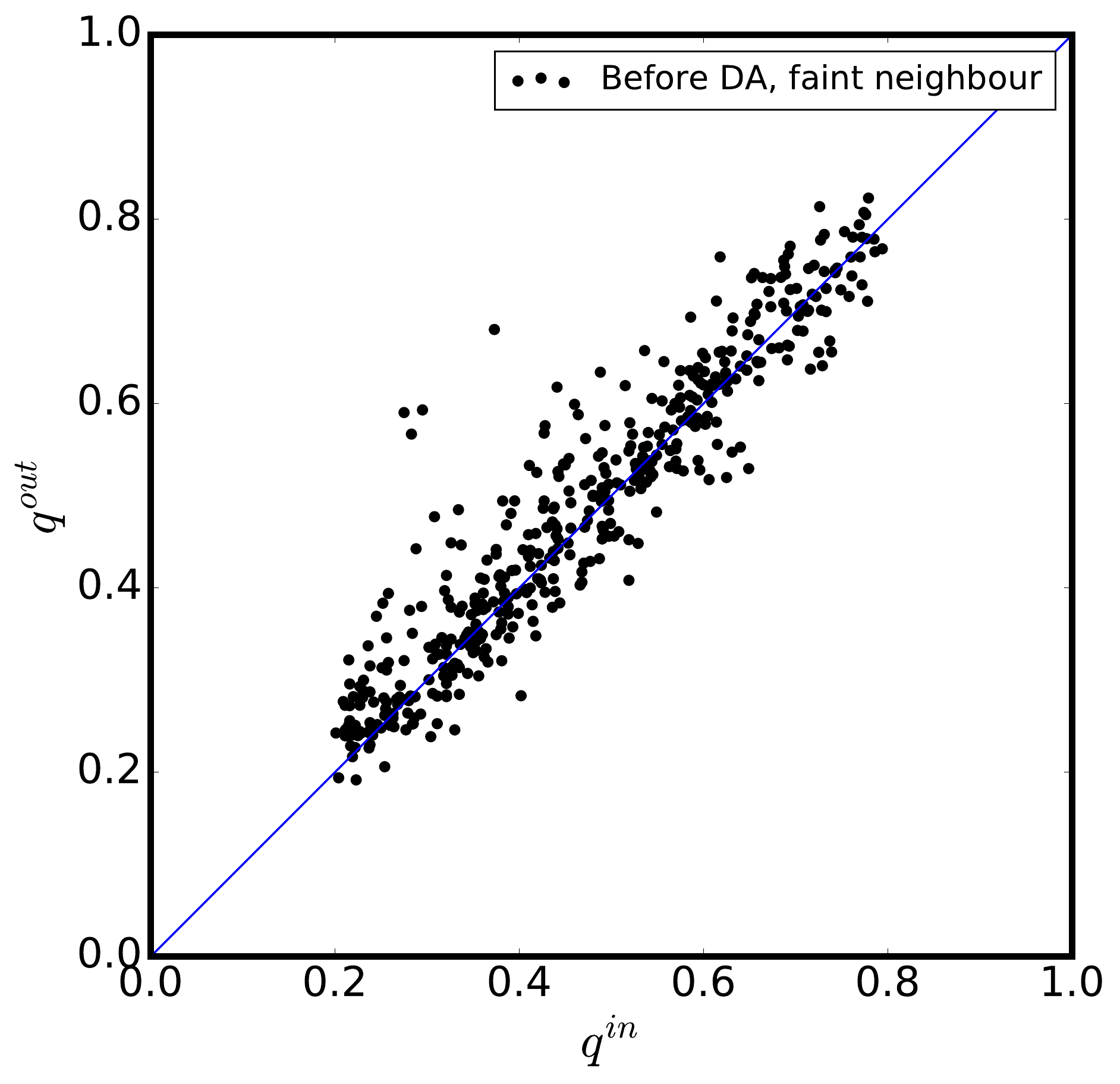}
}  
\subfloat[BDA isolated galaxies]{
  \includegraphics[width=60mm]{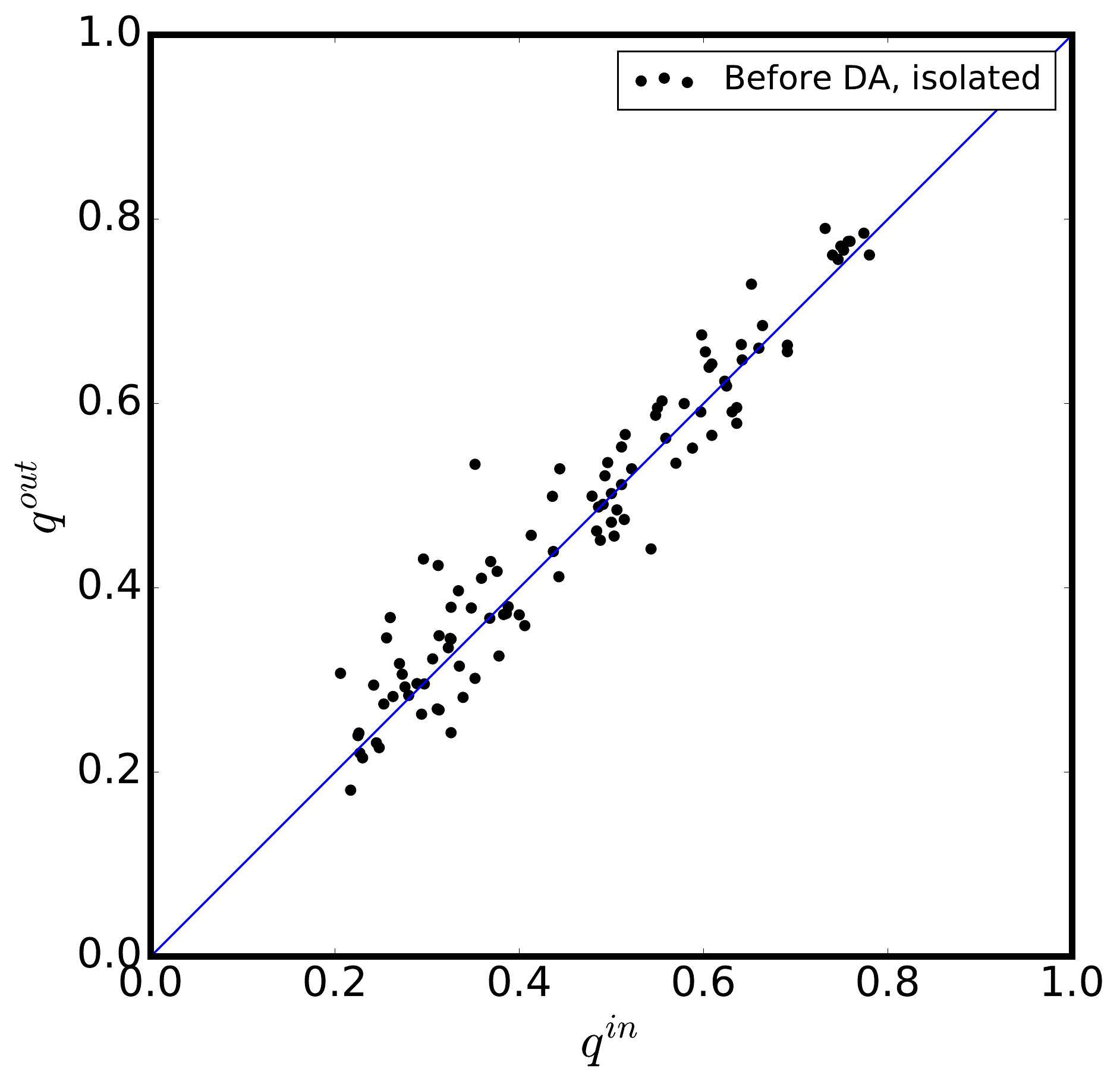}
}  

\caption{We show the results of our CNNs in fitting the axis ratio of one thousand real galaxies. On the x-axis we plot the parameter estimation given from the vanderWel et al. (2012) catalogue, while on the y-axis we give the parameter values estimated with DeepLeGATo. The upper panels show the results obtained for the whole sample before \textit{domain adaptation} step (a) and after the \textit{domain adaptation} (b). The three bottom panels show: c) the first from the left, the results obtained on the 142 galaxies whose companion has at least the 50\% of their flux; d)  the results for the 450 galaxies whose companion have less than the 10\% of the flux of the galaxy; e) the results for the 103 isolated galaxies of our test-sample, i.e. without companion within the stamp}
\label{scatter4} 
\end{figure*}

In this appendix we complete the analysis presented in section \ref{directApp},  and we show the results of the application of \textit{DeepLeGATo} to a test sample of real galaxies. In particular we show the one-to-one scatter-plot comparison of our estimations with those of the \cite{vanderWel_2012} catalogue before and after \textit{domain adaptation}. The sample fitted before \textit{domain adaptation} is presented the the Figures \ref{scatter2}, \ref{scatter3} and \ref{scatter4}.

\end{document}